\documentclass{aa}  

\usepackage{graphicx}
\usepackage{txfonts}
\begin{document}

   \title{Interstellar detection and chemical modeling of \textit{iso}-propanol\\ and its \textit{normal} isomer}

   \subtitle{}

   \author{A. Belloche \inst{1}
          \and R.~T. Garrod \inst{2}
          \and O. Zingsheim \inst{3}
          \and H.~S.~P.~M{\"u}ller \inst{3}
          \and K.~M. Menten \inst{1}
          }

   \institute{Max-Planck-Institut f\"{u}r Radioastronomie, 
              Auf dem H\"{u}gel 69, 53121 Bonn, Germany\\ 
              \email{belloche@mpifr-bonn.mpg.de}
         \and Departments of Chemistry and Astronomy, University of Virginia, 
              Charlottesville, VA 22904, USA        
         \and I. Physikalisches Institut, Universit{\"a}t zu K{\"o}ln, 
              Z{\"u}lpicher Str. 77, 50937 K{\"o}ln, Germany
             }

   \date{Received March 17, 2022; accepted April 20, 2022}

 
  \abstract
   {The detection of a branched alkyl molecule in the high-mass star forming 
protocluster Sgr~B2(N) permitted by the advent of the Atacama Large 
Millimeter/submillimeter Array (ALMA) revealed a new dimension of interstellar 
chemistry. Astrochemical simulations subsequently predicted that beyond a 
certain degree of molecular complexity, branched molecules could even dominate 
over their straight-chain isomers.}
   {More generally, we aim at probing further the presence in the interstellar 
medium of complex organic molecules with the capacity to exhibit both a 
\textit{normal} and \textit{iso} form, via the attachment of a functional 
group to either a primary or secondary carbon atom.}
   {We used the imaging spectral line survey ReMoCA performed with ALMA at 
high angular resolution and the results of a recent spectroscopic study of 
propanol to search for the \textit{iso} and \textit{normal} isomers of this 
molecule in the hot molecular core Sgr~B2(N2). We analyzed the interferometric 
spectra under the assumption of local thermodynamical equilibrium. We expanded 
the network of the astrochemical model MAGICKAL to explore the formation 
routes of propanol and put the observational results in a broader 
astrochemical context.}
   {We report the first interstellar detection of \textit{iso}-propanol, 
\textit{i}-C$_3$H$_7$OH, toward a position of Sgr~B2(N2) that shows narrow 
linewidths. We also report the first secure detection of the \textit{normal} 
isomer of propanol, \textit{n}-C$_3$H$_7$OH, in a hot core. 
\textit{iso}-Propanol is found to be nearly as abundant as 
\textit{normal}-propanol, with an abundance ratio of 0.6 similar to the ratio 
of 0.4 that we obtained previously for \textit{iso}- and 
\textit{normal}-propyl cyanide in Sgr~B2(N2) at lower angular resolution with 
our previous ALMA survey, EMoCA. The observational results are in good 
agreement with the outcomes of our astrochemical models, which indicate that 
OH-radical addition to propylene in dust-grain ice mantles, driven by water 
photodissociation, can produce appropriate quantities of \textit{normal}- and 
\textit{iso}-propanol. The \textit{normal}-to-\textit{iso} ratio in Sgr~B2(N2) 
may be a direct inheritance of the branching ratio of this reaction process.}
   {The detection of \textit{normal}- and \textit{iso}-propanol and their 
ratio indicate that the modest preference toward the \textit{normal} form of 
propyl cyanide determined previously may be a more general feature among 
similarly sized interstellar molecules. Detecting other pairs of interstellar 
organic molecules with a functional group attached either to a primary or 
secondary carbon may help in pinning down the processes that dominate in 
setting their \textit{normal}-to-\textit{iso} ratios. Butanol and its isomers 
would be the next obvious candidates in the alcohol family, but their 
detection in hot cores will be challenging.}


   \maketitle
%

\section{Introduction}
\label{s:introduction}

The detection of a branched alkyl molecule, \textit{iso}-propyl cyanide 
(\textit{i}-C$_3$H$_7$CN), in the interstellar medium with the Atacama Large 
Millimeter/submillimeter Array (ALMA) opened a new window into the chemistry
that takes place in star forming regions \citep[][]{Belloche14}. The 
production of such a branched molecule appears to require the addition of a 
functional group to a nonterminal carbon in the chain forming its backbone.
This detection was made in the frame of our earlier imaging spectral line
survey called Exploring Molecular Complexity with ALMA (EMoCA) that targeted
the high-mass star forming protocluster Sagittarius~(Sgr)~B2(N) 
\citep[][]{Belloche16}. The analysis of the data surprisingly revealed that 
\textit{iso}-propyl cyanide is nearly as abundant as its straight-chain 
isomer, \textit{normal}-propyl cyanide (\textit{n}-C$_3$H$_7$CN), which was 
detected earlier with the IRAM 30~m telescope \citep[][]{Belloche09}, with an 
abundance ratio between the two isomers of 0.4, determined based on our EMoCA
data \citep[][]{Belloche14}. These findings suggested that branched 
carbon-chain molecules may be generally abundant in the interstellar medium.

Since the discovery of \textit{iso}-propyl cyanide, many new complex organic
molecules (COMs), which are carbon-bearing molecules containing at least six 
atoms per definition \citep[][]{Herbst09}, have been reported in the 
interstellar medium, including two polycyclic aromatic hydrocarbons 
\citep[][]{McGuire21}, but no other branched molecule has been identified. The 
astrochemical model MAGICKAL that we employed to interpret the observational 
results obtained for propyl cyanide suggested that for the next, more complex 
member of the alkyl cyanide family, butyl cyanide (C$_4$H$_9$CN), the branched 
isomers should even dominate over the straight chain form \citep[][]{Garrod17}. 
This was one of motivations for us to perform a new imaging spectral line 
survey of Sgr~B2(N) with ALMA at higher angular resolution and with a higher 
sensitivity compared to EMoCA. One of the first results of this new survey, 
called Re-exploring Molecular Complexity with ALMA (ReMoCA), was the first 
interstellar detection of urea, NH$_2$C(O)NH$_2$, and the confirmation of the 
interstellar detection of N-methylformamide, CH$_3$NHCHO 
\citep[][]{Belloche19}, which had initially been tentatively detected with 
EMoCA \citep[][]{Belloche17}.

While we have not been able to identify the isomers of butyl cyanide in the
ReMoCA survey so far, we have systematically searched for all COMs that have
been spectroscopically characterized in the laboratory and for which we had 
access to spectroscopic predictions in electronic format, in particular 
through spectroscopic databases such as the Cologne Database for Molecular 
Spectroscopy\footnote{https://cdms.astro.uni-koeln.de/} 
\citep[CDMS,][]{Mueller05}. Among these COMs, the alkanol family is of 
particular interest. Methanol, CH$_3$OH, and ethanol, C$_2$H$_5$OH, have long
been known to exist in the interstellar medium. Both were first detected
toward Sgr~B2 at low angular resolution \citep[][]{Ball70,Zuckerman75}. Our 
detailed study of alkanols with the EMoCA survey at $1.6\arcsec$ resolution 
revealed that ethanol is 20 times less abundant than methanol in the secondary 
hot molecular core of Sgr~B2(N), called Sgr~B2(N2) \citep[][]{Mueller16}. 
However, the EMoCA survey did not allow us to detect the next, more complex 
member of the alkanol family, propanol (C$_3$H$_7$OH), neither in its 
\textit{normal} form, nor in its \textit{iso} form. We found that 
\textit{normal}-propanol (\textit{n}-C$_3$H$_7$OH), also called propan-1-ol, 
and \textit{iso}-propanol (\textit{i}-C$_3$H$_7$OH), also called propan-2-ol, 
are at least 8 and 22 times less abundant than ethanol in Sgr~B2(N2) at the 
scale traced with EMoCA.

Here we present the search for both \textit{normal}- and \textit{iso}-propanol
toward Sgr~B2(N2) in the ReMoCA survey. \textit{iso}-Propanol is not a branched
alkyl molecule (it contains only three carbon atoms, which can only form a 
straight chain), but there are similarities between propanol and propyl cyanide
in the sense that both molecules have a functional group (-OH or -CN) that can
be attached to a terminal or middle atom of the carbon backbone. This motivates
a comparison of these two families of COMs. The article is structured in the
following way. Section~\ref{s:observations} describes the observational setup, 
the method used to analyze the interferometric spectra, and the origin of the 
spectroscopic predictions employed to identify the detected lines and perform 
the radiative transfer calculations. Section~\ref{s:results} explains how we 
selected the position of the emission in the interferometric map for which the
spectra were extracted and reports the detection of 
both \textit{iso}- and \textit{normal}-propanol, along with the results we 
obtained for methanol and ethanol. Section~\ref{s:chemistry} presents the 
results of our astrochemical model with an expanded network that includes 
propanol. We discuss the results in Sect.~\ref{s:discussion} and lay out our 
conclusions in Sect.~\ref{s:conclusions}.

\section{Observations and radiative transfer modeling}
\label{s:observations}

\subsection{ALMA observations}
\label{ss:alma}

The ReMoCA survey was performed toward Sgr~B2(N) with ALMA in its cycle 4. 
Details about the observations and data reduction were published in 
\citet{Belloche19}. We summarize here only the main characteristics of the 
survey. The phase center was set at the equatorial position 
($\alpha, \delta$)$_{\rm J2000}$= 
($17^{\rm h}47^{\rm m}19{\fs}87, -28^\circ22'16{\farcs}0$). This position is 
located halfway between the two hot molecular cores Sgr~B2(N1) and Sgr~B2(N2). 
The frequency range from 84.1~GHz to 114.4~GHz was fully covered at a spectral 
resolution of 488~kHz (1.7 to 1.3~km~s$^{-1}$) with five frequency tunings 
labeled S1--S5 in increasing frequency order. Each spectral setup consisted of 
four spectral windows labeled W0--W3, with W0--W1 covering the lower sideband 
and W2--W3 covering the upper sideband. The angular resolution ranges from 
$\sim$0.3$\arcsec$ to $\sim$0.8$\arcsec$ (half-power beam width of synthesized 
beam, HPBW) with a median 
value of 0.6$\arcsec$ that corresponds to $\sim$4900~au at the distance of 
Sgr~B2 \citep[8.2~kpc,][]{Reid19}. The survey achieved a sensitivity per 
spectral channel between 0.35~mJy~beam$^{-1}$ and 1.1~mJy~beam$^{-1}$ (rms) 
depending on the setup, with a median value of 0.8~mJy~beam$^{-1}$. 
This corresponds to a an rms brightness temperature noise level of 0.27~K for 
a 0.6$\arcsec$ HPBW at a frequency of 100~GHz. These 
sensitivities were measured with the program \textit{go noise}, which is part 
of the GREG software in the GILDAS 
package\footnote{See http://www.iram.fr/IRAMFR/GILDAS.} and fits the 
distribution of pixel intensities in each channel map. In this work, we used an 
improved version of the splitting of the line and continuum emission as we 
described in \citet{Melosso20}.

\subsection{Radiative transfer modeling}
\label{ss:weeds}

The observed spectra were modeled assuming local thermodynamic equilibrium 
(LTE) with the astronomical software Weeds \citep[][]{Maret11} which is part
of the GILDAS package. The LTE assumption is justified by the high densities 
of the regions where hot-core emission is detected in Sgr~B2(N) 
\citep[$>1 \times 10^{7}$~cm$^{-3}$, see][]{Bonfand19}. Weeds computes the
radiative transfer by accounting for the line optical depth and the finite 
angular resolution of the observations. We derived a best-fit synthetic 
spectrum for each molecule separately, and then added the contributions of all 
identified molecules together. We used a set of five parameters to model the 
contribution of each species: size of the emitting region ($\theta_{\rm s}$), 
column density ($N$), temperature ($T_{\rm rot}$), linewidth ($\Delta V$), and 
velocity offset ($V_{\rm off}$) with respect to the assumed systemic velocity of 
the source.

\subsection{Spectroscopy}
\label{ss:spectro}

For the radiative transfer calculations of the molecules analyzed in 
Sect.~\ref{s:results}, we used spectroscopic predictions that we retrieved 
from the CDMS or that some of us produced in a companion article 
\citep[][]{Zingsheim22}. We used version 3 of the CDMS entry 32504 of methanol 
which is largely based on the study of \citet{Xu08}. Additional data of 
methanol in the range of our survey were taken from \citet{Lees68}, 
\citet{Pickett81}, \citet{Sastry84}, \citet{Herbst84}, \citet{Anderson90}, 
and \citet{Mueller04}. The partition function includes energies up to 
$\varv_{\rm t} = 3$, which should be appropriate up to $\sim$200~K and still 
quite good at 300~K.

For ethanol, we used version 1 of the CDMS entry 46524 that is mainly based on
\citet{Pearson08}, with modifications provided by \citet{Mueller16}. These 
modifications were motivated by the EMoCA survey and resolve severe intensity 
issues. Additional laboratory data to \citet{Pearson08} were contributed by 
\citet{Pearson95,Pearson96,Pearson97}. Contributions from ethanol in excited 
vibrational states were evaluated from the fundamental vibrations in 
\citet{Durig11} employing the harmonic oscillator approximation. This 
approximation is usually the best in the absence of detailed information on 
the energies of overtone and combination states.

We used the CDMS entries 60518 and 60519 (both version 1) for the 
\textit{gauche} and \textit{anti} conformers of \textit{iso}-propanol, 
respectively. Both of them are largely based on \citet{Maeda06b}. 
Contributions of excited vibrational states are based on \citet{Dobrowolski08}.

Calculations of the rotational spectra of the five conformers 
\textit{Gauche-anti} (\textit{Ga}), \textit{Gauche-gauche} (\textit{Gg}), 
\textit{Gauche-gauche'} (\textit{Gg'}), \textit{Anti-anti} (\textit{Aa}), and 
\textit{Anti-gauche} (\textit{Ag}) of \textit{normal}-propanol were carried 
out in the context of our companion spectroscopic study 
\citep[][]{Zingsheim22}, which describes the \textit{Aa} and \textit{Ag} 
conformers. The spectroscopic predictions of \textit{Ga}, \textit{Gg}, and 
\textit{Gg’} are based on \citet{Kisiel10} with a large fraction of the 
\textit{Ga} data from \citet{Maeda06a}. The partition function of 
\textit{normal}-propanol was evaluated as described by \citet{Zingsheim22}. 
Briefly, the partition function was summed up over the \textit{Ga}, 
\textit{Gg}, and \textit{Gg’} conformers, whose energies are accurately known 
\citep[][]{Kisiel10}. Conformational contributions of \textit{Aa} and 
\textit{Ag} were evaluated using their calculated energies 
\citep[][]{Kisiel10}. Contributions from excited vibrational states were 
estimated from the fundamental vibrations in \citet{Fukushima68} employing the 
harmonic oscillator approximation as before.

\section{Results}
\label{s:results}

\subsection{Source selection}
\label{ss:source}

\begin{figure*}
\centerline{\resizebox{0.9\hsize}{!}{\includegraphics[angle=0]{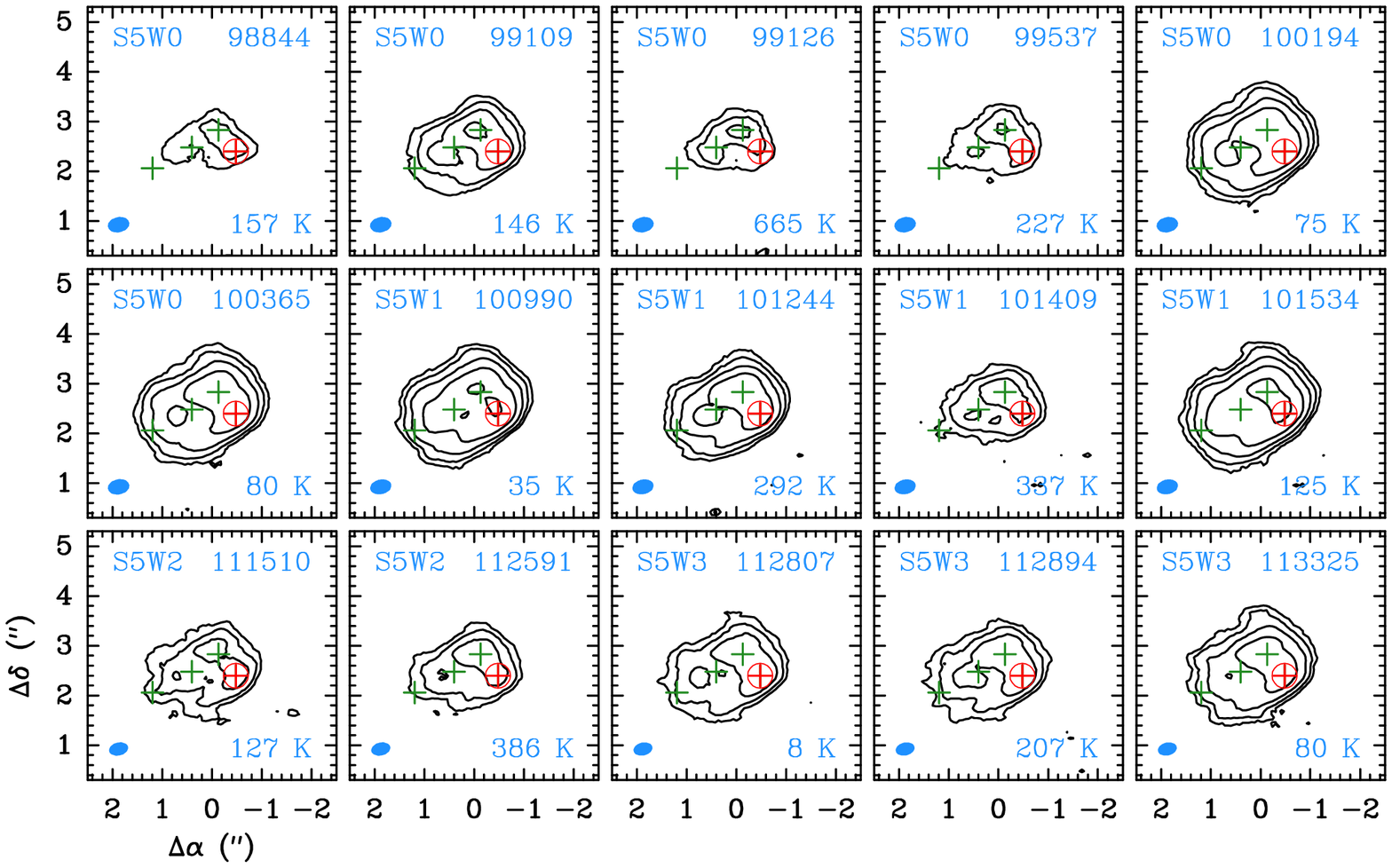}}}
\caption{Integrated intensity maps of selected transitions of ethanol detected 
toward Sgr~B2(N2) in the ReMoCA survey. The equatorial offsets are defined 
with respect to the phase center of the survey (see its coordinates in
Sect.~\ref{ss:alma}). In each panel, the contours start at 6$\sigma$ with 
$\sigma$ the noise level of the integrated intensity map, and they increase by 
a factor of two at each step. The red cross indicates the position of 
Sgr~B2(N2b) and the red circle shows the intrinsic full-width at half maximum 
assumed to model the emission of Sgr~B2(N2b). The green crosses from right to 
left mark the positions of the dust continuum sources AN02, AN03, and AN06 
obtained with ALMA by \citet{SanchezMonge17}. The blue ellipse shows the 
half-power beam width of the ReMoCA survey. The setup (S) and its spectral 
window (W) are given in the top left corner of each panel with the same 
numbering as in Table 2 of \citet{Belloche19}. The frequency (in MHz) and 
upper-level energy (in temperature unit) of the ethanol line are indicated in 
the top right and bottom right corners, respectively. The integration ranges 
were optimized for Sgr B2(N2b) and do not necessarily include all the ethanol 
emission toward the other positions.}
\label{f:maps}
\end{figure*}

We used several prominent and uncontaminated lines of various COMs to explore 
the velocity structure of Sgr~B2(N2), which is resolved with the ReMoCA 
observations. 
A detailed account of this investigation will be reported elsewhere (Belloche 
et al. in prep.). An outcome of this exploration is that the emission from the
south-western part of Sgr~B2(N2) is characterized by narrow lines, with a full
width at half maximum (FWHM) as narrow as $\sim$2~km~s$^{-1}$. The narrowest 
lines are seen at the border of the molecular emission detected with ALMA and 
are thus weak. While narrow linewidths are key to reduce the spectral 
confusion that affects the spectra of Sgr~B2(N) even in the 3~mm wavelength 
range, we also need positions with sufficiently strong emission in order to 
search for faint signals of low-abundance COMs. As a compromise, we selected 
the position indicated with a red cross in Fig.~\ref{f:maps}, which shows 
integrated intensity maps of selected transitions of ethanol that are not 
contaminated by other species and were observed with the highest angular 
resolution in the ReMoCA survey (setup S5). This position has equatorial 
offsets of ($-0.48\arcsec$, $+2.40\arcsec$) with respect to the phase center, 
i.e. an equatorial position ($\alpha, \delta$)$_{\rm J2000}$= 
($17^{\rm h}47^{\rm m}19{\fs}83, -28^\circ22'13{\farcs}6$). We call this position
Sgr~B2(N2b). The COM lines detected toward Sgr~B2(N2b) have a FWHM of 
3.5~km~s$^{-1}$. The systemic velocity toward Sgr~B2(N2b) is 
$V_{\rm lsr} = 74.2$~km~s$^{-1}$. The green crosses displayed in 
Fig.~\ref{f:maps} mark the peak positions of the continuum sources identified 
by \citet{SanchezMonge17} at 1.2~mm with ALMA.

\subsection{Column densities of methanol and ethanol}
\label{ss:methanolethanol}

Methanol and ethanol are both clearly detected in their vibrational ground 
state toward Sgr~B2(N2b), and lines from the former are also detected in its 
first two torsionally excited states (see 
Figs.~\ref{f:spec_ch3oh_ve0}--\ref{f:spec_c2h5oh_ve0}). Rotational lines from 
within the third torsionally excited state of methanol contribute to the 
observed spectrum but none is sufficiently free of contamination from other 
species to secure the identification of this state. We selected the 
transitions that have optical depths lower than 2 and are not too contaminated 
by emission from other species to build population diagrams for both molecules 
(see Figs.~\ref{f:popdiag_ch3oh} and \ref{f:popdiag_c2h5oh}). In both diagrams 
the data indicate a single temperature component. Linear fits to these 
diagrams yield rotational temperatures of $138.4 \pm 1.7$~K and 
$133.9 \pm 1.2$~K for methanol and ethanol, respectively, as reported in 
Table~\ref{t:popfit}. We computed LTE synthetic spectra of methanol and 
ethanol assuming temperatures of 140~K and 135~K, respectively. Given the 
relatively compact morphology of the ethanol emission shown in 
Fig.~\ref{f:maps}, we assumed a size of the emitting region toward Sgr~B2(N2b) 
of 0.5$\arcsec$ and then adjusted the column density until a good match 
between data and model was achieved for the lines that are not too optically 
thick. Very optically thick lines cannot be properly modeled with our simple 
approach that does not account for the structure of the emission along the line 
of sight. Besides, a few lines of methanol are masers \citep[such as the 
$5_{-1}-4_0$ $E$ line at 84521~MHz, see, e.g.,][]{Mueller04} or show a 
combination of absorption and emission and cannot be modeled adequately under 
the LTE approximation.

\begin{table}
 \begin{center}
 \caption{
 Rotational temperature of methanol, ethanol, \textit{g}-\textit{i}-propanol, \textit{a}-\textit{i}-propanol, \textit{Gg'}-\textit{n}-propanol, and \textit{Ag}-\textit{n}-propanol derived from their population diagrams toward Sgr~B2(N2b).
}
 \label{t:popfit}
 \vspace*{0.0ex}
 \begin{tabular}{lll}
 \hline\hline
 \multicolumn{1}{c}{Molecule} & \multicolumn{1}{c}{States\tablefootmark{(a)}} & \multicolumn{1}{c}{$T_{\rm fit}$\tablefootmark{(b)}} \\ 
  & & \multicolumn{1}{c}{\small (K)} \\ 
 \hline
CH$_3$OH & $\varv=0$, $\varv_{\rm t}=1$, $\varv_{\rm t}=2$, $\varv_{\rm t}=3$ & 138.4 (1.7) \\ 
C$_2$H$_5$OH & $\varv=0$ & 133.9 (1.2) \\ 
\textit{g}-\textit{i}-C$_3$H$_7$OH & $\varv=0$ &    80 (220) \\ 
\textit{a}-\textit{i}-C$_3$H$_7$OH & $\varv=0$ &   110 (220) \\ 
\textit{Gg'}-\textit{n}-C$_3$H$_7$OH & $\varv=0$ & \hspace*{0.5ex}   69 (62) \\ 
\textit{Ag}-\textit{n}-C$_3$H$_7$OH & $\varv=0$ & \hspace*{0.5ex}   70 (34) \\ 
\hline 
 \end{tabular}
 \end{center}
 \vspace*{-2.5ex}
 \tablefoot{
 \tablefoottext{a}{Vibrational states that were taken into account to fit the population diagram.}
 \tablefoottext{b}{The standard deviation of the fit is given in parentheses. As explained in Sect.~3 of \citet{Belloche16} and in Sect.~4.4 of \citet{Belloche19}, this uncertainty is purely statistical and should be viewed with caution. It may be underestimated.}
 }
 \end{table}

The results of our LTE modeling are reported in Table~\ref{t:coldens}. We find 
an abundance ratio of methanol to ethanol of 19 toward Sgr~B2(N2b), in very 
good agreement with the ratio of 20 obtained at lower angular resolution 
($\sim$1.6$\arcsec$) with the EMoCA survey toward Sgr~B2(N2) 
\citep[][]{Mueller16}.

The CDMS spectroscopic entry of ethanol (and its associated partition 
function) contains rotational transitions from its two conformers, \textit{anti}
and \textit{gauche}. It accounts for their conformational energy difference
($\Delta E/k=58$~K), the \textit{gauche} conformer being higher in energy than 
the \textit{anti} conformer. This means that our synthetic spectrum of ethanol 
assumes the relative populations of its conformers to be in thermodynamic
equilibrium. The population diagram shown in Fig.~\ref{f:popdiag_c2h5oh}
displays the two conformers in different colors (black for \textit{anti} and 
blue for \textit{gauche}). The fact that the black and blue data points in 
this figure are well fitted with a single straight line clearly demonstrates 
that the assumption of thermodynamic equilibrium between conformers is valid 
for the high-density conditions of Sgr~B2(N2b). Therefore, a single set of LTE 
parameters is sufficient to describe the level populations of ethanol in all
its conformational states and it is adequate to fit its complete spectrum with
this single set of parameters.

\subsection{Detection of \textit{iso}-propanol}
\label{ss:i-propanol}

In order to search for both the \textit{gauche} and \textit{anti} conformers 
of \textit{iso}-propanol, we assumed the same size of the emitting region, 
rotational temperature, linewidth, and velocity offset as for ethanol. The 
only parameter that was variable was the total column density of the molecule. 
Given that the spectroscopic predictions of 
\textit{g}-\textit{i}-C$_3$H$_7$OH and \textit{a}-\textit{i}-C$_3$H$_7$OH 
account for their relative energy, the two conformers can be modeled 
separately using the same column density parameter. It corresponds to the 
total column density of the molecule after accounting for the vibrational 
correction to the partition function that is purely rotational in the CDMS 
entries of these species. As was shown in 
Sect.~\ref{ss:methanolethanol} for ethanol, assuming that the conformers of a 
given molecule have relative populations consistent with thermodynamic 
equilibrium is valid for Sgr~B2(N2b). Therefore, it is natural to expect the
emission of both conformers of \textit{iso}-propanol to be described with a 
single set of LTE parameters applied to the molecule as a whole. As a result,
we did not fit the two conformers independently of each other but rather 
produced consistent synthetic spectra using a single set of LTE parameters.

\begin{figure}
\centerline{\resizebox{1.0\hsize}{!}{\includegraphics[angle=0]{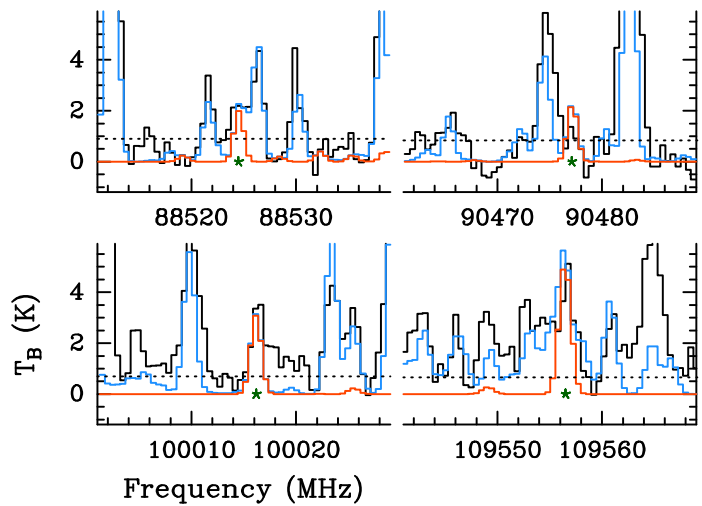}}}
\caption{Spectral lines of \textit{gauche} \textit{iso}-propanol detected in 
the ReMoCA survey toward Sgr~B2(N2b). The best-fit LTE synthetic spectrum of 
\textit{g}-\textit{i}-C$_3$H$_7$OH is displayed in red and overlaid on the 
observed spectrum of Sgr~B2(N2b) shown in black. The blue synthetic spectrum 
contains the contributions of all molecules identified in our survey so far, 
including propanol. The dotted line indicates the $3\sigma$ noise level. The
green stars mark the lines listed as detected in 
Table~\ref{t:propanol_det_int}.}
\label{f:c3h7oh-i-g_det}
\end{figure}

\begin{figure}
\centerline{\resizebox{1.0\hsize}{!}{\includegraphics[angle=0]{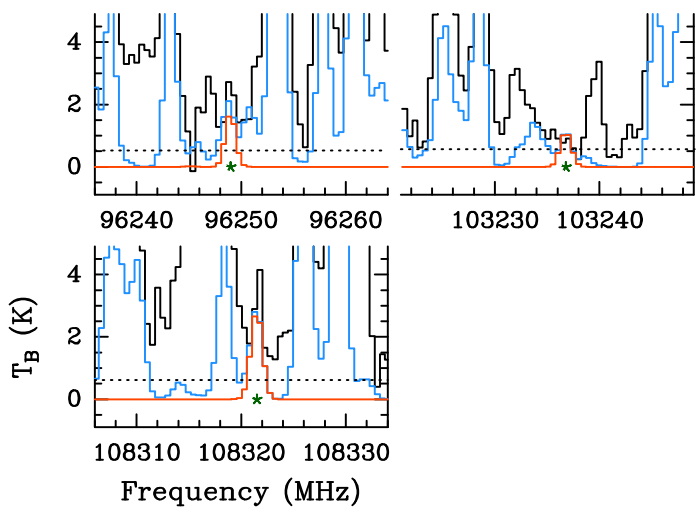}}}
\caption{Spectral lines of \textit{anti} \textit{iso}-propanol detected in the 
ReMoCA survey toward Sgr~B2(N2b). The best-fit LTE synthetic spectrum of 
\textit{a}-\textit{i}-C$_3$H$_7$OH is displayed in red and overlaid on the 
observed spectrum of Sgr~B2(N2b) shown in black. The blue synthetic spectrum 
contains the contributions of all molecules identified in our survey so far, 
including propanol. The dotted line indicates the $3\sigma$ noise level. The
green stars mark the lines listed as detected in 
Table~\ref{t:propanol_det_int}.}
\label{f:c3h7oh-i-a_det}
\end{figure}

The transitions of the \textit{gauche} and 
\textit{anti} conformers that are covered by the ReMoCA survey and are 
expected to contribute significantly to the spectrum of Sgr~B2(N2b) on the 
basis of our LTE model are displayed in Figs.~\ref{f:spec_c3h7oh-i-g_ve0} and 
\ref{f:spec_c3h7oh-i-a_ve0}, respectively. Transitions that are too
heavily blended with much stronger emission from other molecules and therefore 
cannot contribute to the identification of \textit{iso}-propanol are not shown 
in these figures. A close inspection of these figures 
reveals a handful of lines that appear to be detected for each conformer. They 
are marked with green stars in these figures, and are shown specifically in 
Figs.~\ref{f:c3h7oh-i-g_det} and \ref{f:c3h7oh-i-a_det} for the 
\textit{gauche} and \textit{anti} conformers, respectively.

Four spectral lines of \textit{g}-\textit{i}-C$_3$H$_7$OH and three spectral 
lines of \textit{a}-\textit{i}-C$_3$H$_7$OH are sufficiently strong and 
sufficiently free of contamination from other species to be considered as 
detected. Their spectroscopic parameters are listed in 
Table~\ref{t:propanol_det_int}, along with their integrated intensities and 
signal-to-noise ratios. This gives a total number of detected lines of seven 
for \textit{iso-propanol}. Given that our analysis carefully accounts for the 
contamination from all other species identified so far toward Sgr B2(N2b) (the 
blue spectra shown in all figures), we consider that the detection of these 
seven spectral lines, whose relative intensities can be reproduced with a 
single combination of column density and rotational temperature, is sufficient 
to claim a secure detection of \textit{iso}-propanol in Sgr~B2(N2b). This is 
to our knowledge the first interstellar detection of this molecule.

We built population diagrams for both conformers, using the detected 
transitions as well as a few other transitions that are a bit more 
contaminated by the emission of other species that can be removed on the basis 
of our full LTE model of Sgr~B2(N2b). These diagrams are shown in 
Figs.~\ref{f:popdiag_c3h7oh-i-g} and \ref{f:popdiag_c3h7oh-i-a}. In both cases,
the covered range of upper-level energy is unfortunately too narrow to set 
meaningful constraints on the rotational temperature. The formal results of 
the fits to these diagrams are reported in Table~\ref{t:popfit} but the 
uncertainties are much too high to provide any robust insight into the 
rotational temperature of \textit{iso}-propanol. This justifies our choice to 
fix the rotational temperature to the value derived for ethanol in order to 
model the emission of \textit{iso}-propanol.

We obtained a total column density of $1.3 \times 10^{17}$~cm$^{-2}$ for 
\textit{iso}-propanol, which indicates that \textit{iso}-propanol is $\sim$600 
and $\sim$30 times less abundant than methanol and ethanol toward Sgr~B2(N2b), 
respectively (Table~\ref{t:coldens}). Our earlier detailed study of alkanols 
in Sgr~B2(N2) at lower angular resolution with EMoCA resulted in a 
nondetection of \textit{iso}-propanol, with an upper limit indicating that 
\textit{iso}-propanol was at least $\sim$20 times less abundant than ethanol 
\citep[][]{Mueller16}. This earlier nondetection is fully consistent with the 
ratio that we now obtain with ReMoCA. This shows that the high sensitivity and 
angular resolution of the ReMoCA survey was essential in revealing the 
presence of \textit{iso}-propanol in the interstellar medium.

A small discrepancy affects the spectral line displayed in the top right panel 
of Fig.~\ref{f:c3h7oh-i-g_det}. In addition to the fact that this discrepancy 
is at the 2$\sigma$ level only, it seems that the level of the baseline may 
have been slightly overestimated because a number of channels in the displayed 
frequency range have negative intensities that do not look random. Therefore, 
we consider that this discrepancy is insignificant and does not affect our 
identification of the \textit{gauche} conformer of \textit{iso}-propanol.

\begin{table*}[!ht]
 \begin{center}
 \caption{
 Parameters of our best-fit LTE model of methanol, ethanol, \textit{normal}-propanol, and \textit{iso}-propanol toward Sgr~B2(N2b).
}
 \label{t:coldens}
 \vspace*{-1.2ex}
 \begin{tabular}{lcrcccccccr}
 \hline\hline
 \multicolumn{1}{c}{Molecule} & \multicolumn{1}{c}{Status\tablefootmark{(a)}} & \multicolumn{1}{c}{$N_{\rm det}$\tablefootmark{(b)}} & \multicolumn{1}{c}{Size\tablefootmark{(c)}} & \multicolumn{1}{c}{$T_{\mathrm{rot}}$\tablefootmark{(d)}} & \multicolumn{1}{c}{$N$\tablefootmark{(e)}} & \multicolumn{1}{c}{$F_{\rm vib}$\tablefootmark{(f)}} & \multicolumn{1}{c}{$F_{\rm conf}$\tablefootmark{(g)}} & \multicolumn{1}{c}{$\Delta V$\tablefootmark{(h)}} & \multicolumn{1}{c}{$V_{\mathrm{off}}$\tablefootmark{(i)}} & \multicolumn{1}{c}{$\frac{N_{\rm ref}}{N}$\tablefootmark{(j)}} \\ 
  & & & \multicolumn{1}{c}{\small ($''$)} & \multicolumn{1}{c}{\small (K)} & \multicolumn{1}{c}{\small (cm$^{-2}$)} & & & \multicolumn{1}{c}{\small (km~s$^{-1}$)} & \multicolumn{1}{c}{\small (km~s$^{-1}$)} & \\ 
 \hline
 CH$_3$OH, $\varv=0$$^\star$ & d & 48 &  0.5 &  140 &  8.0 (19) & 1.00 & -- & 3.5 & 0.0 &       1 \\ 
 \hspace*{7.8ex} $\varv_{\rm t}=1$ & d & 15 &  0.5 &  140 &  8.0 (19) & 1.00 & -- & 3.5 & 0.0 &       1 \\ 
 \hspace*{7.8ex} $\varv_{\rm t}=2$ & d & 6 &  0.5 &  140 &  8.0 (19) & 1.00 & -- & 3.5 & 0.0 &       1 \\ 
 \hspace*{7.8ex} $\varv_{\rm t}=3$ & t & 0 &  0.5 &  140 &  8.0 (19) & 1.00 & -- & 3.5 & 0.0 &       1 \\ 
\hline 
 C$_2$H$_5$OH, $\varv=0$ & d & 271 &  0.5 &  135 &  4.2 (18) & 1.24 & -- & 3.5 & 0.0 &      19 \\ 
\hline 
 \textit{Ga}-\textit{n}-C$_3$H$_7$OH, $\varv=0$ & n & 0 &  0.5 &  135 &  2.3 (17) & 1.57 & 1.44 & 3.5 & 0.0 &     350 \\ 
 \textit{Gg}-\textit{n}-C$_3$H$_7$OH, $\varv=0$ & n & 0 &  0.5 &  135 &  2.3 (17) & 1.57 & 1.44 & 3.5 & 0.0 &     350 \\ 
 \textit{Gg'}-\textit{n}-C$_3$H$_7$OH, $\varv=0$ & d & 8 &  0.5 &  135 &  2.3 (17) & 1.57 & 1.44 & 3.5 & 0.0 &     350 \\ 
 \textit{Aa}-\textit{n}-C$_3$H$_7$OH, $\varv=0$ & n & 0 &  0.5 &  135 &  2.3 (17) & 1.57 & 1.44 & 3.5 & 0.0 &     350 \\ 
 \textit{Ag}-\textit{n}-C$_3$H$_7$OH, $\varv=0$ & d & 5 &  0.5 &  135 &  2.3 (17) & 1.57 & 1.44 & 3.5 & 0.0 &     350 \\ 
\hline 
 \textit{g}-\textit{i}-C$_3$H$_7$OH, $\varv=0$ & d & 4 &  0.5 &  135 &  1.3 (17) & 1.42 & -- & 3.5 & 0.0 &     630 \\ 
 \textit{a}-\textit{i}-C$_3$H$_7$OH, $\varv=0$ & d & 3 &  0.5 &  135 &  1.3 (17) & 1.42 & -- & 3.5 & 0.0 &     630 \\ 
\hline 
 \end{tabular}
 \end{center}
 \vspace*{-2.5ex}
 \tablefoot{
 \tablefoottext{a}{d: detection, t: tentative detection, n: nondetection.}
 \tablefoottext{b}{Number of detected lines \citep[conservative estimate, see Sect.~3 of][]{Belloche16}. One line of a given species may mean a group of transitions of that species that are blended together.}
 \tablefoottext{c}{Source diameter (\textit{FWHM}).}
 \tablefoottext{d}{Rotational temperature.}
 \tablefoottext{e}{Total column density of the molecule. $x$ ($y$) means $x \times 10^y$. An identical value for all listed vibrational/torsional states of a molecule means that LTE is an adequate description of the vibrational/torsional excitation.}
 \tablefoottext{f}{Correction factor that was applied to the column density to account for the contribution of vibrationally excited states, in the cases where this contribution was not included in the partition function of the spectroscopic predictions.}
 \tablefoottext{g}{Correction factor that was applied to the column density to account for the contribution of other conformers in the cases where this contribution could be estimated but was not included in the partition function of the spectroscopic predictions.}
 \tablefoottext{h}{Linewidth (\textit{FWHM}).}
 \tablefoottext{i}{Velocity offset with respect to the assumed systemic velocity of Sgr~B2(N2b), $V_{\mathrm{sys}} = 74.2$ km~s$^{-1}$.}
 \tablefoottext{j}{Column density ratio, with $N_{\rm ref}$ the column density of the previous reference species marked with a $\star$.}
 }
 \end{table*}

\begin{table*}
 \begin{center}
 \caption{
 Spectroscopic parameters and integrated intensities of transitions of \textit{iso}-propanol and \textit{normal}-propanol detected toward Sgr~B2(N2b) in the ReMoCA survey.
}
 \label{t:propanol_det_int}
 \vspace*{0.0ex}
 \begin{tabular}{crcccccrrccc}
 \hline\hline
 \multicolumn{1}{c}{Transition} & \multicolumn{1}{c}{Frequency} & \multicolumn{1}{c}{$\Delta f$\tablefootmark{(a)}}  & \multicolumn{1}{c}{$A_{\rm ul}$\tablefootmark{(b)}} & \multicolumn{1}{c}{$E_{\rm u}$\tablefootmark{(c)}}  & \multicolumn{1}{c}{$g_{\rm u}$\tablefootmark{(d)}} & \multicolumn{1}{c}{$I_{\rm obs}$\tablefootmark{(e)}} & \multicolumn{1}{c}{$I_{\rm mod}$\tablefootmark{(f)}} & \multicolumn{1}{c}{$I_{\rm all}$\tablefootmark{(g)}} & \multicolumn{1}{c}{$\underline{S_{\rm obs}}$\tablefootmark{(h)}} & \multicolumn{1}{c}{$\underline{S_{\rm dec}}$\tablefootmark{(i)}} & \multicolumn{1}{c}{$\underline{S_{\rm mod}}$\tablefootmark{(j)}} \\ 
 \multicolumn{1}{c}{$J_{K_a,K_c}$} & \multicolumn{1}{c}{\small (MHz)} & \multicolumn{1}{c}{\small (kHz)} & \multicolumn{1}{c}{\small (10$^{-5}$ s$^{-1}$)} & \multicolumn{1}{c}{\small (K)} & & \multicolumn{1}{c}{\small (K km s$^{-1}$)} & \multicolumn{2}{c}{\small (K km s$^{-1}$)} & \multicolumn{1}{c}{\hspace*{-2ex}$N$} & \multicolumn{1}{c}{\hspace*{-2ex}$N$} & \multicolumn{1}{c}{\hspace*{-2ex}$N$} \\ 
 \hline \\[-1.8ex]
\multicolumn{10}{c}{\textit{g}-\textit{i}-C$_3$H$_7$OH, $\varv=0$} \\[0.5ex] 
13$_{1,12, 0}$ -- 12$_{1,11, 1}$ & 88524.536 &   6 &  0.43 &    48.2 & 27 &  10.6(9) &   7.1 &   9.1 &    12 &    10 &   8.3 \\ 
13$_{2,12, 0}$ -- 12$_{2,11, 1}$ & 88524.536 &   6 &  0.43 &    48.2 & 27 & -- & -- & -- & -- & -- & -- \\ 
14$_{0,14, 0}$ -- 13$_{0,13, 1}$ & 90477.172 &   7 &  0.51 &    50.4 & 29 &   5.3(9) &   8.6 &   9.1 &   5.9 &   5.4 &   9.5 \\ 
14$_{1,14, 0}$ -- 13$_{1,13, 1}$ & 90477.172 &   7 &  0.51 &    50.4 & 29 & -- & -- & -- & -- & -- & -- \\ 
15$_{0,15, 0}$ -- 14$_{0,14, 1}$ & 100016.165 &   7 &  0.68 &    57.5 & 31 &  13.8(7) &  11.0 &  11.4 &    21 &    20 &    16 \\ 
15$_{1,15, 0}$ -- 14$_{1,14, 1}$ & 100016.165 &   7 &  0.68 &    57.5 & 31 & -- & -- & -- & -- & -- & -- \\ 
16$_{0,16, 0}$ -- 15$_{0,15, 1}$ & 109556.465 &   7 &  0.91 &    65.0 & 33 &  18.8(6) &  18.5 &  23.3 &    32 &    24 &    32 \\ 
16$_{1,16, 0}$ -- 15$_{1,15, 1}$ & 109556.465 &   7 &  0.91 &    65.0 & 33 & -- & -- & -- & -- & -- & -- \\ 
\hline \\[-1.8ex] 
\multicolumn{10}{c}{\textit{a}-\textit{i}-C$_3$H$_7$OH, $\varv=0$} \\[0.5ex] 
9$_{1,8}$ -- 8$_{2,7}$ & 96248.969 &   4 &  0.80 &   144.9 & 19 &  12.1(5) &   6.3 &   9.5 &    22 &    17 &    12 \\ 
9$_{2,8}$ -- 8$_{1,7}$ & 96248.969 &   4 &  0.80 &   144.9 & 19 & -- & -- & -- & -- & -- & -- \\ 
9$_{2,7}$ -- 8$_{3,6}$ & 103236.772 &   3 &  0.86 &   147.4 & 19 &   4.6(5) &   4.2 &   4.4 &   8.5 &   8.1 &   7.8 \\ 
9$_{3,7}$ -- 8$_{2,6}$ & 103236.795 &   3 &  0.86 &   147.4 & 19 & -- & -- & -- & -- & -- & -- \\ 
11$_{0,11}$ -- 10$_{1,10}$ & 108321.482 &   4 &  1.32 &   152.0 & 23 &  14.3(6) &  10.2 &  10.9 &    26 &    24 &    18 \\ 
11$_{1,11}$ -- 10$_{0,10}$ & 108321.482 &   4 &  1.32 &   152.0 & 23 & -- & -- & -- & -- & -- & -- \\ 
\hline \\[-1.8ex] 
\multicolumn{10}{c}{\textit{Gg'}-\textit{n}-C$_3$H$_7$OH, $\varv=0$} \\[0.5ex] 
9$_{1,8, 2}$ -- 8$_{1,7,2
}$ & 85495.771 &   2 &  0.65 &    94.4 & 76 &   4.1(8) &   3.3 &   3.3 &   5.0 &   4.9 &   3.9 \\ 
10$_{0,10, 2}$ -- 9$_{0,9,2
}$ & 88171.422 &   3 &  0.72 &    96.9 & 84 &   5.6(9) &   3.8 &   4.4 &   6.5 &   5.9 &   4.5 \\ 
10$_{2,9, 2}$ -- 9$_{2,8,2
}$ & 91872.157 &   2 &  0.78 &    99.4 & 84 &   7.4(9) &   4.6 &   5.9 &   8.5 &   6.9 &   5.2 \\ 
10$_{6,5, 2}$ -- 9$_{6,4,2
}$ & 93322.983 &   2 &  0.55 &   114.4 & 84 &   6.4(9) &   5.8 &   6.4 &   7.5 &   6.8 &   6.8 \\ 
10$_{6,4, 2}$ -- 9$_{6,3,2
}$ & 93323.050 &   2 &  0.55 &   114.4 & 84 & -- & -- & -- & -- & -- & -- \\ 
11$_{2,10, 2}$ -- 10$_{2,9, 2}$ & 100777.598 &   2 &  1.04 &   104.3 & 92 &   5.7(7) &   5.6 &   5.7 &   7.6 &   7.6 &   7.6 \\ 
11$_{5,7, 2}$ -- 10$_{5,6, 2}$ & 102774.297 &   2 &  0.88 &   114.2 & 92 &   5.4(5) &   4.0 &   5.4 &    10 &   7.4 &   7.3 \\ 
11$_{1,10, 2}$ -- 10$_{1,9, 2}$ & 103092.965 &   2 &  1.14 &   103.9 & 92 &   8.0(5) &   4.9 &   4.9 &    17 &    17 &    11 \\ 
12$_{4,9, 2}$ -- 11$_{4,8, 2}$ & 112758.561 &   2 &  1.12 &   115.6 & 100 &  14.3(26) &   8.7 &  12.0 &   5.6 &   4.3 &   3.4 \\ 
\hline \\[-1.8ex] 
\multicolumn{10}{c}{\textit{Ag}-\textit{n}-C$_3$H$_7$OH, $\varv=0$} \\[0.5ex] 
13$_{0,13, 3}$ -- 12$_{0,12, 3}$ & 94296.615 &  40 &  0.61 &    97.0 & 54 &   5.0(8) &   4.4 &   4.6 &   5.9 &   5.6 &   5.2 \\ 
13$_{0,13, 2}$ -- 12$_{0,12, 2}$ & 94297.251 &  40 &  0.61 &    97.0 & 54 & -- & -- & -- & -- & -- & -- \\ 
13$_{6,7, 0}$ -- 12$_{6,6, 0}$ & 94959.841 &  40 &  0.49 &   136.0 & 54 &   3.5(7) &   2.5 &   2.6 &   4.9 &   4.8 &   3.5 \\ 
13$_{6,8, 0}$ -- 12$_{6,7, 0}$ & 94959.841 &  40 &  0.49 &   136.0 & 54 & -- & -- & -- & -- & -- & -- \\ 
14$_{5,9, 0}$ -- 13$_{5,8, 0}$ & 102271.836 &  40 &  0.68 &   129.0 & 58 &   6.4(5) &   3.4 &   5.4 &    12 &   8.0 &   6.3 \\ 
14$_{5,10, 0}$ -- 13$_{5,9, 0}$ & 102271.836 &  40 &  0.68 &   129.0 & 58 & -- & -- & -- & -- & -- & -- \\ 
14$_{2,12, 3}$ -- 13$_{2,11, 3}$ & 102855.519 &  40 &  0.78 &   106.4 & 58 &   3.4(5) &   2.1 &   2.3 &   7.2 &   6.9 &   4.5 \\ 
15$_{5,10, 1}$ -- 14$_{5,9, 1}$ & 109618.830 &  40 &  0.86 &   134.3 & 62 &   8.0(7) &   6.3 &   9.2 &    12 &   7.8 &   9.6 \\ 
15$_{5,11, 1}$ -- 14$_{5,10, 1}$ & 109618.830 &  40 &  0.86 &   134.3 & 62 & -- & -- & -- & -- & -- & -- \\ 
 \hline
 \end{tabular}
 \end{center}
 \vspace*{-2.5ex}
 \tablefoot{
 \tablefoottext{a}{Frequency uncertainty.}
 \tablefoottext{b}{Einstein coefficient for spontaneous emission.}
 \tablefoottext{c}{Upper-level energy.}
 \tablefoottext{d}{Upper-level degeneracy.}
 \tablefoottext{e}{Integrated intensity of the observed spectrum in brightness temperature scale. The statistical standard deviation is given in parentheses in unit of the last digit.}
 \tablefoottext{f}{Integrated intensity of the synthetic spectrum of propanol.}
 \tablefoottext{g}{Integrated intensity of the model that contains the contribution of all identified molecules, including propanol.}
 \tablefoottext{h}{Signal-to-noise ratio of $I_{\rm obs}$.}
 \tablefoottext{i}{Signal-to-noise ratio of the integrated intensity, $I_{\rm dec}$, decontaminated from the contribution of molecules other than propanol, i.e. $I_{\rm dec} = I_{\rm obs}-(I_{\rm all}-I_{\rm mod})$, computed with the uncertainty of $I_{\rm obs}$.}
 \tablefoottext{j}{Signal-to-noise ratio of $I_{\rm mod}$ computed with the uncertainty of $I_{\rm obs}$. In the last six columns, a value followed by a dash in the next row represents the value obtained for a group of transitions that are not resolved in the astronomical spectrum.}
 }
 \end{table*}

\subsection{Detection of \textit{normal}-propanol}

\begin{figure}
\centerline{\resizebox{1.0\hsize}{!}{\includegraphics[angle=0]{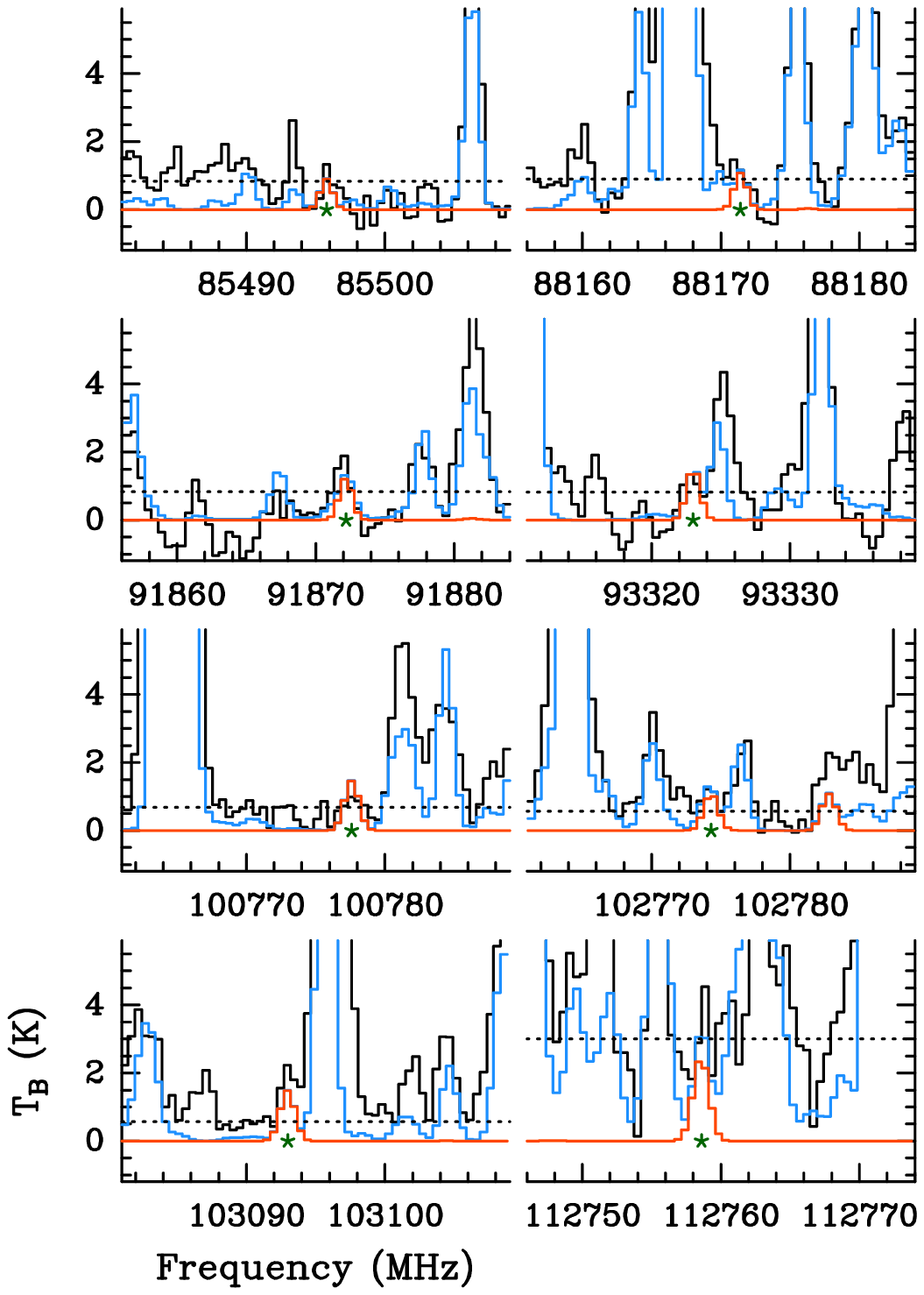}}}
\caption{Spectral lines of \textit{Gauche-gauche'} \textit{normal}-propanol 
detected in the 
ReMoCA survey toward Sgr~B2(N2b). The best-fit LTE synthetic spectrum of 
\textit{Gg'}-\textit{n}-C$_3$H$_7$OH is displayed in red and overlaid on the 
observed spectrum of Sgr~B2(N2b) shown in black. The blue synthetic spectrum 
contains the contributions of all molecules identified in our survey so far, 
including propanol. The dotted line indicates the $3\sigma$ noise level. The
green stars mark the lines listed as detected in 
Table~\ref{t:propanol_det_int}.}
\label{f:c3h7oh-n-Ggp_det}
\end{figure}

\begin{figure}
\centerline{\resizebox{1.0\hsize}{!}{\includegraphics[angle=0]{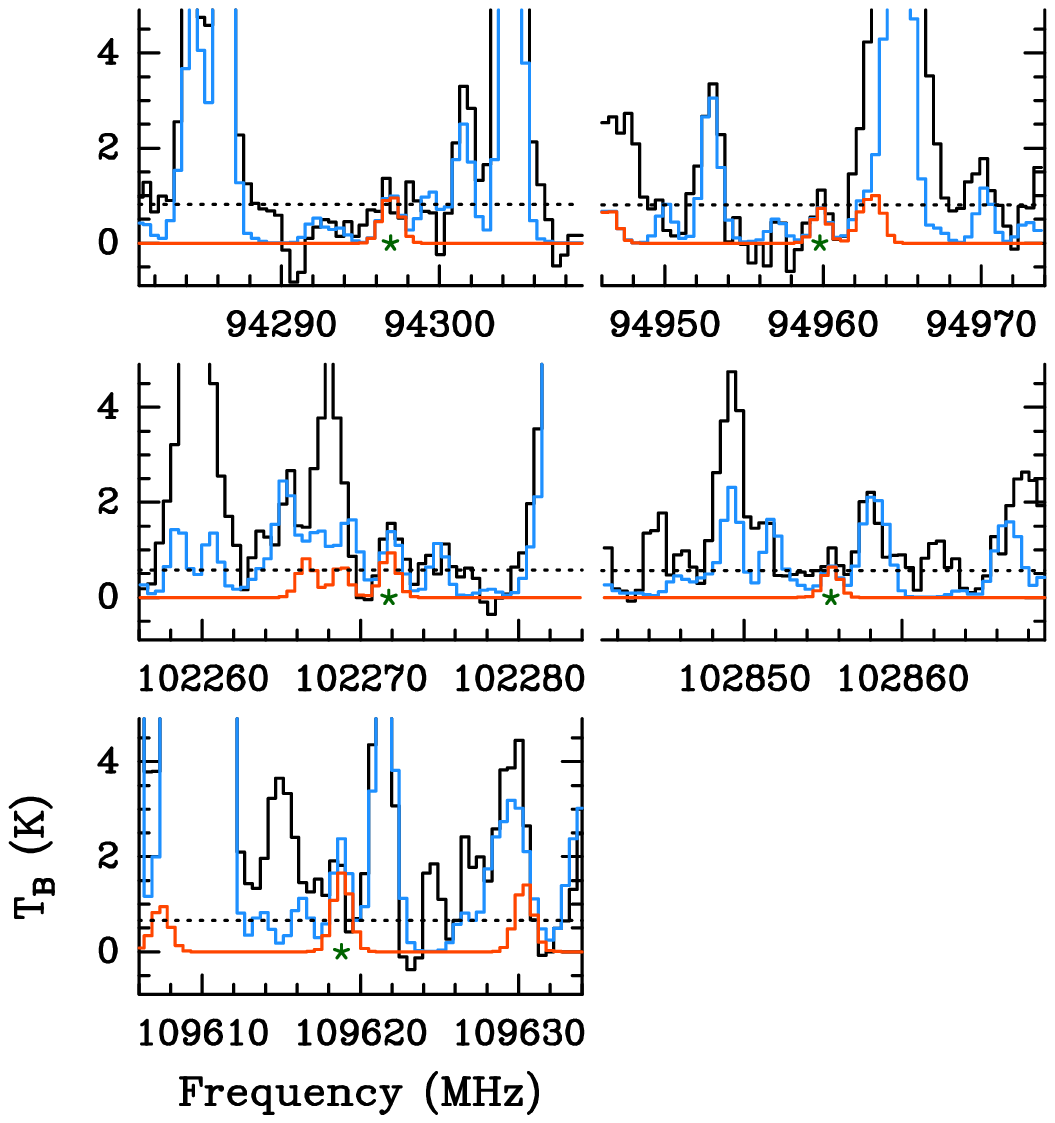}}}
\caption{Spectral lines of \textit{Anti-gauche} \textit{normal}-propanol 
detected in the ReMoCA survey toward Sgr~B2(N2b). The best-fit LTE synthetic 
spectrum of \textit{Ag}-\textit{n}-C$_3$H$_7$OH is displayed in red and 
overlaid on the observed spectrum of Sgr~B2(N2b) shown in black. The blue 
synthetic spectrum contains the contributions of all molecules identified in 
our survey so far, including propanol. The dotted line indicates the $3\sigma$ 
noise level. The green stars mark the lines listed as detected in 
Table~\ref{t:propanol_det_int}.}
\label{f:c3h7oh-n-Ag_det}
\end{figure}

We proceeded in the same way as for \textit{iso}-propanol in order to search 
for the five conformers of \textit{normal}-propanol. In particular, 
like for ethanol (Sect.~\ref{ss:methanolethanol}) and \textit{iso}-propanol 
(Sect.~\ref{ss:i-propanol}), we assumed that the relative populations of these
five conformers are in thermodynamic equilibrium and that the emission of the
molecule can be described with a single set of LTE parameters.
Figures~\ref{f:spec_c3h7oh-n-Ga_ve0}--\ref{f:spec_c3h7oh-n-Ag_ve0} show the 
transitions of the \textit{Gauche-anti}, \textit{Gauche-gauche}, 
\textit{Gauche-gauche'}, \textit{Anti-anti}, and \textit{Anti-gauche} 
conformers of \textit{normal}-propanol that are covered by the ReMoCA survey 
and are expected to contribute significantly to the spectrum of Sgr~B2(N2b) on 
the basis of our LTE model. Here again, transitions that are too
heavily blended with much stronger emission from other molecules and therefore 
cannot contribute to the identification of \textit{iso}-propanol are not shown 
in these figures. A close inspection of these figures reveals that a 
handful of spectral lines of the \textit{Gauche-gauche'} and 
\textit{Anti-gauche} conformers are detected toward Sgr~B2(N2b). They are 
marked with green stars in these figures and displayed separately in 
Figs.~\ref{f:c3h7oh-n-Ggp_det} and \ref{f:c3h7oh-n-Ag_det}, respectively. All 
the transitions of the other conformers are either too weak or too much 
contaminated by the emission of other species to be clearly identified.

Eight spectral lines of \textit{Gg'}-\textit{n}-C$_3$H$_7$OH and five spectral 
lines of \textit{Ag}-\textit{n}-C$_3$H$_7$OH are sufficiently strong and 
sufficiently free of contamination from other species to be considered as 
detected. Their spectroscopic parameters are listed in 
Table~\ref{t:propanol_det_int}, along with their integrated intensities and 
signal-to-noise ratios. This gives a total number of detected lines of 13 for 
\textit{normal}-propanol. As for \textit{iso}-propanol, we consider this 
number of detected lines as sufficient to claim a robust detection of 
\textit{normal}-propanol toward Sgr~B2(N2b). This is to our knowledge the 
first secure detection of \textit{normal}-propanol toward a hot 
core\footnote{The first interstellar detection of \textit{normal}-propanol was 
reported by \citet{JimenezSerra22} toward the molecular cloud G+0.693--0.027
during the refereeing process of this article. This source is thought to be 
a region dominated by shocks and is not a hot core.}.

As for \textit{iso}-propanol, the population diagrams built for 
\textit{normal}-propanol are too scarce to provide any meaningful constraints 
on the rotational temperature of its emission (see 
Figs.~\ref{f:popdiag_c3h7oh-n-Ggp}--\ref{f:popdiag_c3h7oh-n-Ag} and 
Table~\ref{t:popfit}). This again justifies our choice to use the same 
temperature as derived for ethanol in order to model the emission of 
\textit{normal}-propanol.

We obtained a total column density of $2.8 \times 10^{17}$~cm$^{-2}$ for 
\textit{normal}-propanol, which indicates that \textit{normal}-propanol is 
$\sim$350 and $\sim$20 times less abundant than methanol and ethanol toward 
Sgr~B2(N2b), respectively (Table~\ref{t:coldens}). Our earlier detailed study 
of alkanols at lower angular resolution with EMoCA resulted in a nondetection 
of \textit{normal}-propanol in Sgr~B2(N2), with an upper limit indicating that 
\textit{normal}-propanol was at least $\sim$8 times less abundant than ethanol 
\citep[][]{Mueller16}. Like in the case of \textit{iso}-propanol, this earlier
nondetection is fully consistent with the ratio that we now obtain with ReMoCA.

The synthetic spectra computed for the other three conformers of 
\textit{normal}-propanol with the column density derived above 
are, under the assumption of thermodynamic equilibrium, fully 
consistent with the ReMoCA spectrum (see Figs.~\ref{f:spec_c3h7oh-n-Ga_ve0}, 
\ref{f:spec_c3h7oh-n-Gg_ve0}, and \ref{f:spec_c3h7oh-n-Aa_ve0}). All 
transitions of these conformers that are not too weak unfortunately fall at 
frequencies where other molecules emit more strongly, which prevents the 
identification of these conformers in the spectrum toward Sgr~B2(N2b).
The synthetic spectrum of \textit{normal}-propanol at the frequencies 
of several transitions of the \textit{Gauche-anti} conformer shown in 
Fig.~\ref{f:spec_c3h7oh-n-Ga_ve0} is close to the signal detected at these 
frequencies. A column density of \textit{normal}-propanol twice as high as 
the value determined above would not be consistent anymore (see, e.g., the
transitions at 94747, 99937, 102182, or 106120~MHz). Similarly, a column 
density of \textit{normal}-propanol twice as high as the value determined 
above would, for some transitions of the \textit{Gauche-gauche} conformer 
shown in Fig.~\ref{f:spec_c3h7oh-n-Gg_ve0}, not be consistent with the 
observed spectrum anymore (transitions at, e.g., 100696~MHz and 106248~MHz). 
The case of the 
\textit{Anti-anti} conformer is less constraining, with an upper limit set by 
the transition at 95506~MHz in Fig.~\ref{f:spec_c3h7oh-n-Aa_ve0} about three 
times as high as the value derived above.

\subsection{Abundance ratio of \textit{iso}-propanol and 
\textit{normal}-propanol}
\label{ss:ratio}

The column densities reported in Table~\ref{t:coldens} imply that 
\textit{iso}-propanol is twice less abundant than \textit{normal}-propanol 
toward Sgr~B2(N2b) ([\textit{iso}]/[\textit{normal}] = 0.6). This is close to 
the isomeric ratio [\textit{iso}]/[\textit{normal}] = 0.4 that we obtained for 
propyl cyanide, C$_3$H$_7$CN, at lower angular resolution toward Sgr~B2(N2) 
with the EMoCA survey \citep[][]{Belloche14}.

\section{Astrochemical modeling}
\label{s:chemistry}

In order to understand the relative ratios of the \textit{iso} and \textit{normal} forms of propanol detected in our survey, we have run astrochemical models simulating the chemistry occurring during the cold collapse of the core and the subsequent warm-up to typical hot-core temperatures. These simulations use the three-phase, gas-grain astrochemical kinetics code MAGICKAL \citep{Garrod13}. The models presented here build upon the hot-core models of \citet{Garrod22}, using those authors' \textit{final} model setup. Along with more typical gas-phase processes, the code employs the most up-to-date treatment of grain-surface and bulk-ice chemistry. This includes not only diffusive surface reaction processes, but -- crucially -- a range of non-diffusive reaction mechanisms that can occur both on the grain/ice surfaces and within the bulk ice mantles \citep{Jin20}.

Diffusive reactions occur when one or other reactant is able to diffuse thermally on the grain/ice surface, allowing them to meet and react. In the non-diffusive case, some other process brings the reactants together to allow the same reaction to occur. Thus, in general, it is only the mechanism by which a meeting occurs that differentiates diffusive from non-diffusive reactions, rather than the underlying reaction network \citep[with the limited exception of the so-called three-body excited formation mechanism, 3-BEF, see][]{Jin20}. However, the reaction rates of non-diffusive reactions take on an entirely different form from their diffusive counterparts, and can allow reactions between species of low thermal mobility to occur even at very low temperatures.

Two main non-diffusive mechanisms considered in the model are: (i) photodissociation-induced (PDI) reactions, in which the photodissociation of a grain-surface/ice molecule in the presence of some other chemical species results in the latter being able to react with one of the photo-products; and (ii) three-body (3-B) reactions, in which a preceding two-body reaction on the grain occurs in the presence of some other chemical species with which the reaction product may itself react. In the latter case, the preceding reaction could itself be diffusive or non-diffusive. The PDI and 3-B processes may take place both on the grain/ice surfaces and within the bulk ice mantle. The Eley-Rideal process is also included in the model, whereby an atom or molecule adsorbing onto the grain surface from the gas phase immediately meets its reaction partner upon adsorption; however, this mechanism is generally of small significance.

Whether occurring as diffusive or non-diffusive processes, reactions with activation energy barriers may experience low efficiency if the reactants are able to diffuse away from each other before reaction occurs. However, in cases where the reactants are completely immobile, a slow, barrier-mediated reaction may occur with high efficiency -- for example, when bulk-ice photodissociation leads to a photo-product radical being trapped in the ice next to a stable molecule, with which it may react after overcoming an activation energy barrier.

One further noteworthy adjustment made in the \citet{Garrod22} model is the removal of the bulk-diffusion process for species other than H and H$_2$; aside from these two, bulk species are deemed to be too bulky to diffuse via interstitial hopping within the ice structure. Consequently, atomic H and H$_2$ are the only species whose diffusion may lead to reactions in the bulk. Non-diffusive mechanisms nevertheless allow reactions in the bulk ice to proceed.

The model of \citet{Garrod22} included an expanded reaction network, which we further extend to include propanol. We also update several of the reactions relating to the chemistry of \textit{normal}- and \textit{iso}-propyl cyanide \citep[][]{Garrod17}. The new network is described in more detail in Sect.~\ref{network}.

The physical conditions used in the models follow past simulations, beginning with a cold collapse stage in which the gas density, $n_{\rm{H}}$, increases from $3 \times 10^{3}$ to $2 \times 10^{8}$ cm$^{-3}$ over a period of approximately 1 Myr, under free-fall collapse. Due to the gradual increase in visual extinction (3 to 500 magnitudes), the dust temperature falls from $\sim$14.7~K to a fixed lower limit of 8~K, following the relationship of \citet{Garrod11}. The collapse is halted when the desired density is reached; a warm-up stage then ensues at fixed gas density, in which the coupled gas and dust temperatures gradually rise to a final value of 400~K, at which point the model ends. We use the same three warm-up timescales as in past models: \textit{fast}, \textit{medium}, and \textit{slow}, corresponding to characteristic timescales of $5 \times 10^{4}$, $2 \times 10^{5}$, and $1 \times 10^{6}$ yr taken to reach a representative hot-core temperature of 200~K \citep[more detail is provided by][]{Garrod06,Garrod13}. We assume here that the \textit{slow} model is most relevant to Sgr B2(N), based on its better match to observational data in past comparisons.

\subsection{Reaction network} 
\label{network}

We constructed a new sub-network of chemical reactions for propanol that was added to our existing chemical network derived from \citet{Garrod22}. This new network is not the first to consider propanol in the ISM; \citet{Charnley95} presented a gas-phase chemical model of hot cores in which both $n$-C$_3$H$_7$OH and $i$-C$_3$H$_7$OH were presumed to form on dust grains, before being thermally released into the gas phase where they were destroyed. Those authors' network thus seems to have included only gas-phase, destructive mechanisms for propanol. More recently, \citet{Manigand21} included a generic form of propanol in their gas-grain network that did not differentiate between the {\em normal} and {\em iso} forms. Here we present the first model to incorporate a full gas-phase and dust-grain chemistry with active formation and destruction routes for both forms of propanol.

The new network is informed by that constructed for \textit{normal}- and \textit{iso}-propyl cyanide by \citet{Belloche14} and refined and extended by \citet{Garrod17} to include the butyl cyanides and the pentanes. The new network also incorporates several new ethanol-related reactions introduced into the network by Jin et al. (in prep.), which mainly concern the hydrogenation state of the C-C-O molecular backbone on grain surfaces.

Construction of the network follows the general approach laid out by \citet{Garrod08} and followed in later publications; \textit{normal}- and \textit{iso}-propanol are assumed to form on dust grains, primarily through the barrierless addition of radicals on the grain/ice surfaces or within the bulk ice. The chemistry of most of the relevant radicals was already present in earlier networks, but several additional radicals have been included along with the two forms of propanol, and these also have a full complement of their own gas-phase and dust-grain chemical processes. The network differentiates between structurally isomeric radicals, in which the radical site may be present on different atoms, e.g., the primary radical 
$\overset{\centerdot}{\mathrm{C}}$H$_2$CH$_2$OH, 
versus the secondary radical 
CH$_3\overset{\centerdot}{\mathrm{C}}$HOH; 
the addition of a methyl (CH$_3$) radical to each of these would lead to $n$-C$_3$H$_7$OH and $i$-C$_3$H$_7$OH, respectively. The addition of OH to the isomeric radicals of formula C$_3$H$_7$ may similarly produce one or other of the propanol structures. A unique mechanism for $n$-propanol is also provided by the addition of the radicals C$_2$H$_5$ and CH$_2$OH.

The radicals involved in this chemistry may be produced by atomic addition to smaller molecules, such as by the hydrogenation of vinyl alcohol, C$_2$H$_3$OH, by atomic H on grain surfaces, leading to $\overset{\centerdot}{\mathrm{C}}$H$_2$CH$_2$OH 
and 
CH$_3\overset{\centerdot}{\mathrm{C}}$HOH. Stable molecules may also be photodissociated by external UV photons or by the cosmic ray-induced UV field to produce these and other radicals. All the new chemical species are provided photodissociation mechanisms from both UV sources, with the products assumed to be radical-radical or radical-atom pairs.

Abstraction of a hydrogen atom from a stable molecule by a reactive radical (producing another radical) is also possible, although these reactions typically have activation energy barriers. As a result, the most effective examples are those in which the barriers are low (e.g., involving attack by the OH radical or an H atom). Due to the low temperatures involved for much of the grain-surface and ice chemistry, H-abstraction reactions are assumed typically to be driven by the quantum tunneling of the abstracted H atom. A rectangular barrier treatment is adopted in such cases, as per past models. Based on the substantial differences in activation energy barrier measured or calculated for these processes, the secondary radicals tend to be favored over primary radicals, due to the lower energies of their structures.

Several other reactions are included that involve the addition of an OH radical to a hydrocarbon. Of particular relevance to propanol formation is the reaction between OH and propylene, C$_3$H$_6$. Evidence from the literature indicates that these reactions are essentially barrierless in the gas phase  \citep{Atkinson97,Thomsen09}, while both experimental and  theoretical analyses indicate that the addition of OH to the terminal carbon occurs at a slightly higher rate than to the central carbon atom \citep{Izsak09,Daranlot10,Loison10}, in ratios ranging from 58:42 to 72:28. In our network, we assume a ratio 3:2 (see Sec. \ref{sec:orientation}) in the production of 
CH$_3\overset{\centerdot}{\mathrm{C}}$HCH$_2$OH 
versus  
$\overset{\centerdot}{\mathrm{C}}$H$_2$CH(OH)CH$_3$, which are the precursor radicals of \textit{normal} and \textit{iso}-propanol, respectively.

Similar mechanisms were studied by \citet{Gannon07} for the gas-phase addition of the CN radical to propylene, who found them to have null or minimal activation energy barriers. These latter were included by \citet{Garrod17} as grain-surface processes. In that network we assumed that the product of the grain-surface reaction would be the lowest-energy radical (rather than two products, as would be expected in the gas-phase low-density limit). Here we assume the same product ratios as with the equivalent OH reaction above. In both the CN and OH reactions, routes for the abstraction of H, to produce HCN or H$_2$O and the C$_3$H$_5$ radical, are also included.

Finally, following the approach of \citet{Garrod22}, we include surface and bulk-ice reactions between the ground-state diradical CH$_2$ (methylene) and the stable molecules C$_2$H$_5$OH and C$_2$H$_5$CN. Although these reactions are assumed to have activation energy barriers, they will result in two radical products which are assumed to recombine to produce a larger molecule with 50\% efficiency. In this way, the CH$_2$ radical may insert itself in a two-step process (H-abstraction followed by CH$_3$ radical addition) into ethanol or ethyl cyanide to produce either the \textit{normal} or \textit{iso} form of propanol or propyl cyanide. There are no measured data apparent in the literature for these particular reactions, although gas-phase data exist for the analogous CH$_2$ + C$_3$H$_8$ reactions (leading to radicals). The gas-phase activation energy barriers are not very well defined for these reactions either, but they seem to indicate that production of the primary C$_3$H$_7$ radical would be favored \citep{Tsang88}, which for the ethanol-related reactions would lead to the preferential production of $n$-propanol.

Tables \ref{tab-addition} and \ref{tab-abstraction} provide details of most of the new, existing, or updated grain-surface/bulk-ice reactions in the network related to propanol and (in a few cases) to propyl cyanide. Table \ref{tab-addition} contains mostly radical-radical or atom-radical addition reactions. Table \ref{tab-abstraction} shows H-abstraction reactions related to H and OH attack; similar abstraction reactions were also included in the network for radicals such as NH$_2$, in keeping with the construction of past networks, but they have a relatively small effect on the results in the present case. All grain-surface and bulk-ice reactions in the network are equally available for all diffusive and non-diffusive reaction mechanisms.

Binding energies and enthalpies of formation used in the model for several existing or new chemical species are shown in Table \ref{tab-quantities}. Binding energies, where not directly measured, are interpolated from known values, following past publications. The assumed binding energy of water (on amorphous water ice) is the value adopted by Jin et al. (in prep.), which is lower than the one employed by \citet{Garrod22}. Binding energies are used to define the thermal and non-thermal desorption rates of these molecules from the grains. For chemical desorption, whose rate calculations require the enthalpies of formation, an efficiency factor \citep{Garrod07} of $a=0.001$ is assumed, to ensure that methanol is not overproduced in the gas phase at low temperatures.

Finally, a selection of (primarily) destruction reactions were included for all new species, encompassing photodissociation via external and CR-induced UV radiation, and ion-molecule reactions with C$^+$, H$^+$, He$^+$, H$_3^+$, H$_3$O$^+$, and HCO$^+$ ions. For reactions with C$^+$, H$^+$, or He$^+$, immediate fragmentation of the neutral molecule ensues. For H$_3^+$, H$_3$O$^+$, and HCO$^+$, reaction leads to proton transfer, producing protonated forms of the neutrals involved. Electronic recombination reactions for these ions were also included, with a default fraction of 5\% of these leading back to the original neutral, while the other branches involve the breakdown of the molecule's heavy-atom backbone. Following \citet{Taquet16}, and as also implemented by \citet{Garrod22}, we include proton transfer reactions between the protonated molecules and ammonia (NH$_3$) assuming 100\% efficiency, for molecules whose proton affinity is less than that of ammonia. This tends to enhance the ability of the neutrals to survive in the gas phase. Further details of the construction process for the gas-phase network are provided by \citet{Garrod08}, \citet{Garrod13}, and others. 

\begin{figure*}
\centerline{
\resizebox{0.45\hsize}{!}{\includegraphics[angle=0]{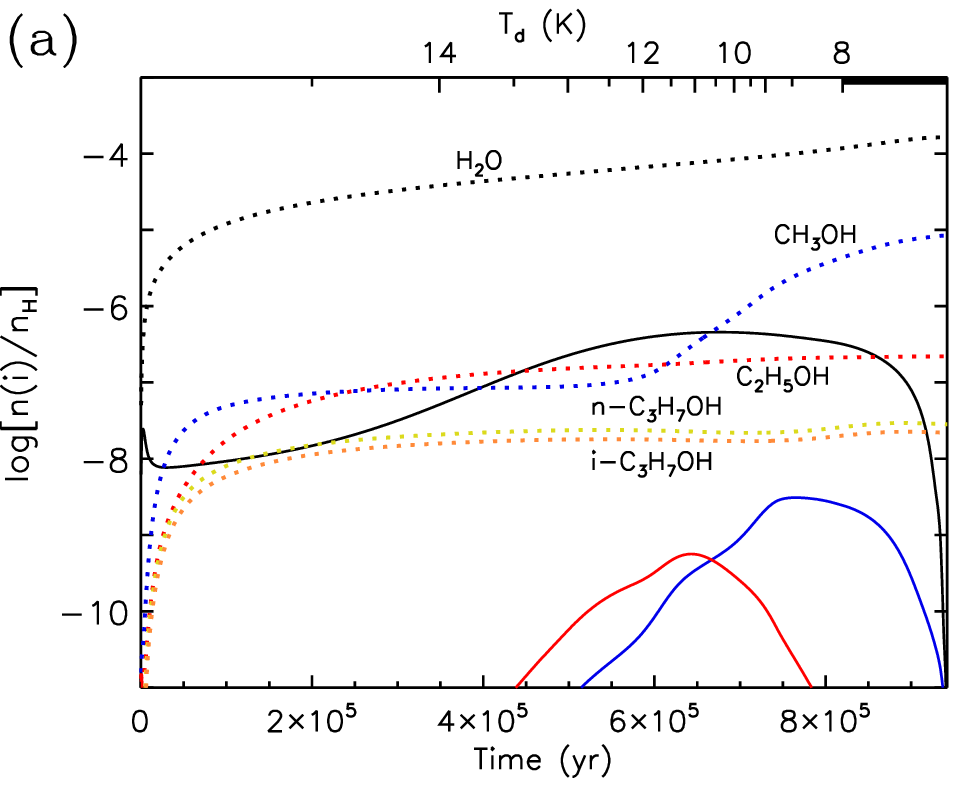}}
\resizebox{0.479\hsize}{!}{\includegraphics[angle=0]{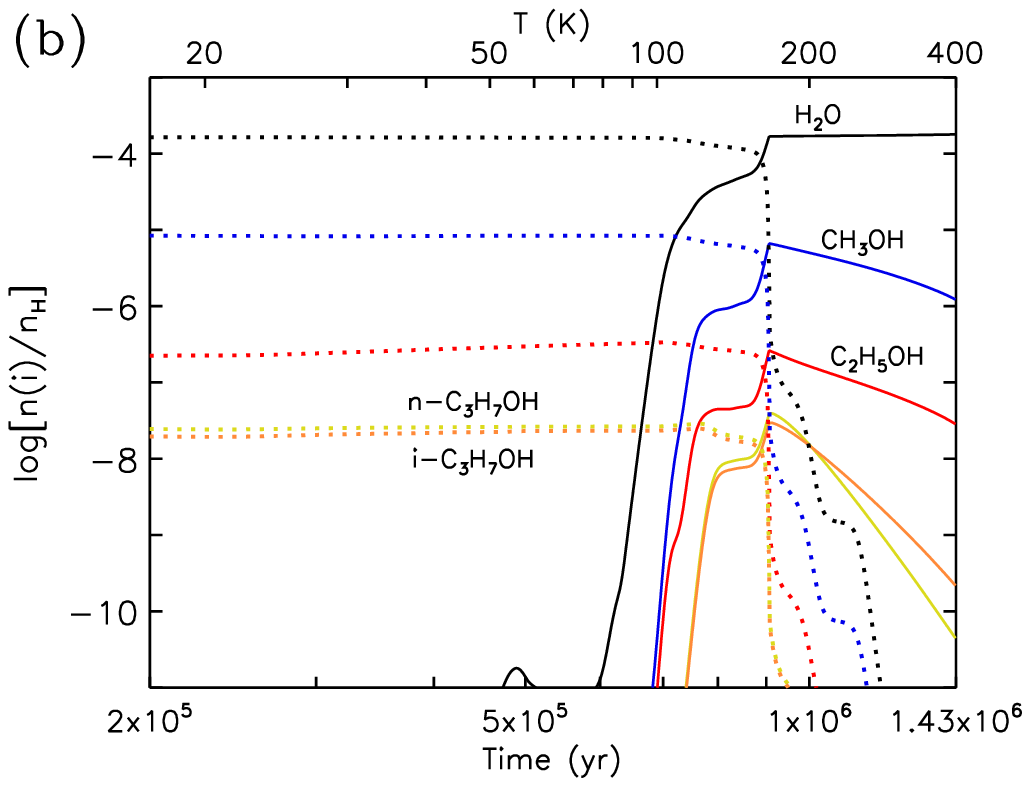}}
}
\caption{Abundances with respect to total H for selected OH-group bearing molecules during the collapse stage and subsequent warm-up using the \textit{slow} warm-up model. Solid lines indicate gas-phase species; dotted lines of the same colours indicate the same species on the dust grains. \textbf{Panel (a)}: Stage 1 (collapse) model results, plotted linearly against time. The dust temperature is shown on the top axis, with an initial value of $\sim$14.7~K, which falls to a minimum of 8~K, while the gas temperature is held at 10~K throughout the collapse model. \textbf{Panel (b)}: Stage 2 (warm up) model results, plotted logarithmically against time. Time values correspond to the time during the warm-up stage, which are plotted beginning $2 \times 10^5$ yr into the warm up. The top axis shows the gas and dust temperatures, which are identical at this point in the model.}
\label{f:mod:OH}
\end{figure*}

\begin{figure*}
\centerline{
\resizebox{0.45\hsize}{!}{\includegraphics[angle=0]{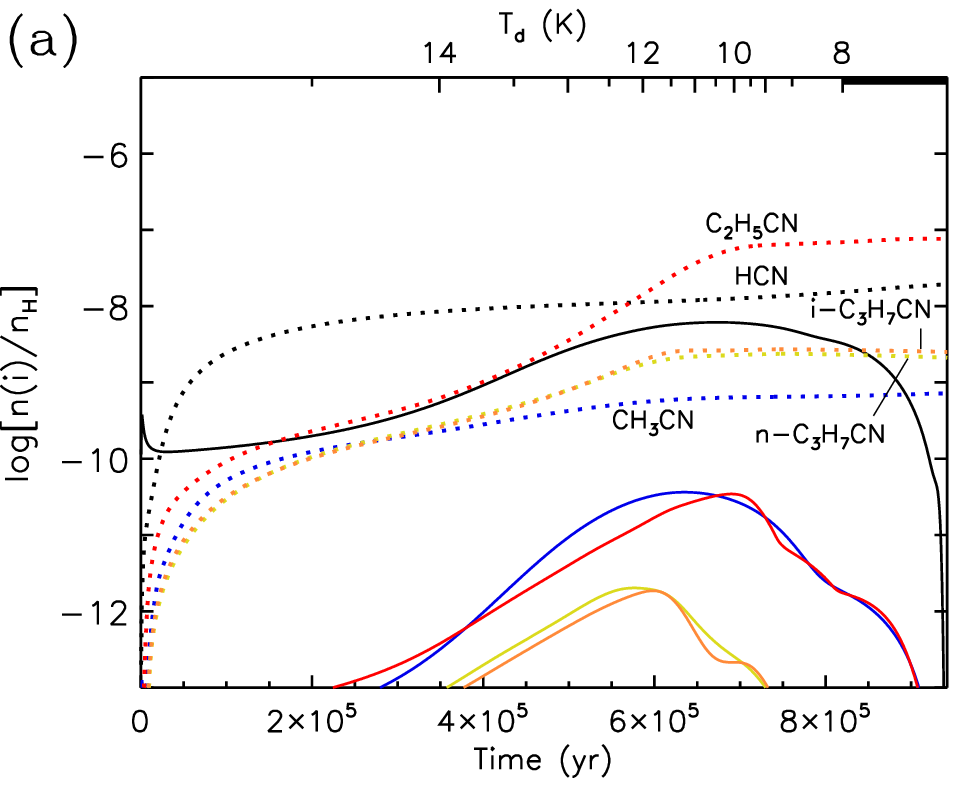}}
\resizebox{0.479\hsize}{!}{\includegraphics[angle=0]{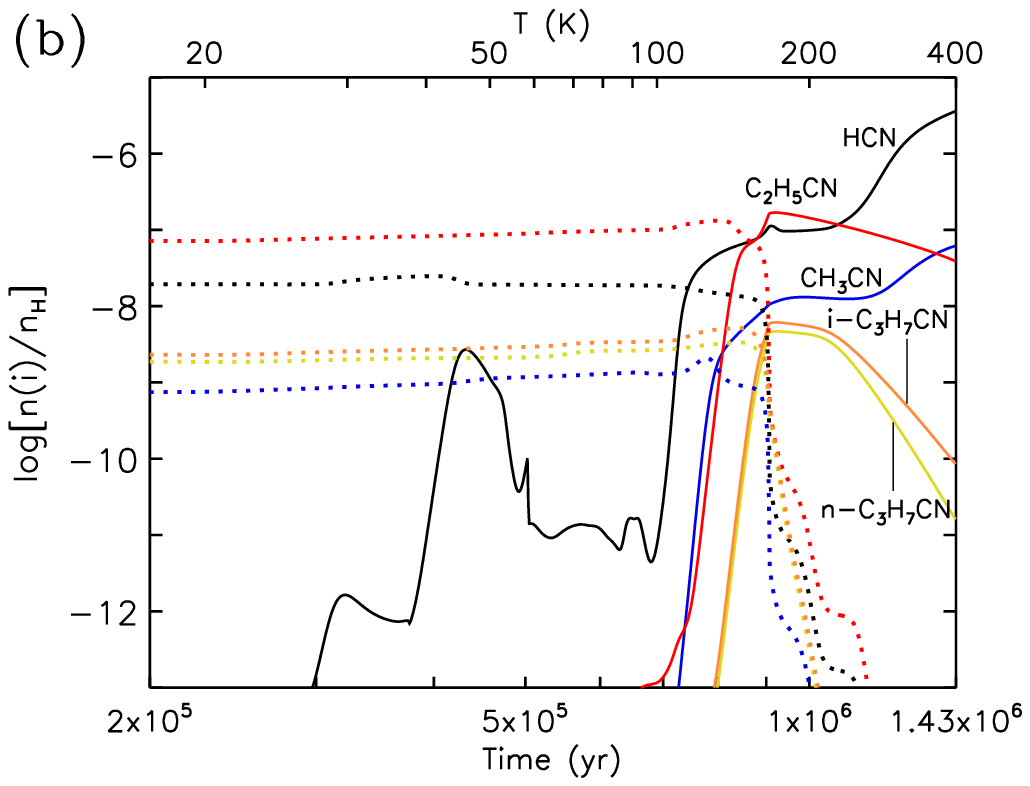}}
}
\caption{Abundances with respect to total H for selected CN-group bearing molecules during the collapse stage and subsequent warm-up using the \textit{slow} warm-up model. Other details as per Fig. \ref{f:mod:OH}.}
\label{f:mod:CN}
\end{figure*}

\subsection{Treatment of orientation-specific reactions on dust grains}
\label{sec:orientation}

Most of the new barrier-mediated reactions occurring on the grains, as well as some barrierless reactions, make use of the new $F_{\rm{dir}}$ and $F_{\rm{comp}}$ parameters introduced by \citet{Garrod22}, as these reactions are important in differentiating the two forms of propanol and of propyl cyanide produced in the models. The cyanide-related reactions that were already present in the network have been updated to include this parameterization in a consistent way. Only a selection of new reactions (i.e. those most relevant to propanol and propyl cyanide) have yet been set up to use this method, which will be developed comprehensively in future work.

The $F_{\rm{dir}}$ and $F_{\rm{comp}}$ parameters allow account to be taken of the directionality of surface and bulk-ice reactions, and of the competition between barrier-mediated reactions that involve identical reactants. $F_{\rm{dir}}$ describes the fraction of orientation space of the two reactants from which a particular reaction is accessible. For example, a particular H atom in a large molecule may be susceptible to abstraction only with a particular orientation of the attacking radical with respect to the molecule. In this way, steric effects related to the specific structure of individual molecules may be considered.

All reaction types may in principle be assigned an $F_{\rm{dir}}$ value in the model, but the $F_{\rm{comp}}$ parameter is needed only for reactant-reactant pairings that have multiple barrier-mediated reactions available, for which the competition between these reactions must be controlled. For example, abstraction of different H atoms from the same molecule may be possible, depending on the orientation of the arriving radical with respect to the molecule (as defined by $F_{\rm{dir}}$), while each abstraction process may have a different barrier. However, under certain orientations multiple abstraction processes may be accessible at once, and would therefore be competitive. A value of $F_{\rm{comp}}=1$ for a particular abstraction reaction would indicate that for all $F_{\rm{dir}}$ values in which that reaction is accessible, it is competing with all other barrier-mediated processes according to their thermal or tunneling rates. A value of $F_{\rm{comp}}=0$ would indicate that no other reaction outcomes are accessible at orientations accessible to the reaction of interest.

A more detailed description of the implementation of these parameters is provided by \citet{Garrod22}. Here, we have chosen a statistical distribution of reaction accessibility, in the absence of other information. That is, the three H atoms on the primary carbon of the ethanol molecule and the two H atoms on the secondary carbon of the same molecule are accessible for abstraction in a ratio of 3:2. For propane, which has six H atoms on the primary carbons, this ratio is 6:2; this is implemented in the model by adopting $F_{\rm{dir}}$ values of 0.75 and 0.25 for reactions 52/53 and 54/55. In the case of abstraction of secondary-carbon H atoms, as well as these reactions having lower accessibility ($F_{\rm{dir}}$) they are also much more likely to be in direct competition with abstraction from the primary carbon, due to crowding. $F_{\rm{comp}}=1$ is therefore chosen for attack on the secondary-carbon H atoms, while $F_{\rm{comp}}=0.25$ is chosen for the primary-carbon H atoms. In fact, due to the lower barriers to attack on the secondary-carbon H atoms, this route will nevertheless tend to win the competition in the fraction of cases ($F_{\rm{dir}}$) in which it is accessible.

As noted by \citet{Garrod22}, the $F_{\rm{dir}}$ values for any particular collection of reaction outcomes do not need to sum to unity, but we indeed choose this to be the case for the generic parameters that we apply to all relevant alcohol- or nitrile-related reactions in the new network. For molecules with three functional groups, such as C$_2$H$_5$OH, C$_2$H$_5$CN, or  C$_3$H$_8$, the central CH$_2$ group is assumed to be accessible to attack in 25\% of cases ($F_{\rm{dir}}$ = 0.25), based on the statistical fraction of H atoms on the central carbon out of the total of 8 present in propane. Hydrogen atoms on a terminal carbon are assumed to be accessible in 37.5\% of cases ($F_{\rm{dir}}$ = 0.375).
Then, regardless of the reaction or functional group, the other end of the molecule is also assumed to be accessible in 37.5\% of cases. The latter includes the abstraction of H from the hydroxyl group in ethanol (reactions 58 and 61). Since H-abstraction cannot occur from the CN group in ethyl cyanide, in 37.5\% of cases the meeting of H or OH with C$_2$H$_5$CN will lead to no reaction accessibility at all.

These same $F_{\rm{dir}}$ values are applied also to the addition reactions of OH and CN to propylene. As noted in Sect. \ref{network}, experimental and theoretical determinations of the ratios of radical production are available in the case of OH + C$_3$H$_6$ (reactions 49 and 50). While the generic statistical values we adopt here for all similar reactions represent crude assumptions ($F_{\rm{dir}}=0.375$ for radical addition to a terminal carbon and $F_{\rm{dir}}=0.25$ to a central carbon), the resulting 3:2 ratio lies close to the experimental and theoretical values.

For molecules with four functional groups, such as $n$-C$_3$H$_7$OH and $i$-C$_3$H$_7$OH, a similar statistical treatment is applied, using C$_4$H$_{10}$ as the model, i.e. attack on terminal functional groups has $F_{\rm{dir}}=0.3$ (3/10), while the inner functional groups split the remainder. This means that the hydrogen atom on the central carbon in $i$-C$_3$H$_7$OH has an accessibility of only 10\%.

Separate from $F_{\rm{dir}}$ values, which determine the fraction of reactant meetings from which a particular reaction is accessible (constituting a pre-reaction efficiency factor applied to the meeting rate), branching ratios may be used to decide on the outcome of otherwise identical meeting and reaction conditions (i.e. post-reaction splitting of the reaction rate). Branching ratios always sum to unity. Here they are applied to several barrierless radical-radical reactions, in cases where H-atom transfer is in competition with addition. The two branches may be considered accessible from the same orientations.

\begin{table*}
 \begin{center}
 \caption{
 Peak gas-phase abundances of selected molecules obtained with the three warm-up timescale models. Abundances are shown as a fraction with respect to H$_2$ at the moment when the molecule reaches its own peak abundance. The temperature of the dust and gas corresponding to the peak abundance value is also shown.
}
 \label{t:modabun}
 \vspace*{0.0ex}
\begin{tabular}{lrrrrrrrr}
 \hline\hline
  & \multicolumn{2}{c}{Fast} && \multicolumn{2}{c}{Medium} && \multicolumn{2}{c}{Slow} \\
\cline{2-3} \cline{5-6} \cline{8-9} \\

Molecule & $n[i]/n[\mathrm{H}_{2}]$ & $T$ (K)  && $n[i]/n[\mathrm{H}_{2}]$ & $T$ (K)  && $n[i]/n[\mathrm{H}_{2}]$ & $T$ (K)  \\
 \hline
CH$_3$OH         & 1.88(-5) & 168 && 1.73(-5) & 167 && 1.33(-5) & 166  \\
C$_2$H$_5$OH     & 5.33(-7) & 241 && 4.89(-7) & 167 && 5.12(-7) & 166  \\
$n$-C$_3$H$_7$OH & 7.12(-8) & 212 && 7.29(-8) & 168 && 7.90(-8) & 167  \\
$i$-C$_3$H$_7$OH & 5.78(-8) & 219 && 6.03(-8) & 169 && 5.92(-8) & 167  \\
CH$_3$CN         & 3.42(-9) & 400 && 1.27(-8) & 398 && 1.23(-7) & 400  \\
C$_2$H$_5$CN     & 1.96(-7) & 170 && 2.33(-7) & 167 && 3.38(-7) & 171  \\
$n$-C$_3$H$_7$CN & 9.25(-9) & 180 && 9.25(-9) & 184 && 9.35(-9) & 171  \\
$i$-C$_3$H$_7$CN & 1.10(-8) & 180 && 1.18(-8) & 181 && 1.23(-8) & 171  \\
 \hline 
 \end{tabular}
 \end{center}
 \vspace*{-2.5ex}
 \tablefoot{$x$ ($y$) means $x \times 10^y$.}
 \end{table*}

\begin{table*}
\begin{center}
\caption{Ratios of branched-chain molecules with their straight-chain forms, for each model, as well as ratios between larger and smaller homologues; data correspond to the ratios of the peak values given in Table \ref{t:modabun}. Observed ratios are also shown.}
\label{tab-ratios}
\renewcommand{\arraystretch}{1.0}

\begin{tabular}[t]{rcccc}
\hline \hline
Molecular ratio & \multicolumn{1}{c}{[Fast]} & \multicolumn{1}{c}{[Medium]} & \multicolumn{1}{c}{Slow} & Observations \\
\hline
CH$_3$OH / C$_2$H$_5$OH                          &  35.3  &  35.4  &  25.4  & 19   \\
CH$_3$OH / $n$-C$_3$H$_7$OH                      & 264    & 237    & 168    & 350  \\
CH$_3$OH / $i$-C$_3$H$_7$OH         \vspace{1mm} & 325    & 287    & 224    & 630  \\
C$_2$H$_5$OH / $n$-C$_3$H$_7$OH                  &   7.48 &   6.71 &   6.60 & 18   \\
C$_2$H$_5$OH / $i$-C$_3$H$_7$OH     \vspace{1mm} &   9.22 &   8.11 &   8.80 & 33   \\
$n$-C$_3$H$_7$OH / $i$-C$_3$H$_7$OH \vspace{1mm} &   1.23 &   1.21 &   1.33 & 1.8  \\

C$_2$H$_5$CN / CH$_3$CN                          & 57.3   & 18.3   &   2.74  &   2.8 $^a$   \\
C$_2$H$_5$CN / $n$-C$_3$H$_7$CN     \vspace{1mm} & 21.2   & 25.2   &  36.1   &  34 $^{a,b}$ \\
$n$-C$_3$H$_7$CN / $i$-C$_3$H$_7$CN              & 0.842  & 0.785  &   0.762 &  2.5 $^b$   \\

\hline
\end{tabular}
\end{center}
\tablefoot{Observational values are those presented in this work, with the exception of $^a$ \citet{Belloche16}, $^b$ \citet{Belloche14}.
}

\end{table*}

\subsection{Chemical modelling results} \label{model-results}

Figure \ref{f:mod:OH} shows the stage 1 (collapse) and stage 2 (warm up) chemical abundances of methanol, ethanol, $n$-propanol, $i$-propanol, and water, with respect to total hydrogen. Solid lines in each correspond to gas-phase molecules, with dotted lines of the same colours indicating the grain-surface forms of the same species. The stage 1 results (panel a) are plotted linearly with time. It can be seen that methanol production on grains begins early on in the model, but only reaches a value close to its peak abundance in the ice much later on. Ethanol also begins to be formed quite early on, but quickly reaches a level close to its maximum on the grain surfaces. Similar behavior is seen for both forms of propanol, although there is a slight increase toward the end of stage 1.

Methanol is mainly produced by repetitive H-atom addition to CO on the grains, as per past models. Ethanol is initially produced mainly within the bulk ice mantle, through PDI reactions (i.e. non-diffusive reactions driven by photodissociation) between C$_2$H$_5$ and OH (listed as reaction 5 in Table \ref{tab-addition}). The photodissociation is the result of external UV radiation penetrating the initial 3 magnitudes of extinction. Similar reactions drive vinyl alcohol (C$_2$H$_3$OH) production, and the hydrogenation of vinyl alcohol in the bulk ice by mobile H atoms soon takes over as the main ethanol formation mechanism (reactions 1, 2, 10, and 12). Once $A_{\rm{V}}$ exceeds around 5 mag, the external UV field only has weak effect, and the production of solid-phase ethanol is dominated by surface reactions such as methylene addition to methanol (reaction 4) and hydrogenation of ketene (CH$_2$CO) and acetaldehyde (CH$_3$CHO). Once formed on the grain/ice surface, that ethanol is incorporated into the bulk ice as the mantle grows.

Production of \textit{normal}- and \textit{iso}-propanol begins within the bulk ice similarly to ethanol, driven by the PDI reaction of OH with propylene (C$_3$H$_6$) (reactions 49 and 50), followed by hydrogenation of the resultant radical by mobile H (reactions 24 and 33). Being barrierless, these OH reactions lead to a preference (3:2) for $n$-C$_3$H$_7$OH production, based on their $F_{\rm{dir}}$ values. However, a modest amount of propanol is also formed by PDI reactions of OH with either the primary or secondary radical forms of C$_3$H$_7$  (reactions 20 and 29), which are themselves formed by diffusive reactions of atomic H with propylene. These reactions (46 and 47) have activation energy barriers that strongly favor the secondary radical; combined with the $F_{\rm{dir}}$ values that favor the primary radical, this leads to approximately equal production of the two. As a result, at these early times, propanol formation only slightly favors the \textit{normal} form, in a ratio somewhat less than 3:2.

The later rise in solid-phase propanol abundances during stage 1 occurs when the gas density becomes high and dust temperatures fall; production on the surface of the ice takes over (followed by incorporation into the mantle). Again, OH reactions with propylene dominate.

During stage 2 (panel b -- \textit{slow} model results are shown), the abundances of $n$- and $i$-propanol are mostly flat until a temperature of around 100~K is reached; OH reactions with propylene and OH addition to C$_3$H$_7$ radicals driven by photodissociation (PDI) enhance propanol abundances a little during this period. 

As water ice begins to leave the grains via thermal desorption, unreacted propanol-precursor radicals (i.e. various structures of formula C$_3$H$_7$O) trapped in the ice are released onto the grain surface, where they may be more easily hydrogenated, boosting the production of both $n$-C$_3$H$_7$OH and $i$-C$_3$H$_7$OH. The reaction of OH with propylene, driven mainly by the PDI mechanism, also continues to produce such radicals in the ice right up until the complete loss of the ice mantles.

Propanol molecules reach their peak abundances in the gas phase immediately following their complete release from the grains. $n$-C$_3$H$_7$OH initially dominates over $i$-C$_3$H$_7$OH, in the ratio preserved from the solid phase: $\sim$1.6. Table \ref{t:modabun} shows the peak abundances achieved in each of the stage 2 warm-up models, along with the corresponding model temperatures at which each peak value occurs. Following their abundance peaks, the $n$:$i$ propanol ratio falls rapidly; however, this is caused by the slightly greater total CR-induced UV destruction rates for the \textit{normal} form. Since all of these rates are simply estimates, the post-peak drop in $n$:$i$ is not a reliable feature of the models.

\begin{figure*}
\centerline{\resizebox{0.99\hsize}{!}{\includegraphics[angle=0]{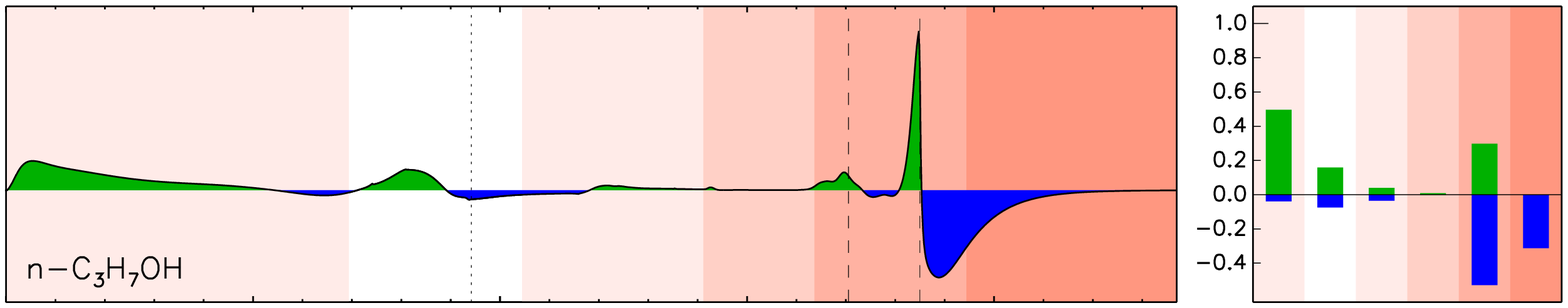}}}
\centerline{\resizebox{0.99\hsize}{!}{\includegraphics[angle=0]{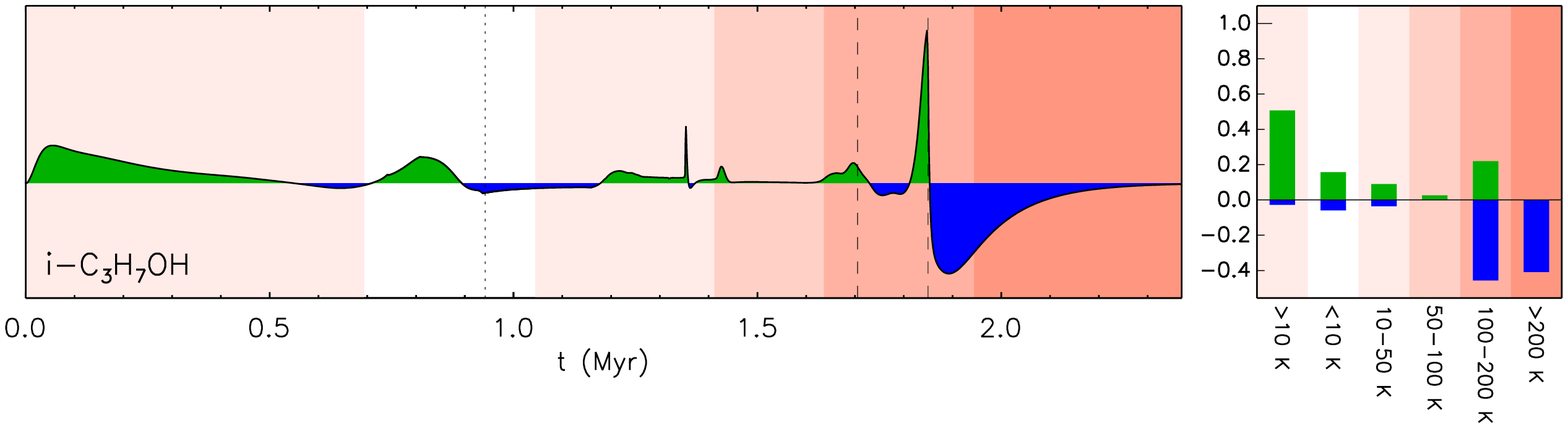}}}
\caption{\textbf{\textit{Left panels:}} Net rate of change (arbitrary units) in the abundances of \textit{normal}- and \textit{iso}-propanol, summed over all chemical phases, during stages 1 and 2. Data correspond to the \textit{slow} warm-up timescale in stage 2; the vertical dotted line indicates the start of the warm-up phase. Net gain is shown in green, net loss in blue. Vertical dashed lines indicate the onset and end-point of water desorption. Background colors indicate the dust temperature regime; from left to right these are: $>$10~K, $<$10~K, 10--50~K, 50--100~K, 100--200~K, 200--400~K. The initial dust temperature is $\sim$14.7~K. \textbf{\textit{Right panels:}} Net rates of change integrated over each temperature range. Positive (formation) and negative (destruction) rates are integrated independently; both are normalized to the total integrated formation rate.}
\label{f:mod:rates}
\end{figure*}

Figure \ref{f:mod:CN} shows the behavior of chemical abundances of various nitriles, including \textit{normal}- and \textit{iso}-propyl cyanide. Like propanol, the propyl cyanides also form on the grains at early times; however, in this case, the production is driven initially by surface reactions. H-abstraction from surface C$_2$H$_5$CN preferentially produces the secondary radical (reaction 63), due to the lower activation energy barrier of this reaction, in spite of its smaller $F_{\rm{dir}}$ value. Production of \textit{iso}-propyl cyanide is therefore marginally higher than that of the \textit{normal} form. 

In contrast to the results of \citet{Garrod17}, the reaction of CN with propylene is not an effective mechanism for propyl cyanide formation in the present models. The removal of bulk diffusion for heavy species (i.e. CN) means that this reaction does not occur rapidly within the ice, even at elevated temperatures, while on the ice surfaces the production rate of the CN radical is not great enough to make this pathway significant.

At times around $5 \times 10^5$ yr into the collapse model, methylene (CH$_2$) insertion into ethyl cyanide allows $n$- and $i$-C$_3$H$_7$CN to grow, with the \textit{normal} form becoming slightly dominant in its abundance. This behavior is then reversed toward the end of the stage 1 model, as H-abstraction from C$_2$H$_5$CN, followed by methyl-group addition, takes over as the main propyl cyanide-forming mechanism on the ice surfaces, which substantially favors the production of the CH$_3\overset{\centerdot}{\mathrm{C}}$HCN radical, thus bumping up the $i$-C$_3$H$_7$CN isomer sufficiently to become slightly more abundant than the $n$ form in the ice mantles.

During the warm-up of the core (panel b), the $n$:$i$ ratio of propyl cyanide formed earlier on is largely preserved, leading to a gas-phase ratio that is close to unity. Note that, in the present model, the gas-phase destruction of propyl cyanide is slower than in the \citet{Garrod17} models, due to proton transfer of protonated propyl cyanide to ammonia, which slows the destructive effects of dissociative electronic recombination.

Table \ref{tab-ratios} shows the ratios of the various alcohols and nitriles obtained from the models and observations; for the nitriles, the observational values derive from \citet{Belloche14} and \citet{Belloche16}. All of the model ratios correspond to the quotient of the peak values given in Table \ref{t:modabun}. Modeled ratios of methanol with the two forms of propanol are broadly in agreement with the observations, i.e. within a factor of around \textbf{2--3} in each case, when comparing the \textit{slow} model results, which were generally found by \citet{Garrod22} to provide the best match to Sgr B2(N) data. 
Ethanol itself is slightly underproduced with respect to methanol, by 
a factor $\sim$1.3, which raises the disagreement in its ratios with 
\textit{normal}- and \textit{iso}-propanol to a factor of something closer to 
3--4.
The modeled $n$:$i$ ratio for propanol is exceedingly close to the observed value. Although the models generally produce a very good match to the observational ratios of the nitriles, the $n$:$i$ ratio for propyl cyanide is close to unity, rather than the observed value of 2.5. The results of the models are discussed further in Sect. \ref{model-discussion}.

\begin{figure}
\centerline{\resizebox{0.99\hsize}{!}{\includegraphics[angle=0]{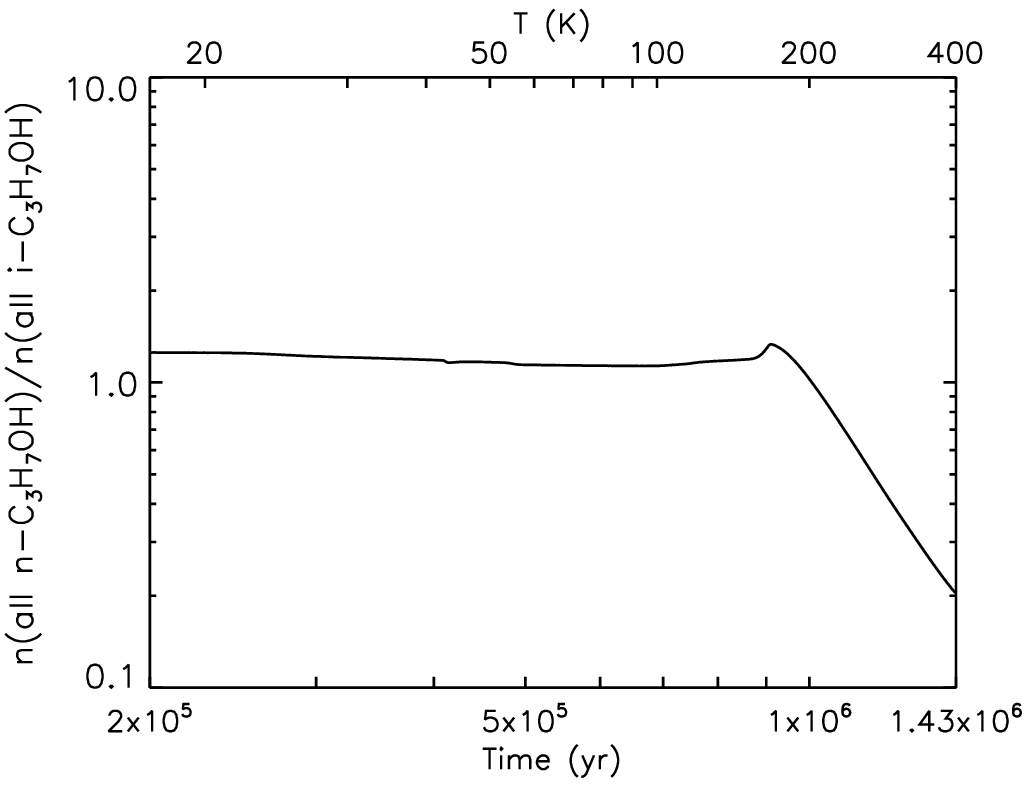}}}
\centerline{\resizebox{0.99\hsize}{!}{\includegraphics[angle=0]{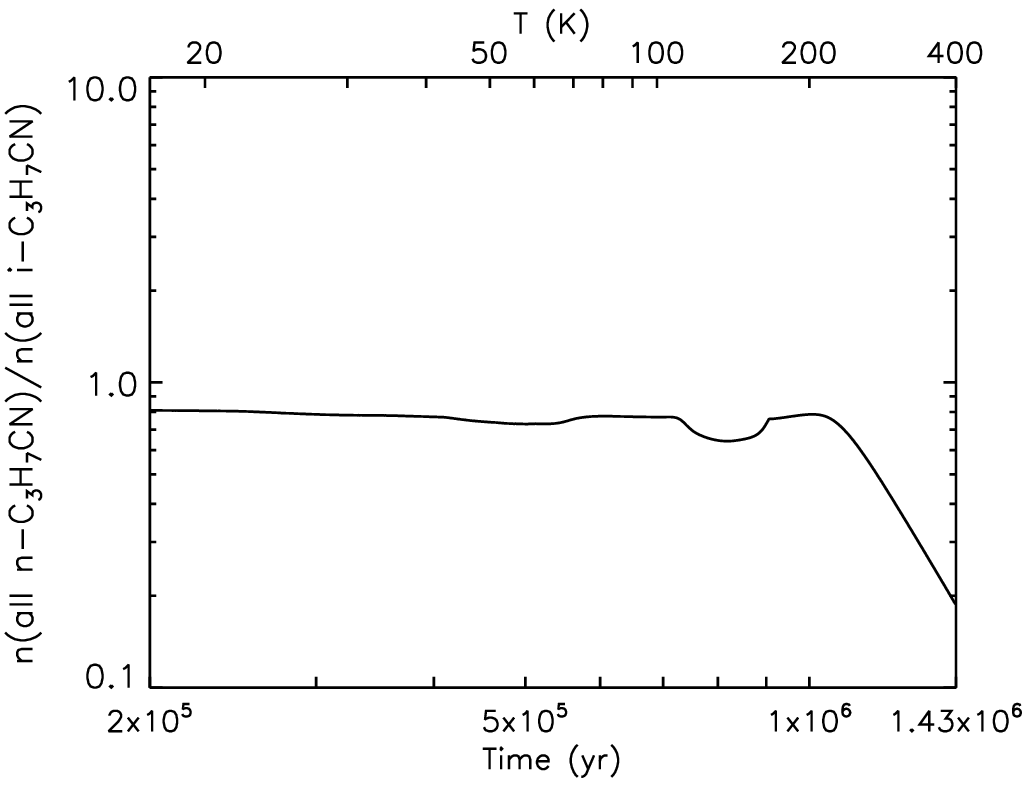}}}
\caption{Ratios of \textit{normal}- to \textit{iso}-propanol (upper panel) and \textit{normal}- to \textit{iso}-propyl cyanide (lower panel), summed over all phases, during the warm-up (stage 2) model.}
\label{f:mod:ratio}
\end{figure}

\section{Discussion}
\label{s:discussion}

\subsection{Comparison of Sgr~B2(N2) results to other sources}

In their article on the detection of \textit{trans} ethyl methyl ether
(\hbox{\textit{t}-C$_2$H$_5$OCH$_3$}), which is an isomer of propanol, toward 
Orion~KL with ALMA and the IRAM 30~m telescope, \citet{Tercero15} also 
reported on a search for the \textit{Gauche}-\textit{anti} conformer of 
\textit{normal}-propanol. They obtained a lower limit on the abundance 
ratio of the vibrational ground state of ethanol to the vibrational ground 
state of \textit{Ga}-\textit{n}-C$_3$H$_7$OH of 60. Accounting for vibrational 
and conformational corrections to the partition function at the temperature that
they determined for this source (100~K), this nondetection translates into a 
lower limit on the abundance ratio of ethanol to \textit{normal}-propanol of 14 
\citep[see][]{Mueller16}. This indicates that, if present in Orion~KL, 
\textit{normal}-propanol is not more abundant than in Sgr~B2(N2b) with respect
to ethanol. However, given that ethanol is $\sim$30 times less abundant than
methanol in Orion~KL \citep[][]{Kolesnikova14} while their ratio is only a 
factor 19 in Sgr~B2(N2b) (Table~\ref{t:coldens}), we can conclude that 
\textit{normal}-propanol is, relative to 
methanol, more abundant in Sgr B2(N2b) than in Orion~KL.

Nondetections of both \textit{normal-} and \textit{iso}-propanol were also
reported toward the Class~0 protostar IRAS~16293-2422B on the basis of the 
Protostellar Interferometric Line Survey (PILS) performed with ALMA 
\citep[][]{Manigand21}. A similar column density upper limit was obtained for 
both isomers ($3 \times 10^{15}$~cm$^{-2}$). Considering the column 
densities of methanol and ethanol derived with the PILS survey toward the same 
source by \citet{Jorgensen18} ($1 \times 10^{19}$ and 
$2.3 \times 10^{17}$~cm$^{-2}$, respectively), the upper limits obtained by 
\citet{Manigand21} imply that both isomers of propanol are at least 3300 and 
80 times less abundant than methanol and ethanol in this source, respectively. 
This means that propanol is, relative to methanol, at least one order of 
magnitude less abundant in IRAS~16293-2422B than in Sgr~B2(N2b). 
\citet{Manigand21} applied conformation corrections to the partition function 
in order to derive their column density upper limits, but they do not mention 
any vibrational correction. At the temperature they considered (100~K), this 
correction factor is smaller than 1.5 and would not significantly reduce the 
large difference between the two sources. This difference for propanol is 
surprising given the strong correlation that was noticed between the chemical 
composition of IRAS~16293-2422B and Sgr~B2(N2) for oxygen-bearing species 
\citep{Jorgensen20}. The correlation between the two sources may break down
beyond a certain degree of chemical complexity.

\subsection{Observed isomeric ratios of propanol and propyl cyanide}

It is striking that the isomeric ratio [\textit{iso}]/[\textit{normal}] that 
we obtained in Sgr~B2(N2) is similar for both propanol and propyl cyanide 
(0.6 versus 0.4). Besides, the abundance ratios of ethanol to 
\textit{normal}-propanol and of ethyl cyanide to \textit{normal}-propyl cyanide 
are similar within a factor of $\sim$2 (15 versus 34). This probably indicates 
similarities in the chemical processes that form these two families of 
molecules, as we discuss further in the next section.

\subsection{Discussion of chemical model results} 
\label{model-discussion}

Although the introduction of non-diffusive grain-surface and bulk-ice reaction mechanisms into the model has changed some of the general behavior, the suggestion by \citet{Garrod17} that OH addition to propylene would be an important mechanism for propanol production is borne out here -- it appears to be dominant throughout the chemical evolution. Furthermore, their prediction that this mechanism would lead to an $i$:$n$ ratio ``around unity or a little less'' also stands up well.

In comparing the modeled abundance ratios with the observations, we 
note that our comparison between the peak abundances of different molecules, 
as shown in Table~\ref{tab-ratios}, introduces a degree of error related to 
the relative desorption temperatures of those species and their survival
timescales in the gas phase; the observed column densities may reflect spatial 
variation that the present models are not equipped to reproduce. However, we
would expect these considerations to be much less important when comparing the 
two forms of propanol, whose desorption properties should be much more similar 
than between the different homologs (e.g., methanol or ethanol). The behavior 
of the cyanides should be affected in a similar way. Thus, the 
\textit{n}:\textit{i} ratios for propanol and propyl cyanide might be 
considered the most robust aspect of the model results when comparing with 
observations.

In fact, the models come remarkably close to the observed abundance ratios of all the molecules considered, including the $n$:$i$ ratios of both propanol and propyl cyanide. Furthermore, while the ratios between different molecules are somewhat variable for different model warm-up timescales, the $n$:$i$ ratios are fairly stable, especially for propanol. Part of this stability may be due to the fact that, in general, the dominant production mechanisms for the \textit{iso} or \textit{normal} forms of either propanol or propyl cyanide tend to be the same throughout; the reaction of OH with propylene is responsible for forming much of the $n$- and $i$-C$_3$H$_7$OH, while the abstraction of H from ethyl cyanide, followed by addition of methyl to the resulting radical is responsible for most $n$- and $i$-C$_3$H$_7$CN production.

Much of the formation of propanol and propyl cyanide occurs during the cold stages of the hot core evolution; Figure~\ref{f:mod:rates} shows the net rates of formation of $n$-C$_3$H$_7$OH and $i$-C$_3$H$_7$OH, summed over all phases (gas, surface, and bulk), from the start of the stage 1 model to the end of stage 2. In the right-hand panels are shown the totals of the net formation rates occurring during several different temperature regimes experienced by the hot core \citep[the sum of net positive rates is shown in green, the sum of net negative rates is shown in blue -- see][for further explanation of this style of plot]{Garrod22}. Nearly half of all propanol is formed while the visual extinction is quite low and the dust temperatures are still greater than 10~K, through photodissociation-induced reactions within the young ice mantles caused by the external UV field. As noted in Sect. \ref{model-results}, a modest contribution of reactions between OH and the two C$_3$H$_7$ radicals in the bulk ice lessens the dominance of the \textit{normal} form at this point in the models.

Later in the models, during the period in which the desorption of the ice mantles becomes significant (indicated in Fig. \ref{f:mod:rates} by the two dashed vertical lines), the OH + C$_3$H$_6$ reactions are more dominant. Also, at these higher temperatures, the barrier against the reaction of H with C$_3$H$_6$ to produce the primary form of the C$_3$H$_7$ radical is less important, so that even the OH + C$_3$H$_7$ mechanism tends to favor production of $n$-propanol more strongly. Figure \ref{f:mod:ratio} indeed shows the $n$:$i$ ratio during the stage 2 evolution, where it may be seen that the ratio reaches its peak at the moment of maximum desorption from the grains. It may be considered, then, that if, during the cold collapse stage, the core were to spend less time at low visual extinctions or began forming ice mantles at a somewhat higher extinction, then while the overall production of propanol would be less, its production would be dominated more by the high-temperature, late-stage mechanisms that favor a higher $n$:$i$ ratio.

In the present models, the ultimate reason for the $n$:$i$ ratio in propanol being a little greater than unity seems to originate in the choice of $F_{\rm{dir}}$ values for certain key reactions, in particular the reactions of OH with propylene. Here, production ratios of 3:2 are chosen for radicals related to the $n$ and $i$ forms of propanol, respectively, and this ratio is approximately preserved in propanol abundances throughout the model run. Theoretical modeling on the outcomes of these reactions in the gas phase (under the high-pressure limit, which is appropriate for the solid-phase conditions of interest here), based on transition state theory analysis, indeed indicate ratios similar to these: \citet{Daranlot10} find 58:42, while \citet{Izsak09} find that 65.8\% leads to terminal carbon addition, i.e. close to 2:1. The experimental work of \citet{Loison10} yields a ratio of 72:28. The 3:2 ratio used in our grain-chemistry network is therefore a little conservative, but still within the bounds provided by detailed studies of this system.

The slightly low $n$:$i$ propanol ratio provided by our models, as compared with the observations, may therefore be explained as an underestimate of the chemical-reaction ratio applied in our network. It is interesting to consider that the observed $n$:$i$ ratio of propanol may be a direct reflection of the underlying branching ratio of the OH + C$_3$H$_6$ reaction.

Although the modelled abundances of propyl cyanide are close to the observed values, the deviation in the $n$:$i$ ratio in this case appears a little more stark. The models do not allow for a substantial degree of production of propyl cyanide through CN radical addition to propylene in the same way as propanol is formed. Although this mechanism is not hindered in any way, it relies on the photodissociation of HCN in the ice to produce CN, in the same way as the analogous propanol-forming reaction relies on the photodissociation of water. As may be seen from Figures \ref{f:mod:OH} and \ref{f:mod:CN}, the solid-phase abundance of HCN is around four orders of magnitude lower than that of water. As a result, the reaction of CN with propylene is simply not rapid enough to contend with the competing mechanism relating to H-abstraction from ethyl cyanide on the surface of the ice. Unlike the OH and CN addition processes, this reaction does have barriers that are substantial enough to favor strongly the production of the secondary radical at early times, which leads to the formation of $i$-C$_3$H$_7$CN upon the addition of a methyl radical. However, the greater accessibility (i.e. greater $F_{\rm{dir}}$ value) of the process that leads to the primary radical tends to balance out the effects of its higher barrier, especially when the dust temperatures are close to their minimum, which reduces the possibility that the H atom attacking ethyl cyanide will diffuse away before the reaction can occur. This effect is more pronounced in the bulk ice, where atomic H has a higher barrier to diffusion than on the ice surface.

Ultimately, it is the steric effects involved in the attack of H on C$_2$H$_5$CN on the grains (reactions 62 and 63), which are reproduced by the adoption of the $F_{\rm{dir}}$ and $F_{\rm{comp}}$ parameters in the models, that determine the maximum allowed $n$:$i$ ratio for propyl cyanide (here, 3:2). Under conditions where H atoms have low mobility, reaction will commonly occur in spite of the activation energy barriers, and thus $F_{\rm{dir}}$ will determine the ratio of primary to secondary radicals produced. However, the temperature threshold at which the barriers become unimportant will indeed be a function of those very activation energy values, and these are poorly constrained in the literature; our models use only estimates based on analogous (gas phase) reactions involving propane. If the activation energy for abstraction of H from the secondary carbon in propyl cyanide were nearer to the value for the attack on the primary carbon, then the $n$:$i$ ratio produced by the models would be closer to our chosen statistical limit of 3:2, although this value itself is insufficient to reproduce the observed ratio. Assuming that no other chemical mechanism is strongly contributing to propyl cyanide production, then this could indicate that abstraction of H from the secondary carbon of C$_2$H$_5$CN may be more sterically hindered than we assume here. Such an effect might be more prevalent when manifested on an ice surface or within the ice mantle, wherein any sort of roaming effect for the attacking H atom could be very limited due to the surrounding ice structure.

As noted above, the $n$:$i$ ratios for propanol and propyl cyanide that we observe toward Sgr B2(N) are similar enough that one might suspect similar underlying chemical processes to be forming them. However, unless the chemical models vastly under-produce HCN, and thus CN, in the ices, it seems unlikely that propanol and propyl cyanide share \textit{exactly} analogous formation mechanisms. Nevertheless, similar steric effects may be controlling the reactivity of their precursors in similar ways and to similar degrees. This would be the case if activation energies played only a small role, with the physical accessibility of either the primary or secondary carbon functional group being the main determinant of the outcome of a particular reaction. Such an effect may be more prevalent within and upon low-temperature ices, in which surrounding water molecules may physically inhibit access to particular reaction channels, while the attacking atom or radical may also be trapped or sufficiently immobile such that a reaction would become inevitable, even if an activation energy barrier should be present. In that case, the ratio of the probabilities of attack on a terminal versus central functional group of some arbitrary molecule should assume a value that would somewhat favor reaction at the terminal methyl group (potentially by a factor of say 2), due to the terminal carbon holding a larger number of H atoms and having greater accessibility from different orientations.

It is possible, then, that the similar $n$:$i$ ratios of propanol and propyl cyanide are due to the reactions that form them both lying in this steric limit of reaction accessibility, caused by the reactions occurring in/upon ices. To determine the veracity of this supposition will require detailed chemical calculations in which the relevant reactions are simulated on an ice surface or in the presence of a surrounding ice matrix. Results determined from gas-phase measurements or calculations may not be sufficient to approximate such a slim divergence between the \textit{iso} and \textit{normal} forms of the molecules studied here.

It is valuable to compare our models with other recent treatments of propanol chemistry. The astrochemical models presented by \citet{Manigand21} considered a generic form of propanol, using a network that incorporated the scheme proposed by \citet{Qasim19} for the hydrogenation of grain-surface propenal (C$_2$H$_3$CHO) to propanal (C$_2$H$_5$CHO), and further to \textit{normal}-propanol. They also included a reaction between atomic O and propane (C$_3$H$_8$), producing the radicals OH and C$_3$H$_7$, which could further recombine to form propanol. The latter process in particular was highlighted by Manigand et al. as being a productive route to propanol in their models. The propenal/propanal routes are not included in our own network, as we do not have propenal in the model, and its full inclusion would require a far more extended network. Nor does our model include the apparently more important reaction of atomic oxygen with propane; however, the activation energy barriers for this process are not well defined in the literature (because the fits to experimental gas-phase rates, conducted at room temperature and above, have a strong temperature dependence in the collisional portion). Our own Evans-Polanyi calculations for the H-abstraction process \citep[based on the data provided by][]{Dean99} to form each of the two C$_3$H$_7$ isomers suggests barriers that are rather higher than those employed by Manigand et al. (i.e. 2100--2800~K, vs. 1000~K for each branch), with the secondary radical energetically preferred, as is typical. Thus, this process should be somewhat less efficient than the OH reaction with propylene, all else being equal. Furthermore, the production rate of O in the ice mantles, either by photodissociation of OH or CO$_2$ or through reactive mechanisms, is substantially slower than that of OH, which occurs primarily through photodissociation of water. In our model, it therefore seems unlikely that this process would make a major contribution to \textit{normal}-propanol production. If the propenal/propanal mechanism that was also considered by Manigand et al. were to become an important contributor to \textit{normal}-propanol production, it is unclear how stable the $n$:$i$ ratio might be among different astronomical sources.

Unfortunately, it is difficult to compare directly our chemical models with those of \citet{Manigand21}, as the latter use a purely diffusive grain-surface/mantle chemistry, meaning that the production of COMs is strongly dependent on radical mobilities, which may only become significant at elevated temperatures. Our own models rely on non-diffusive processes to form many complex organic molecules on the grains \citep{Garrod22}, which can occur at very low temperatures and early times.

\subsection{The crucial role of high angular resolution}

Compared to its predecessor EMoCA, the higher angular resolution of the ReMoCA 
survey turned out to be crucial in bringing down the degree of spectral 
confusion that limits our ability to identify low-abundance COMs in the 
emission spectra 
of hot cores. The smaller beam of ReMoCA allowed us to resolve the molecular 
emission of Sgr~B2(N2), the secondary hot core of Sgr B2(N), and thereby 
reveal regions of smaller velocity dispersion. It is striking to realize that 
the continuous gain in our sustained efforts over the past 15 years to 
decipher the molecular content of Sgr~B2(N) has greatly benefitted from a
substantial increase in angular resolution that came along with a decrease of 
the spectral line widths of this star forming region by a factor of 5. Both 
are required to beat the inevitable crowding of emission from various species 
near the confusion limit. This long-term project started with an 
angular resolution of $\sim$30$\arcsec$ using the IRAM 30~m telescope that 
traced lines with a FWHM of 17~km~s$^{-1}$ \citep[][]{Belloche13}, continued 
with ALMA at an angular resolution of 1.6$\arcsec$ that unveiled lines with 
a FWHM of 5~km~s$^{-1}$ \citep[][]{Belloche16}, and now culminates in revealing 
line widths of 3.5~km~s$^{-1}$ thanks to the 0.6$\arcsec$ angular resolution
achieved with the ReMoCA survey, and even 2~km~s$^{-1}$ at the edge of the 
molecular emission detected around Sgr~B2(N2).

While the gain in angular resolution was key in reducing the spectral 
confusion in the Sgr~B2(N) region, the gain in sensitivity allowed by the 
higher number of antennas in ALMA's cycle 4 and the longer integration time 
awarded to the ReMoCA survey also played a crucial role in the detection of 
both isomers of propanol. The abundance ratios derived for the latter with 
respect to ethanol with the ReMoCA survey indicate that the EMoCA survey was 
short of a detection of propanol by a factor of $\sim$2 in sensitivity.

All of the numerous new detections of COMs reported in 2021--2022 (see the 
"Molecules in Space" webpage of the 
CDMS\footnote{https://cdms.astro.uni-koeln.de/classic/molecules}) were 
obtained with single-dish telescopes, thanks to three surveys that achieved 
high sensitivities through huge amounts of observing time and large 
instantaneous bandwidths: the QUIJOTE survey of the dark cloud core TMC1 with 
the Yebes 40~m telescope \citep[e.g.,][]{Cernicharo21}, the GOTHAM survey of 
TMC1 with the Green Bank Telescope \citep[e.g.,][]{McGuire21}, and the survey 
of the molecular cloud G$+$0.693$-$0.027 in the vicinity of Sgr~B2(N) with the 
Yebes 40~m and IRAM 30~m telescopes \citep[e.g.,][]{RodriguezAlmeida21}. 
Despite the obvious strength of these single-dish surveys, which profit from 
the large angular sizes of the emission regions, our detection of 
\textit{iso-}propanol and \textit{normal}-propanol shows that ALMA still has a 
role to play 
in the discovery of COMs, especially in star forming regions at the scales 
where protostellar systems emerge. However, strategies must be found to beat 
the confusion limit that is easily achieved in these hot and dense regions. 
Resolving low-velocity-dispersion regions at high angular resolution is one 
way, which we have explored here; going to lower frequencies with, e.g., 
band 2 of ALMA in the near future may also be promising.

Interestingly, the conformers of \textit{normal}-propanol that contributed to 
its detection toward Sgr~B2(N2b), the \textit{Gauche-gauche'} and 
\textit{Anti-gauche} conformers, are two of its higher-energy conformers 
(though the energy of all conformers is within tens of wavenumbers).
This illustrates that it is difficult to predict which conformer of a given 
molecule should be the most favorable one to achieve a detection in the 
line-crowded spectrum of a hot core, in particular when the energy difference is
small with respect to the excitation (rotation) temperature, and even more so,
if the dipole moment or one of its components is more favorable. The key 
factor turns out to be the degree of contamination of the spectral lines of 
that molecule from emission of other molecules. This suggests that 
spectroscopists should also characterize higher-energy conformers of the COMs 
investigated in their laboratories and not only focus on the lowest-energy 
ones.

\section{Conclusions}
\label{s:conclusions}

The high angular resolution and sensitivity achieved with ALMA in the frame of 
the ReMoCA survey has led to the first interstellar detection of 
\textit{iso}-propanol, and the first robust detection of its isomer, 
\textit{normal}-propanol, in a hot core. These detections were obtained toward 
Sgr~B2(N2b), a new position revealed by the ReMoCA survey and characterized by 
narrow line widths of 3.5~km~s$^{-1}$ in the secondary hot core of Sgr~B2(N). 
The main conclusions of this work are the following:

\begin{enumerate}
 \item Both conformers of \textit{iso}-propanol and two of the higher-energy 
 conformers of \textit{normal}-propanol have contributed to the detections of 
 these isomers toward Sgr~B2(N2b). The transitions of the other three 
 conformers of \textit{normal}-propanol fall at frequencies that are more 
 affected by blending with emission of other species. Therefore, the  
 zero-point energy of a conformer of a given molecule does not prefigure its 
 detectability in the congested spectra of hot cores.
 \item \textit{iso}-Propanol and \textit{normal}-propanol are 600 and 350 
 times less abundant than methanol in Sgr~B2(N2b), respectively. They are 30 
 and 20 times less abundant than ethanol, respectively. 
 \item The abundance of \textit{iso}-propanol relative to 
 \textit{normal}-propanol is 0.6 toward Sgr~B2(N2b), which is strikingly 
 similar to the isomeric ratio of 0.4 obtained earlier for propyl cyanide 
 toward Sgr~B2(N2). This probably indicates similarities in the chemical 
 processes that form these two families of molecules.
 \item The chemical models indicate that propanol is mainly produced within 
 the dust-grain ice mantles, via photodissociation-induced reactions between 
 OH and propylene (C$_3$H$_6$), although other mechanisms also make modest 
 contributions. The observed \textit{normal}-to-\textit{iso} ratio is 
 consistent with a direct preservation of the ratio of OH addition to the 
 terminal versus central carbon atom in propylene.
 
\end{enumerate}

\noindent With the detection of \textit{iso}- and \textit{normal}-propanol, 
two pairs of organic molecules with a functional group attached either to a 
primary or secondary carbon atom have now been detected in the interstellar 
medium, the second pair being \textit{iso}- and \textit{normal}-propyl 
cyanide, C$_3$H$_7$CN \citep[][]{Belloche09,Belloche14}. In the latter case, 
the carbon in the CN group technically  makes \textit{iso}-propyl cyanide a 
branched carbon-chain molecule, while \textit{iso}-propanol is not branched by 
this definition. Nevertheless, in the placement of the -OH or -CN functional 
group, both pairs of molecules are directly comparable.

The similarity of the propanol and propyl cyanide 
\textit{normal}-to-\textit{iso} ratios is not fully accounted for by the 
models presented here, although a perfect match could likely be achieved with 
a small degree of tuning in certain poorly-defined parameters (such as 
activation energies against the abstraction of different H atoms from ethyl 
cyanide). Finding other such pairs of molecules may help in pinning down the 
processes that dominate in setting these ratios in each family of molecules. 
The search for \textit{normal}-butanal (C$_3$H$_7$CHO) and its branched 
isomer, \textit{iso}-butanal, toward Sgr~B2(N) and G+0.693--0.027 has 
unfortunately not been successful so far \citep[][]{SanzNovo22}. A 
nondetection of lactaldehyde (2-hydroxypropanal), CH$_3$CH(OH)CHO, was also 
reported toward several star forming regions \citep[][]{Alonso19}, but no
investigation of its isomer, 3-hydroxypropanal (HOCH$_2$CH$_2$CHO), which 
might be more abundant, has been reported to date. As mentioned in 
Sect.~\ref{s:introduction}, our search for the straight-chain and branched 
forms of butyl cyanide, C$_4$H$_9$CN, in the ReMoCA survey has not been 
successful so far. In the alcohol family, the next candidate would be butanol 
(C$_4$H$_9$OH) and its isomers, surely a challenging search given the 
difficult detection of propanol itself.

\begin{acknowledgements}
This paper makes use of the following ALMA data: 
ADS/JAO.ALMA\#2016.1.00074.S. 
ALMA is a partnership of ESO (representing its member states), NSF (USA), and 
NINS (Japan), together with NRC (Canada), NSC and ASIAA (Taiwan), and KASI 
(Republic of Korea), in cooperation with the Republic of Chile. The Joint ALMA 
Observatory is operated by ESO, AUI/NRAO, and NAOJ. The interferometric data 
are available in the ALMA archive at https://almascience.eso.org/aq/.
Part of this work has been carried out within the Collaborative
Research Centre 956, sub-project B3, funded by the Deutsche
Forschungsgemeinschaft (DFG) -- project ID 184018867. 
RTG acknowledges support from the Astronomy \& Astrophysics program
of the National Science Foundation (grant No. AST 19-06489).

\end{acknowledgements}

\bibliographystyle{aa}
\bibliography{art_propanol_aph}

\begin{appendix}
\label{appendix}
\section{Complementary figures: Spectra}
\label{a:spectra}

Figures~\ref{f:spec_ch3oh_ve0}--\ref{f:spec_c3h7oh-n-Ag_ve0} show the
transitions of CH$_3$OH $\varv = 0$, CH$_3$OH $\varv_{\rm t} = 1$, 
CH$_3$OH $\varv_{\rm t} = 2$, CH$_3$OH $\varv_{\rm t} = 3$, C$_2$H$_5$OH, 
\textit{g}-\textit{i}-C$_3$H$_7$OH, \textit{a}-\textit{i}-C$_3$H$_7$OH,
\textit{Ga}-\textit{n}-C$_3$H$_7$OH, \textit{Gg}-\textit{n}-C$_3$H$_7$OH,
\textit{Gg'}-\textit{n}-C$_3$H$_7$OH, \textit{Aa}-\textit{n}-C$_3$H$_7$OH, and
\textit{Ag}-\textit{n}-C$_3$H$_7$OH, respectively,
that are covered by the ReMoCA survey and contribute significantly to 
the signal detected toward Sgr~B2(N2b). Transitions of a given molecule
that are too heavily blended with much stronger emission from other molecules 
and therefore cannot contribute to the identification of this molecule are not 
shown in these figures.

\begin{figure*}
\centerline{\resizebox{0.82\hsize}{!}{\includegraphics[angle=0]{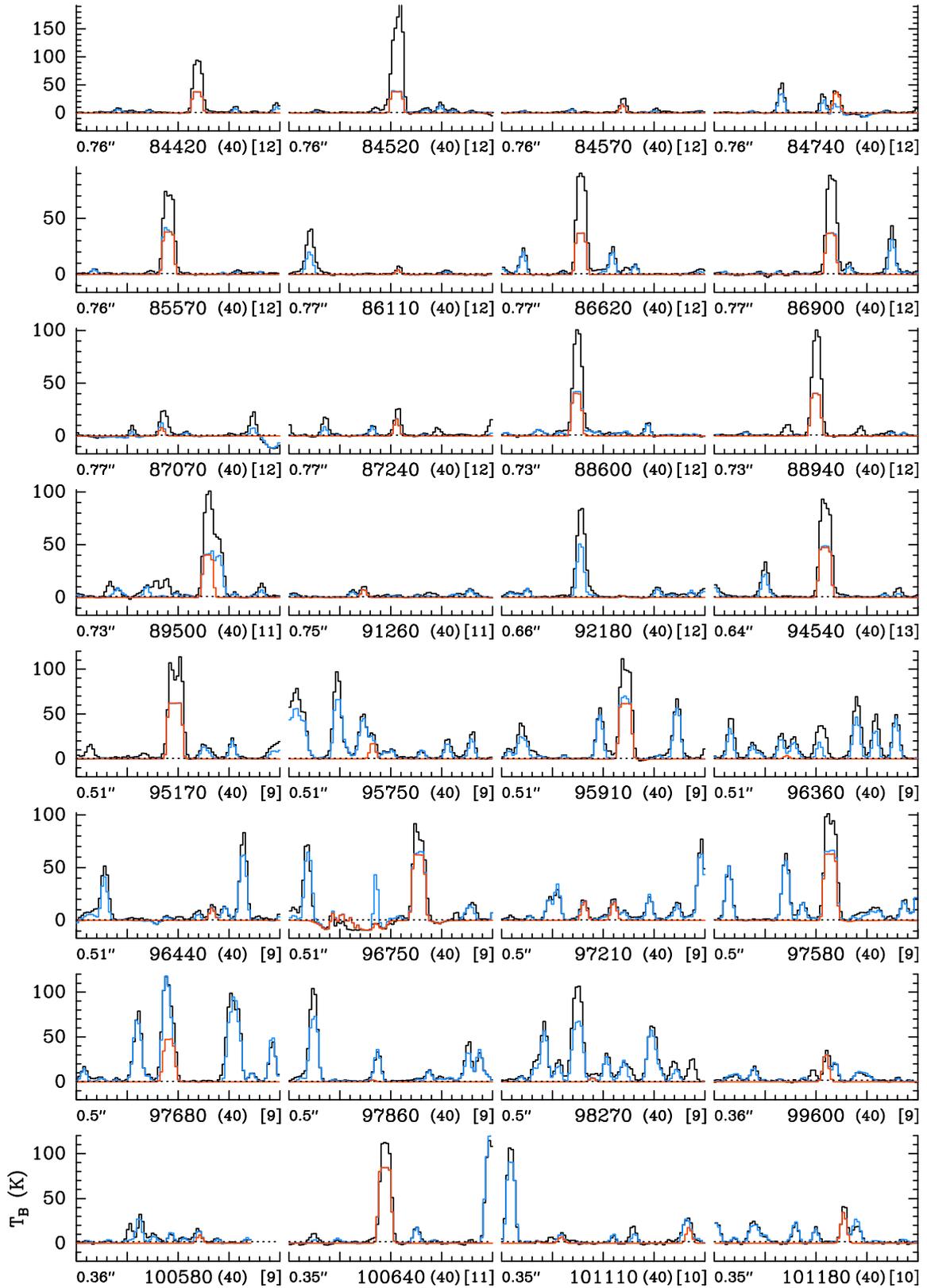}}}
\caption{Transitions of CH$_3$OH, $\varv = 0$ covered by the ReMoCA
survey with ALMA. The best-fit LTE synthetic spectrum of CH$_3$OH, $\varv = 0$ 
is displayed in red and overlaid on the observed spectrum of Sgr~B2(N2b) shown 
in black. The blue synthetic spectrum contains the contributions of all 
molecules identified in our survey so far, including the species shown in red. 
The central frequency is indicated in MHz below each 
panel as well as the half-power beam width on the left, the width of each 
panel in MHz in parentheses, and the continuum level in K of the 
baseline-subtracted spectra in brackets. The y-axis is labeled in 
brightness temperature units (K). The dotted line indicates the $3\sigma$ 
noise level.}
\label{f:spec_ch3oh_ve0}
\end{figure*}

\clearpage
\begin{figure*}
\addtocounter{figure}{-1}
\centerline{\resizebox{0.82\hsize}{!}{\includegraphics[angle=0]{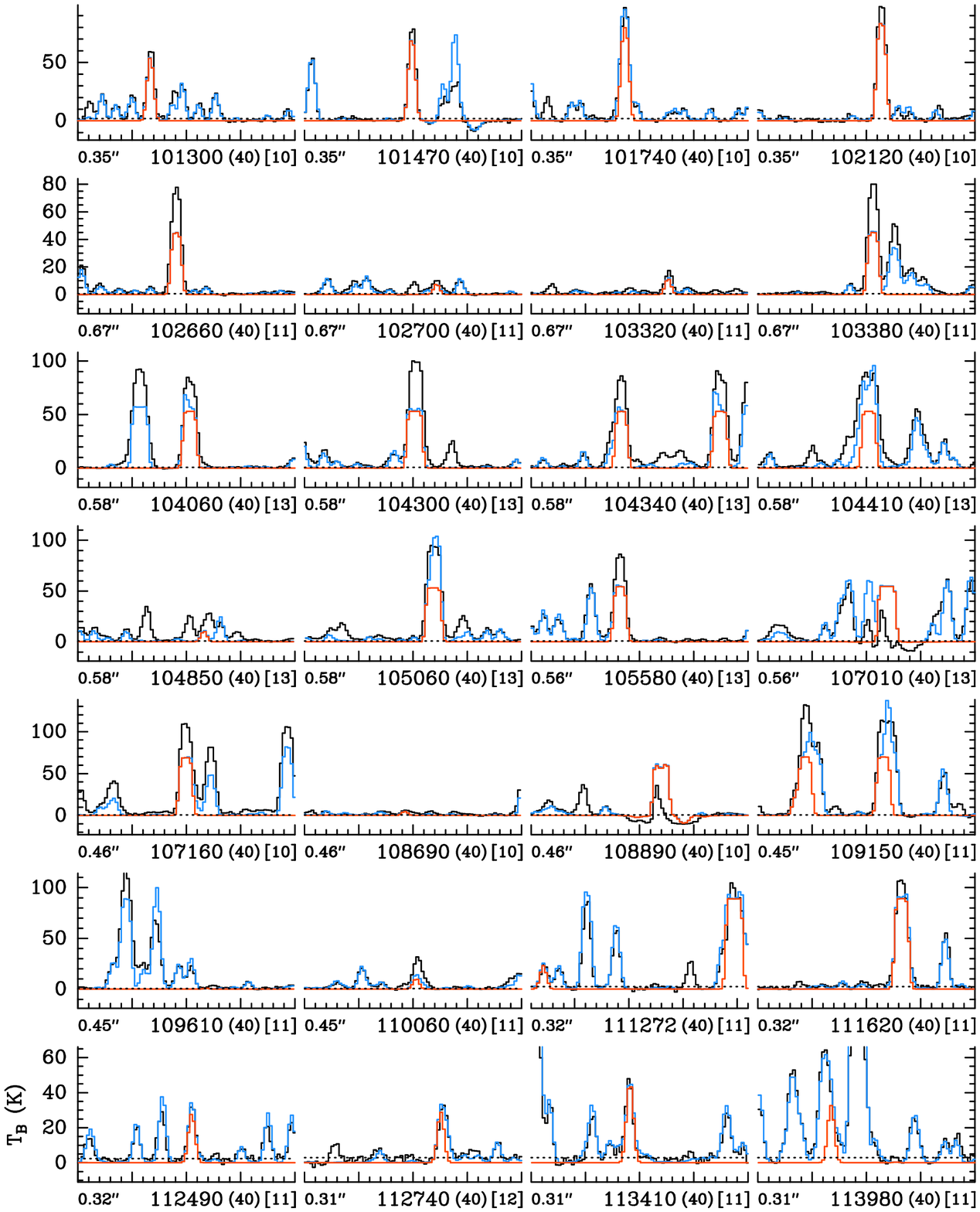}}}
\caption{continued.
}
\end{figure*}

\begin{figure*}
\centerline{\resizebox{0.82\hsize}{!}{\includegraphics[angle=0]{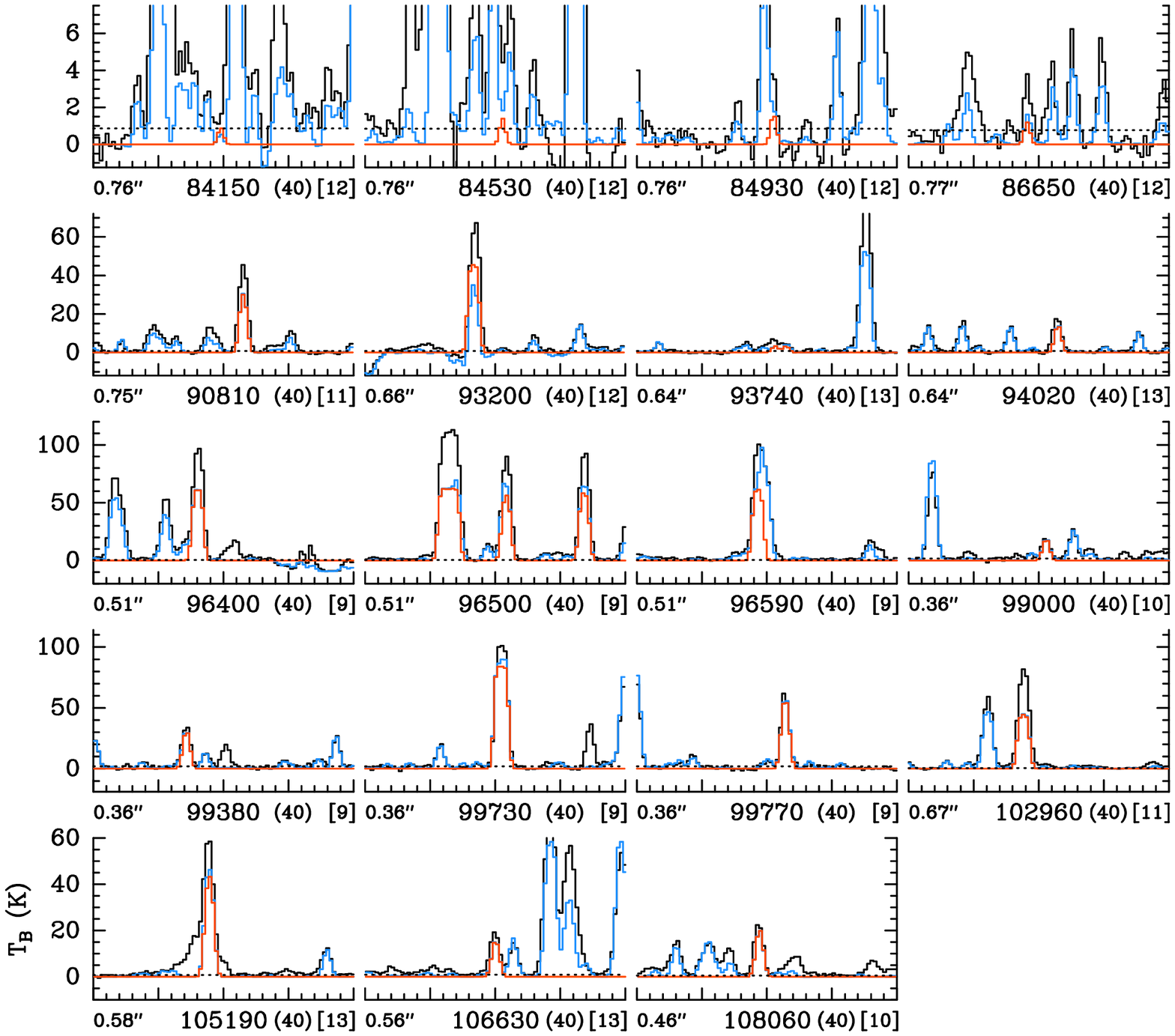}}}
\caption{Same as Fig.~\ref{f:spec_ch3oh_ve0} but for CH$_3$OH, $\varv_{\rm t}$=1.}
\label{f:spec_ch3oh_ve1}
\end{figure*}

\begin{figure*}
\centerline{\resizebox{0.82\hsize}{!}{\includegraphics[angle=0]{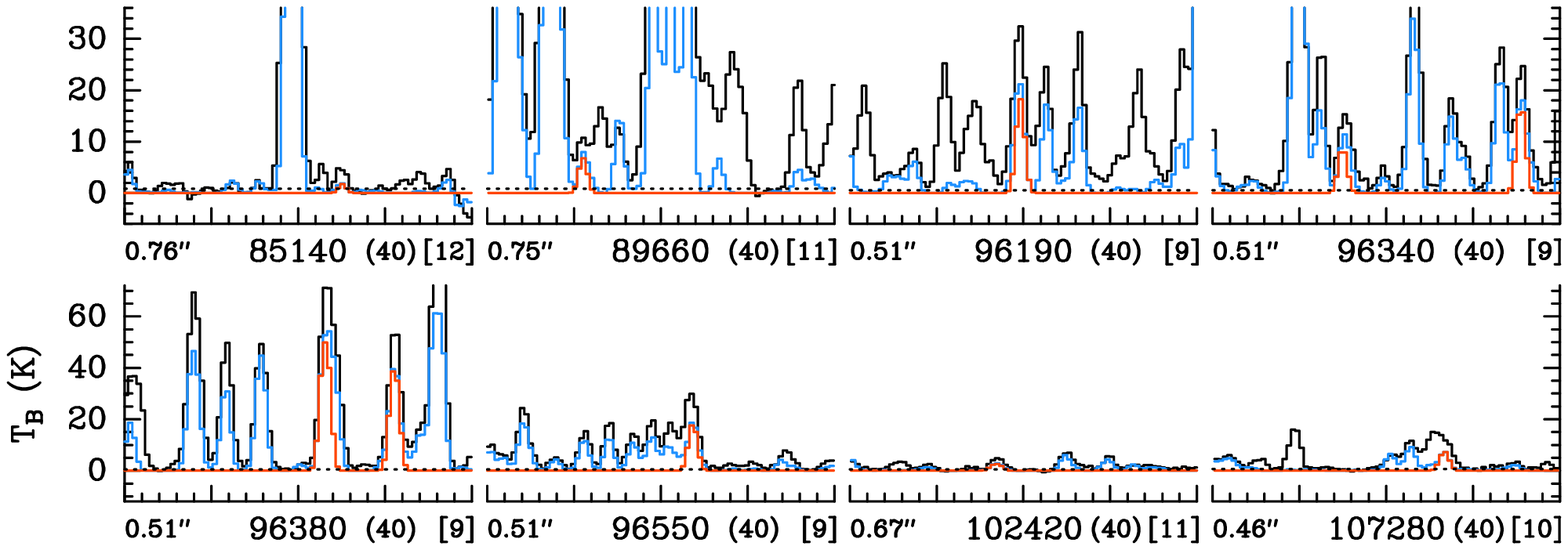}}}
\caption{Same as Fig.~\ref{f:spec_ch3oh_ve0} but for CH$_3$OH, $\varv_{\rm t}$=2.}
\label{f:spec_ch3oh_ve2}
\end{figure*}

\begin{figure*}
\centerline{\resizebox{0.44\hsize}{!}{\includegraphics[angle=0]{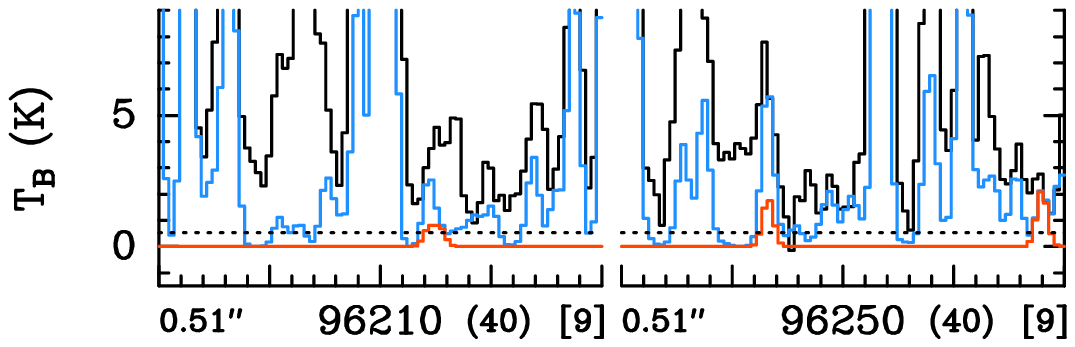}}}
\caption{Same as Fig.~\ref{f:spec_ch3oh_ve0} but for CH$_3$OH, $\varv_{\rm t}$=3.}
\label{f:spec_ch3oh_ve3}
\end{figure*}

\clearpage
\begin{figure*}
\centerline{\resizebox{0.82\hsize}{!}{\includegraphics[angle=0]{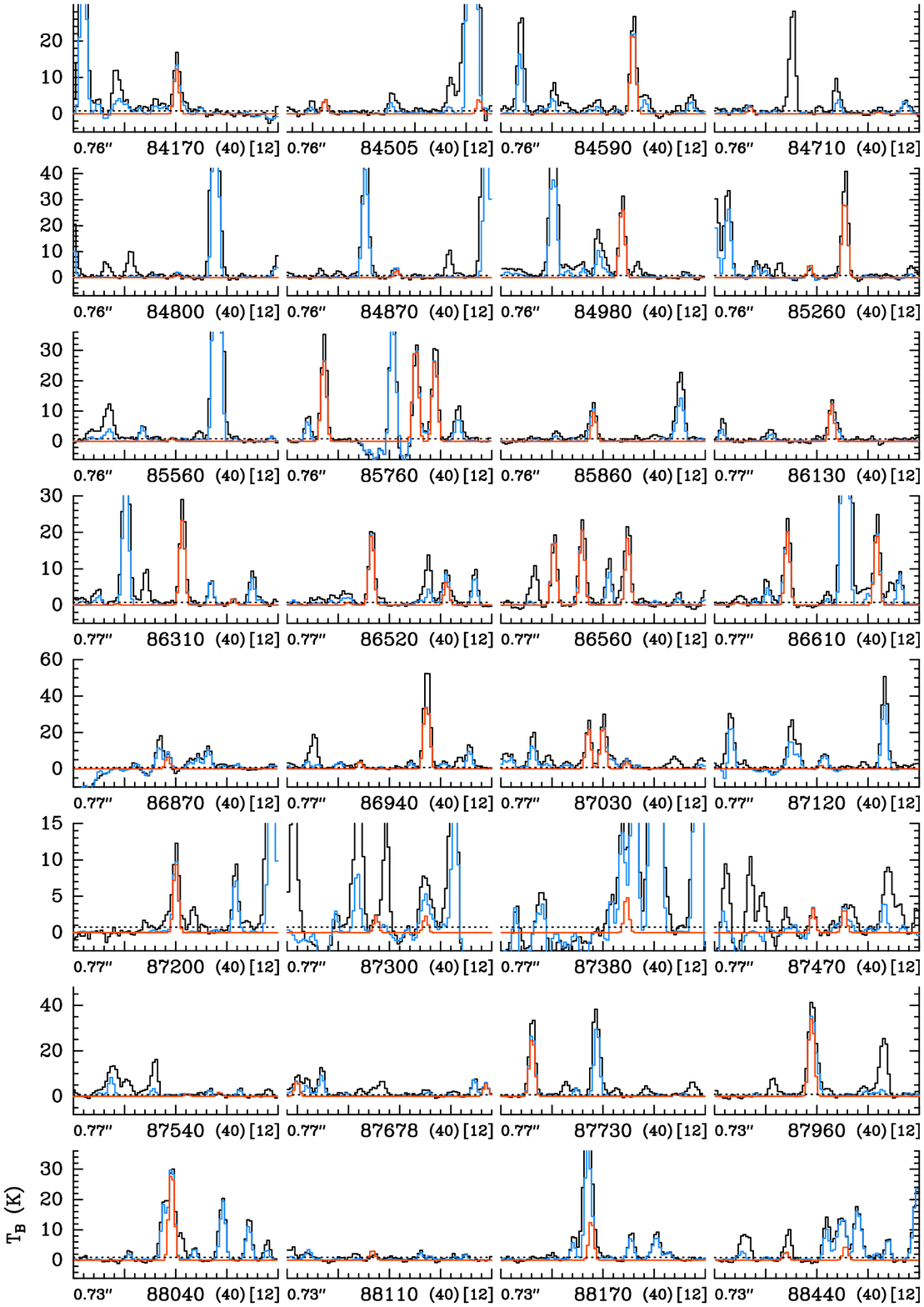}}}
\caption{Same as Fig.~\ref{f:spec_ch3oh_ve0} but for C$_2$H$_5$OH, $\varv=0$.}
\label{f:spec_c2h5oh_ve0}
\end{figure*}

\clearpage
\begin{figure*}
\addtocounter{figure}{-1}
\centerline{\resizebox{0.82\hsize}{!}{\includegraphics[angle=0]{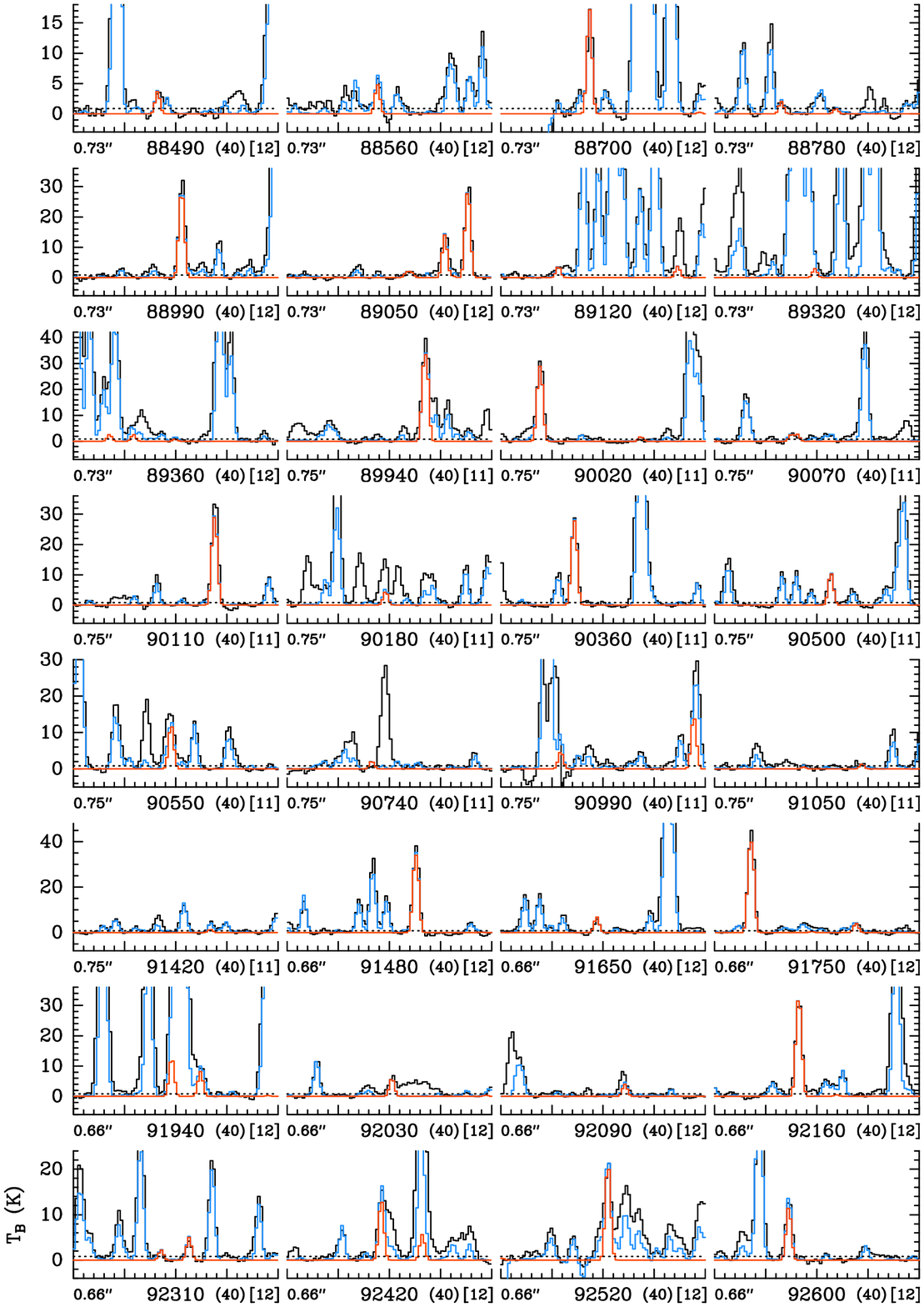}}}
\caption{continued.
}
\end{figure*}

\clearpage
\begin{figure*}
\addtocounter{figure}{-1}
\centerline{\resizebox{0.82\hsize}{!}{\includegraphics[angle=0]{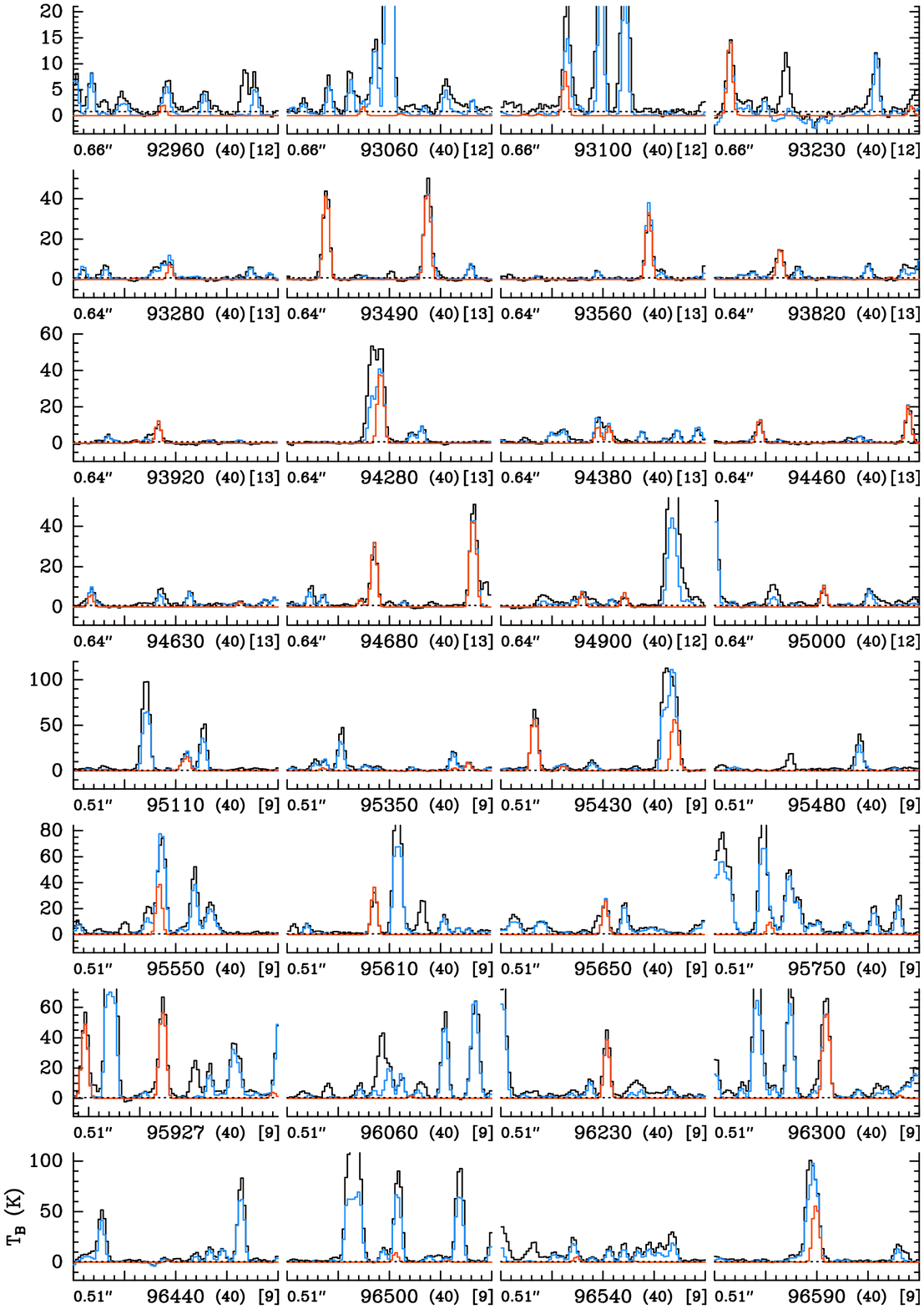}}}
\caption{continued.
}
\end{figure*}

\clearpage
\begin{figure*}
\addtocounter{figure}{-1}
\centerline{\resizebox{0.82\hsize}{!}{\includegraphics[angle=0]{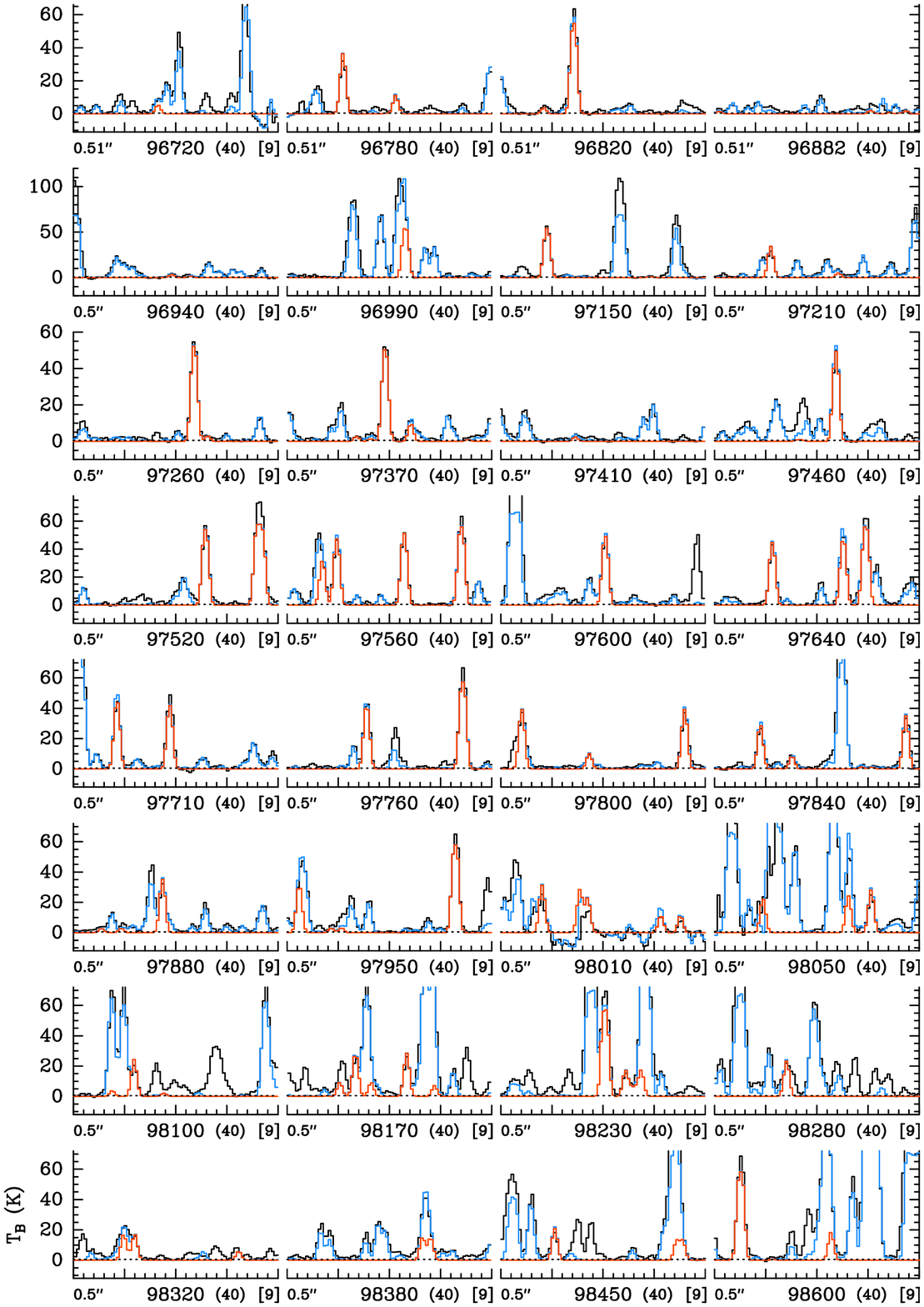}}}
\caption{continued.
}
\end{figure*}

\clearpage
\begin{figure*}
\addtocounter{figure}{-1}
\centerline{\resizebox{0.82\hsize}{!}{\includegraphics[angle=0]{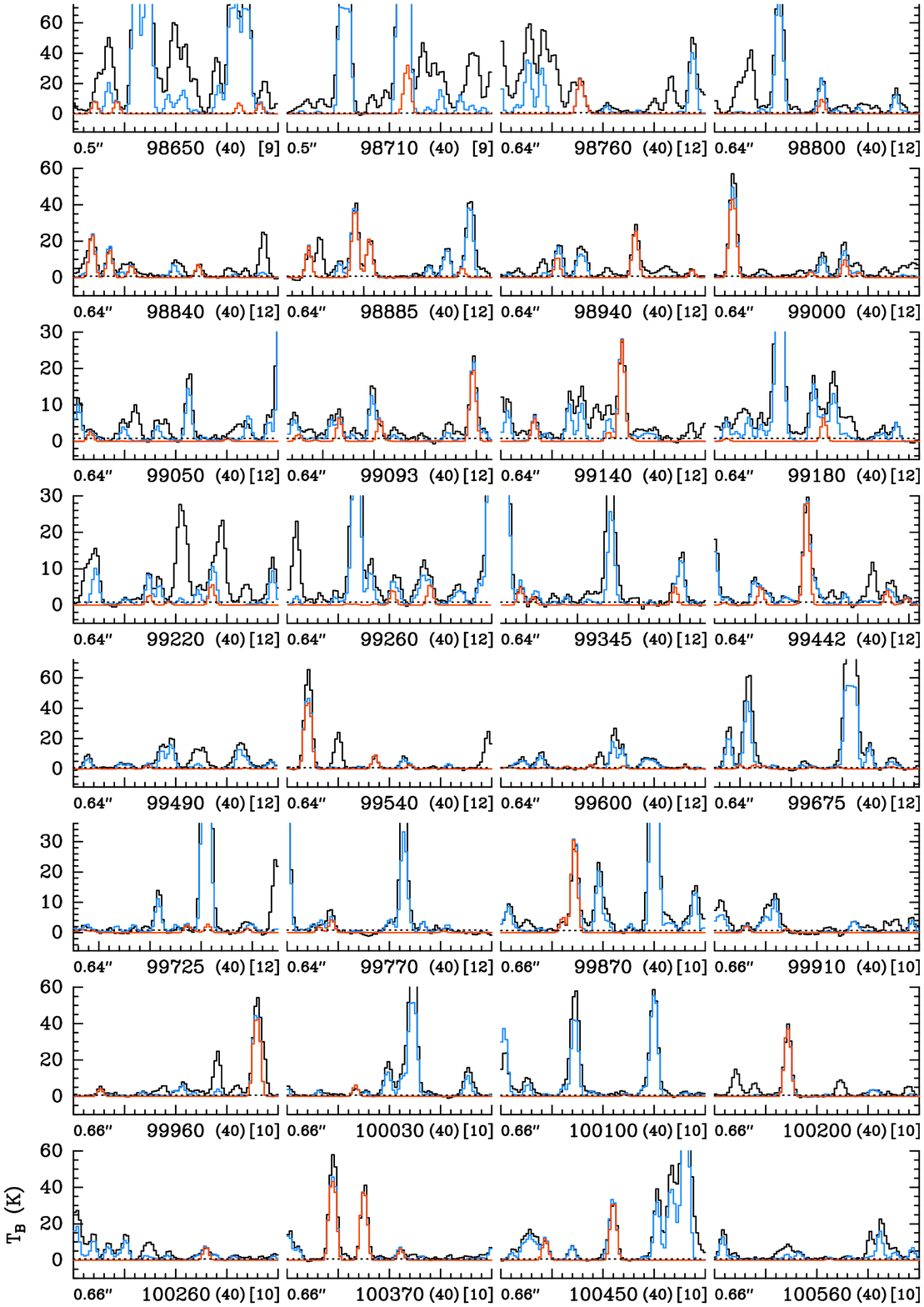}}}
\caption{continued.
}
\end{figure*}

\clearpage
\begin{figure*}
\addtocounter{figure}{-1}
\centerline{\resizebox{0.82\hsize}{!}{\includegraphics[angle=0]{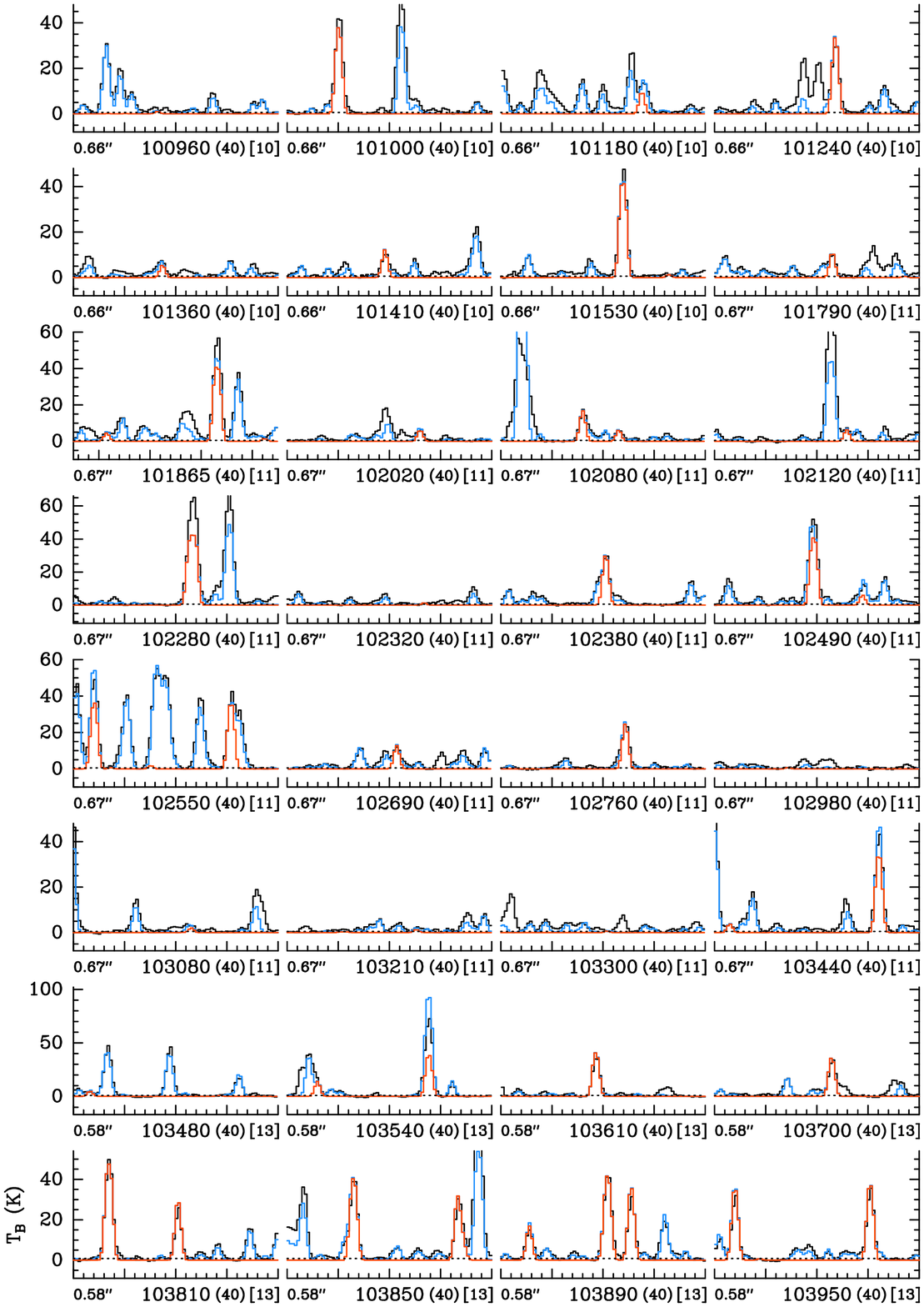}}}
\caption{continued.
}
\end{figure*}

\clearpage
\begin{figure*}
\addtocounter{figure}{-1}
\centerline{\resizebox{0.82\hsize}{!}{\includegraphics[angle=0]{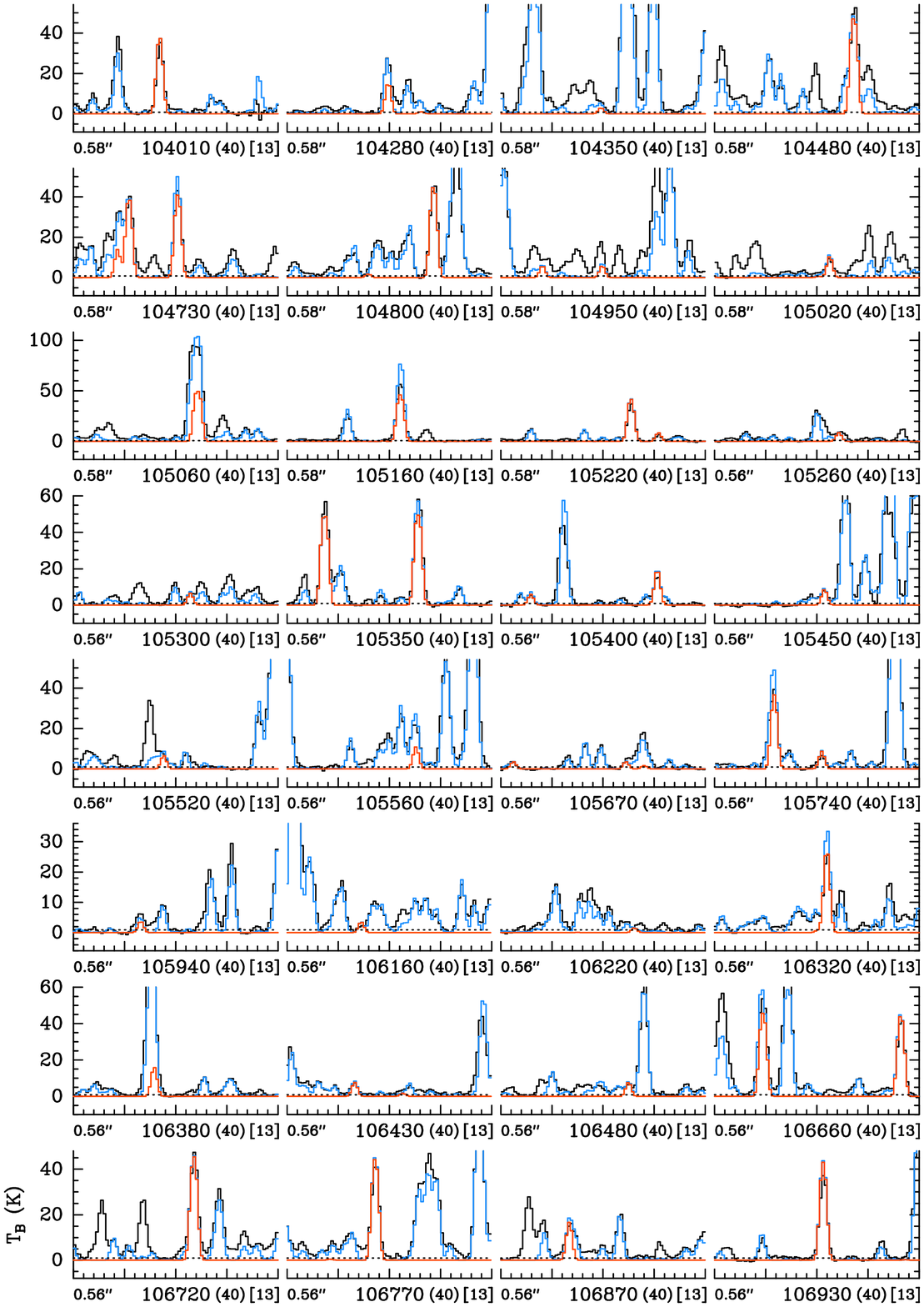}}}
\caption{continued.
}
\end{figure*}

\clearpage
\begin{figure*}
\addtocounter{figure}{-1}
\centerline{\resizebox{0.82\hsize}{!}{\includegraphics[angle=0]{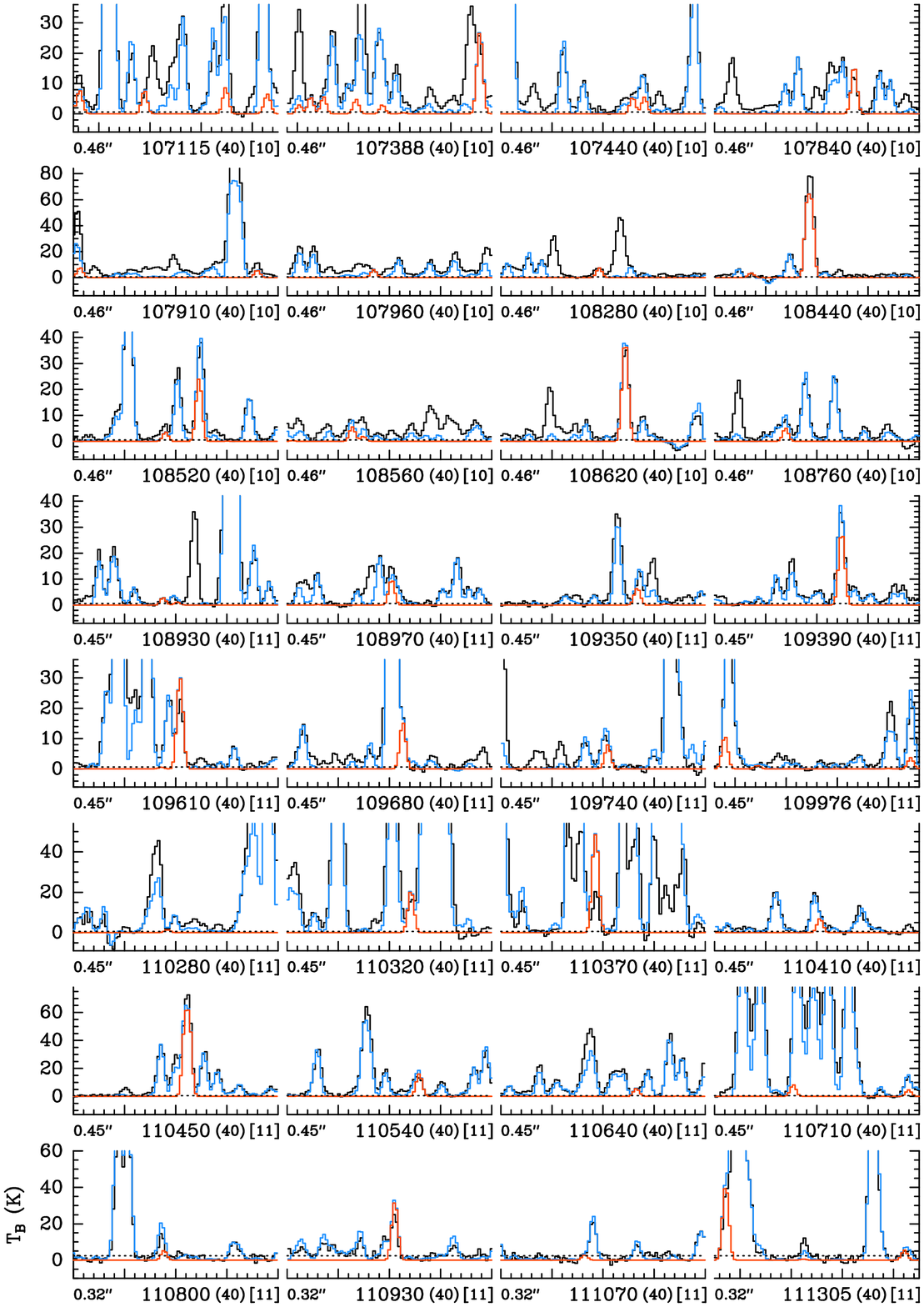}}}
\caption{continued.
}
\end{figure*}

\clearpage
\begin{figure*}
\addtocounter{figure}{-1}
\centerline{\resizebox{0.82\hsize}{!}{\includegraphics[angle=0]{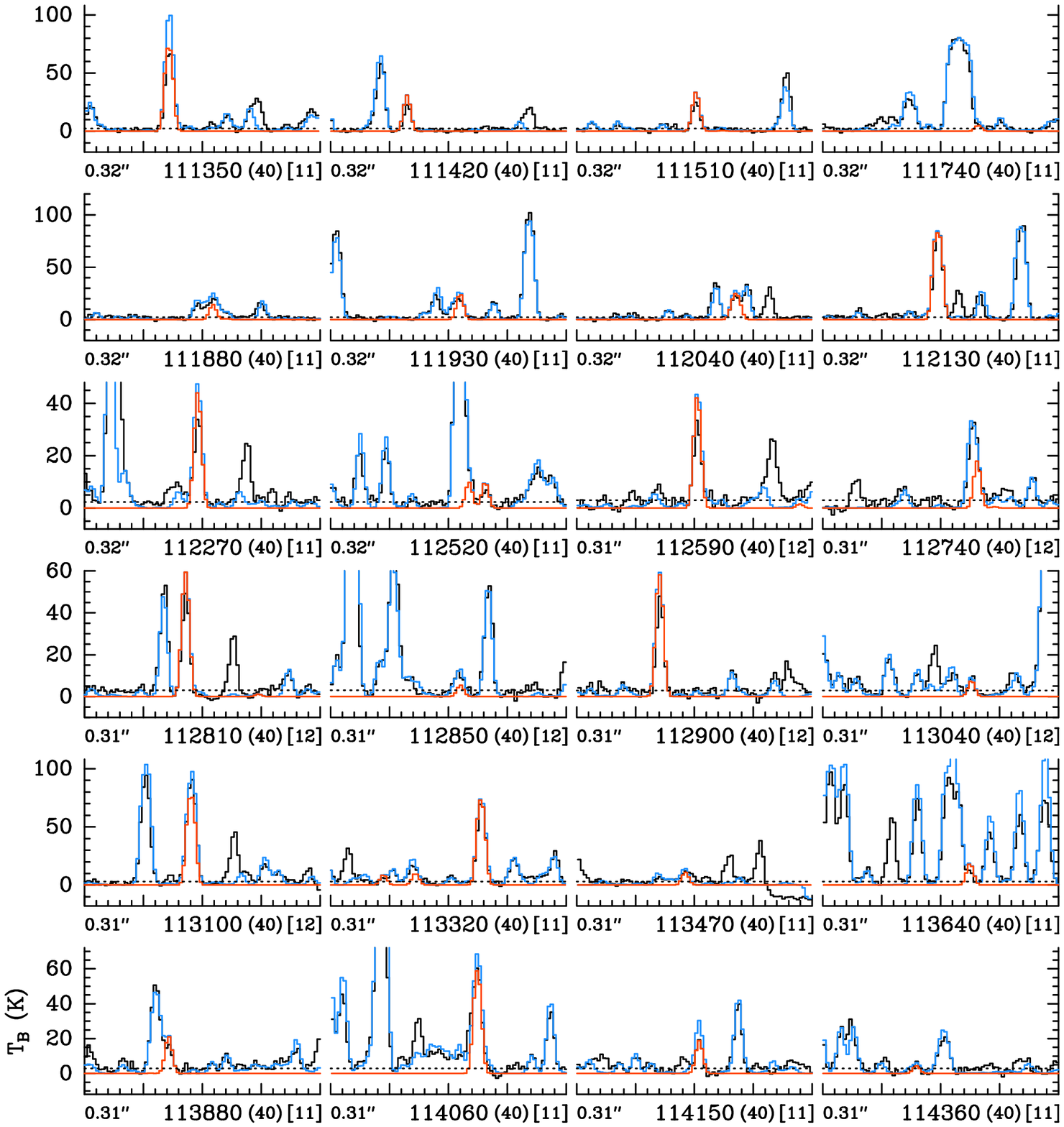}}}
\caption{continued.
}
\end{figure*}

\begin{figure*}
\centerline{\resizebox{0.82\hsize}{!}{\includegraphics[angle=0]{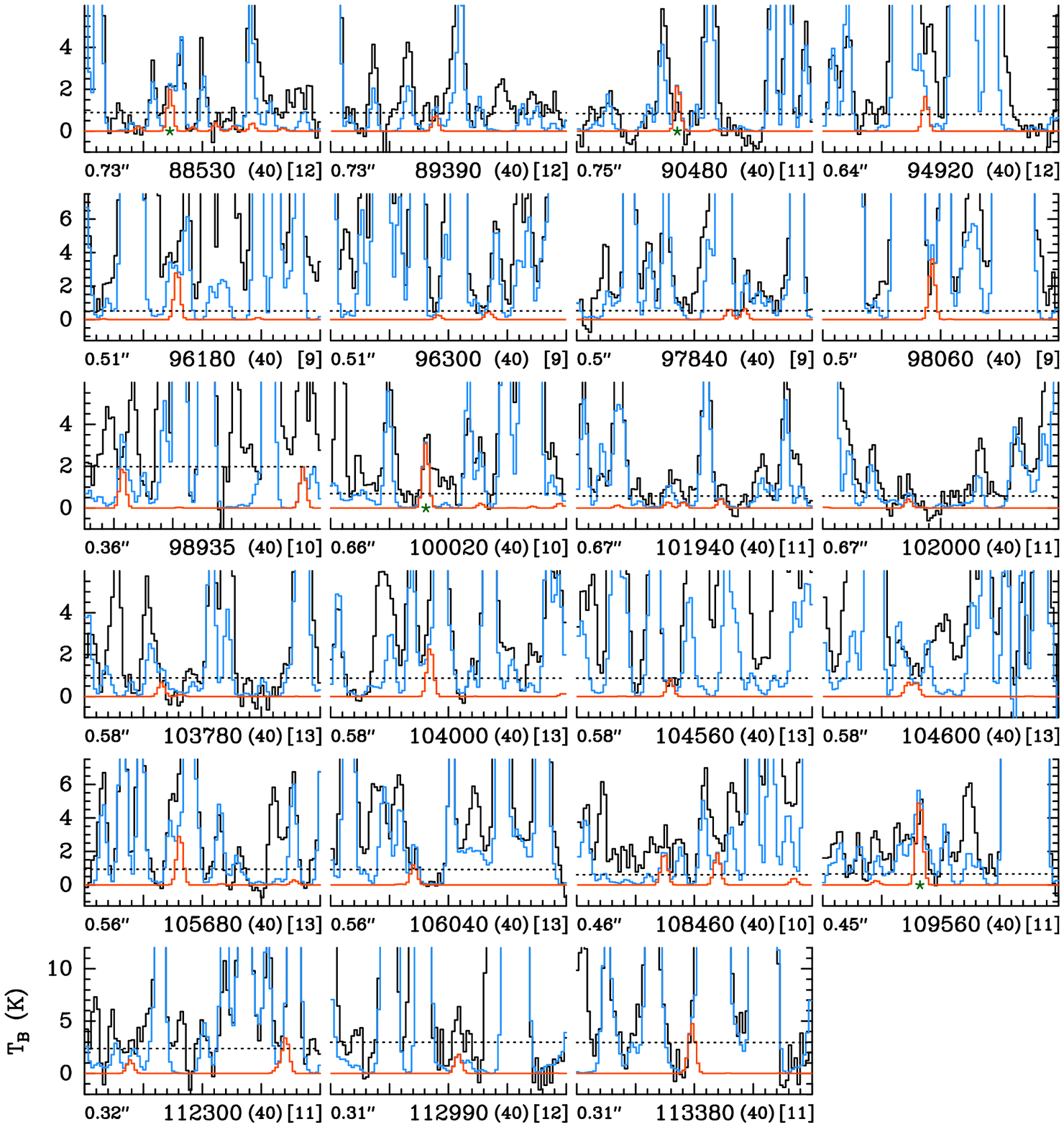}}}
\caption{Same as Fig.~\ref{f:spec_ch3oh_ve0} but for \textit{g}-\textit{i}-C$_3$H$_7$OH, $\varv$=0. The green stars mark the transitions that we consider as detected.}
\label{f:spec_c3h7oh-i-g_ve0}
\end{figure*}

\begin{figure*}
\centerline{\resizebox{0.82\hsize}{!}{\includegraphics[angle=0]{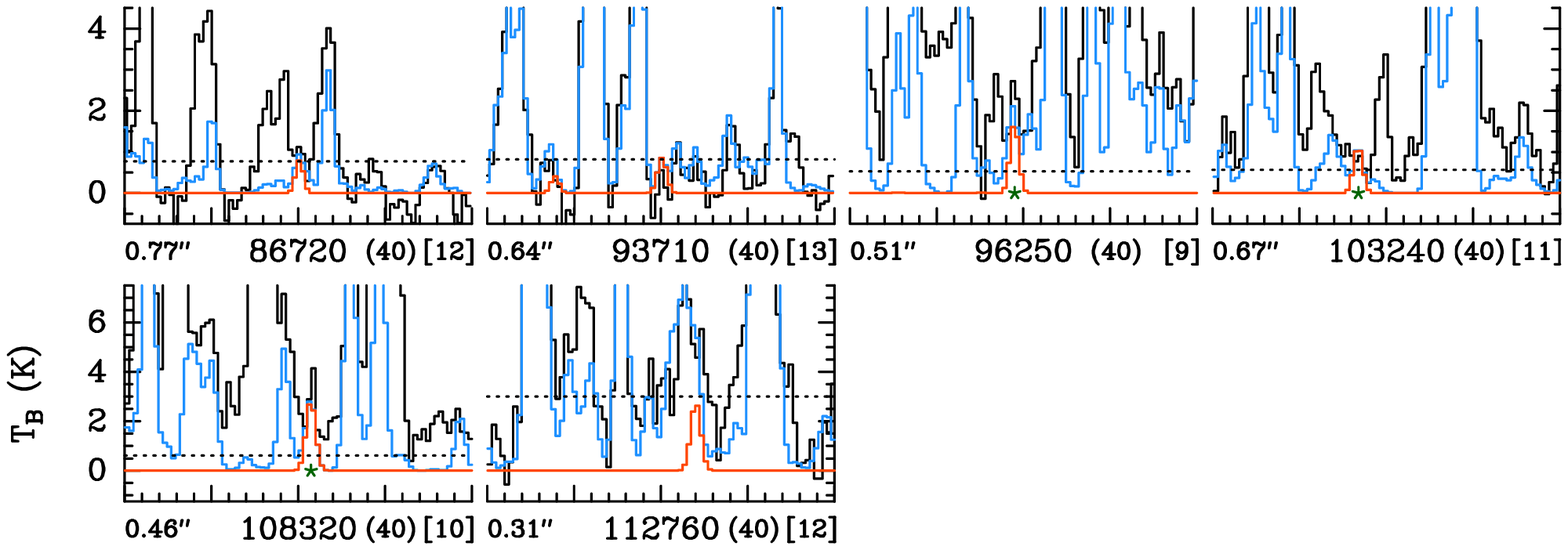}}}
\caption{Same as Fig.~\ref{f:spec_ch3oh_ve0} but for \textit{a}-\textit{i}-C$_3$H$_7$OH, $\varv$=0. The green stars mark the transitions that we consider as detected.}
\label{f:spec_c3h7oh-i-a_ve0}
\end{figure*}

\begin{figure*}
\centerline{\resizebox{0.82\hsize}{!}{\includegraphics[angle=0]{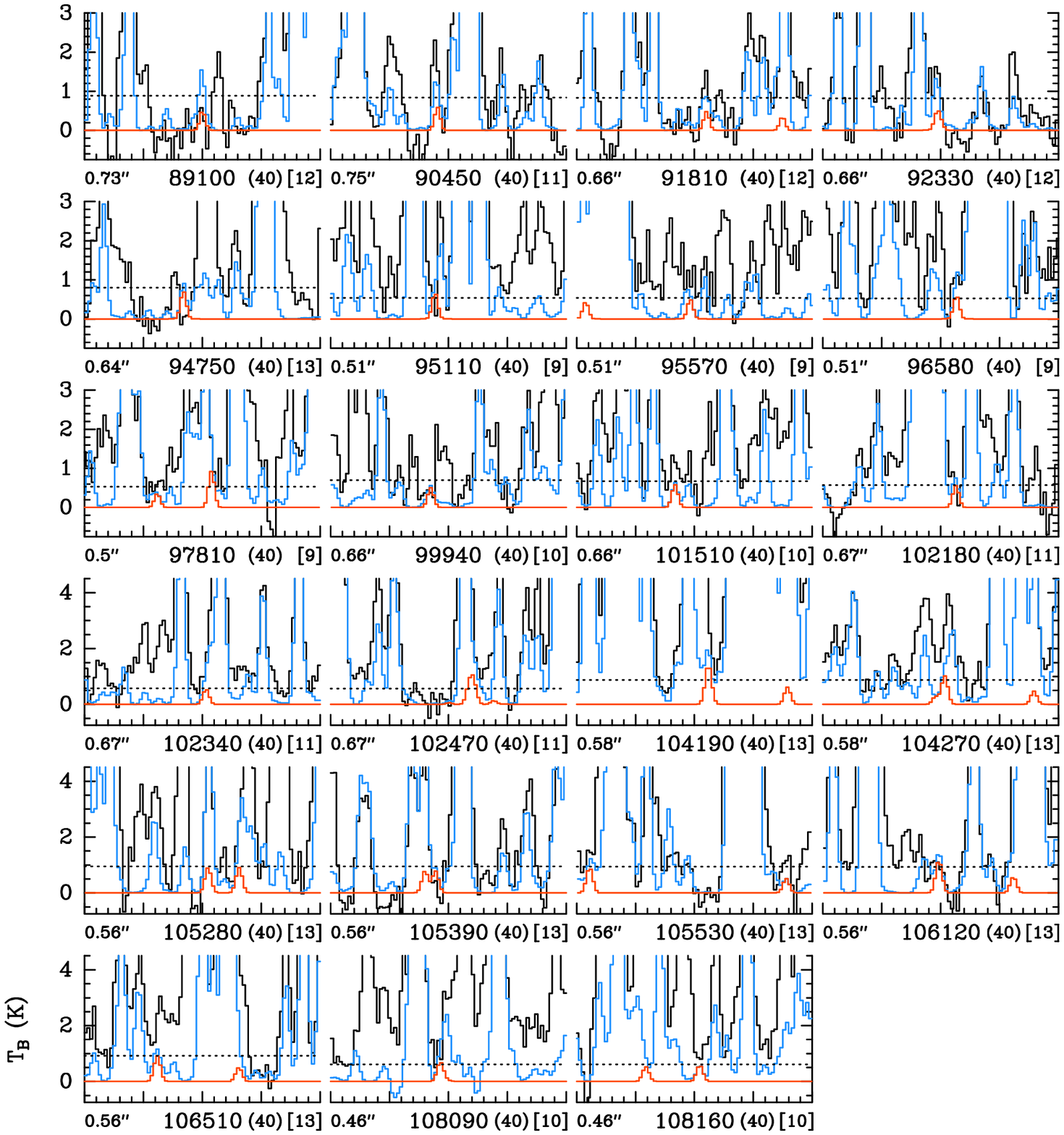}}}
\caption{Same as Fig.~\ref{f:spec_ch3oh_ve0} but for \textit{Ga}-\textit{n}-C$_3$H$_7$OH, $\varv$=0.}
\label{f:spec_c3h7oh-n-Ga_ve0}
\end{figure*}

\begin{figure*}
\centerline{\resizebox{0.82\hsize}{!}{\includegraphics[angle=0]{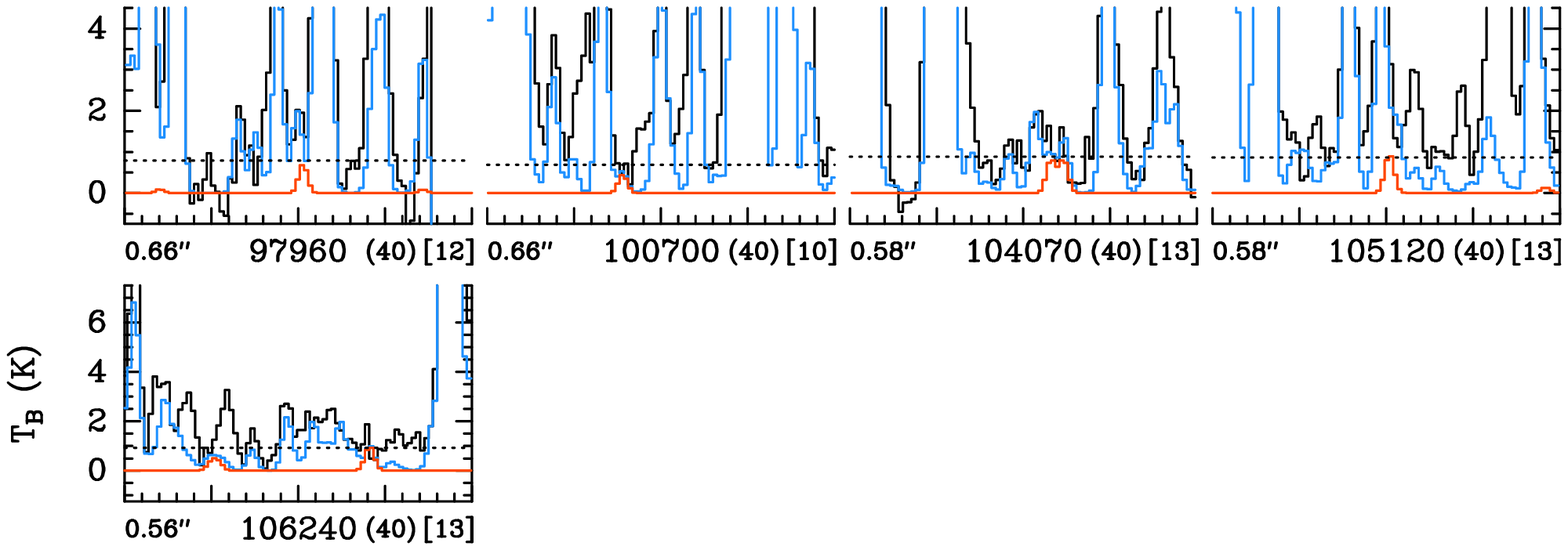}}}
\caption{Same as Fig.~\ref{f:spec_ch3oh_ve0} but for \textit{Gg}-\textit{n}-C$_3$H$_7$OH, $\varv$=0.}
\label{f:spec_c3h7oh-n-Gg_ve0}
\end{figure*}

\begin{figure*}
\centerline{\resizebox{0.82\hsize}{!}{\includegraphics[angle=0]{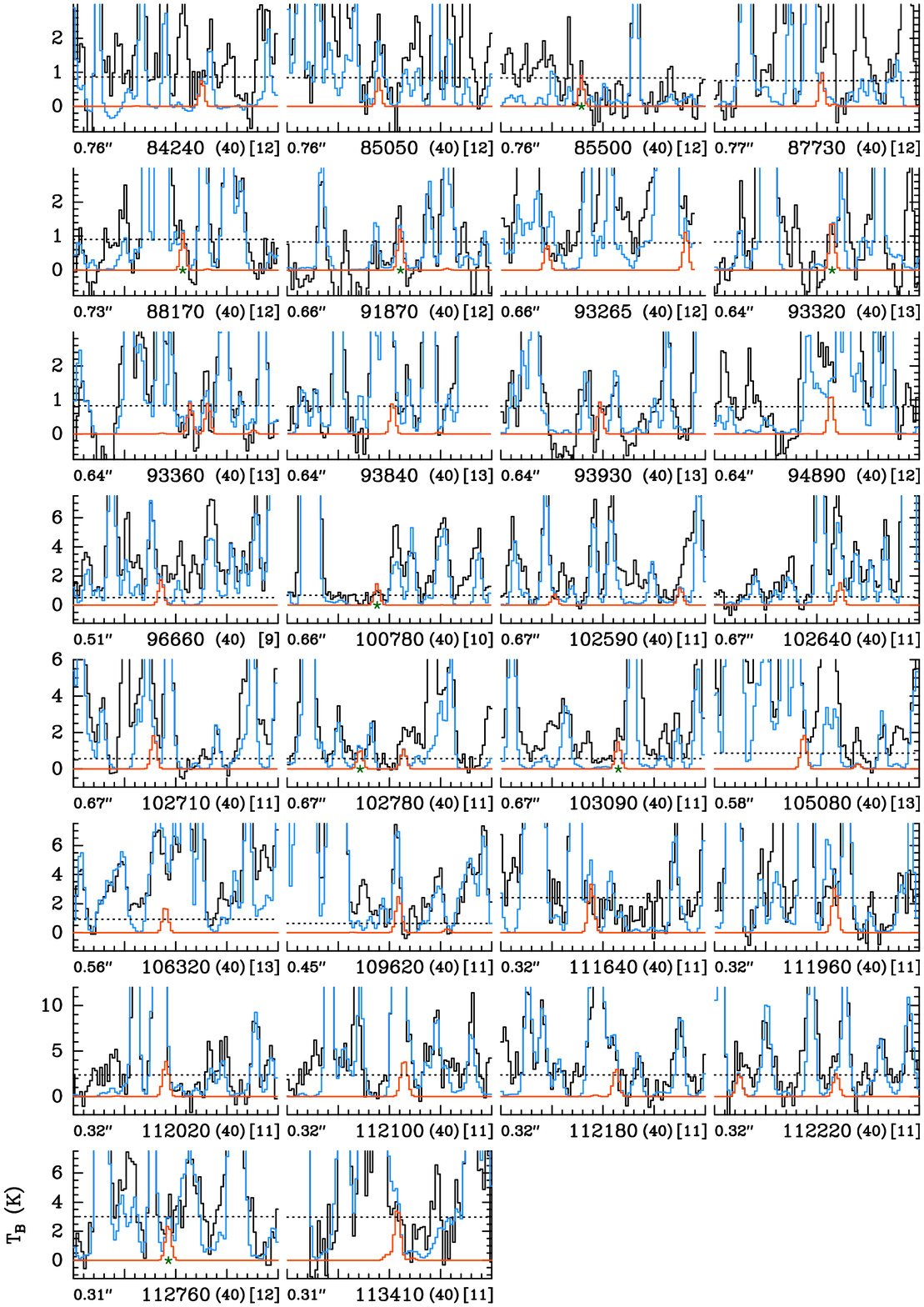}}}
\caption{Same as Fig.~\ref{f:spec_ch3oh_ve0} but for \textit{Gg'}-\textit{n}-C$_3$H$_7$OH, $\varv$=0. The green stars mark the transitions that we consider as detected.}
\label{f:spec_c3h7oh-n-Ggp_ve0}
\end{figure*}

\begin{figure*}
\centerline{\resizebox{0.82\hsize}{!}{\includegraphics[angle=0]{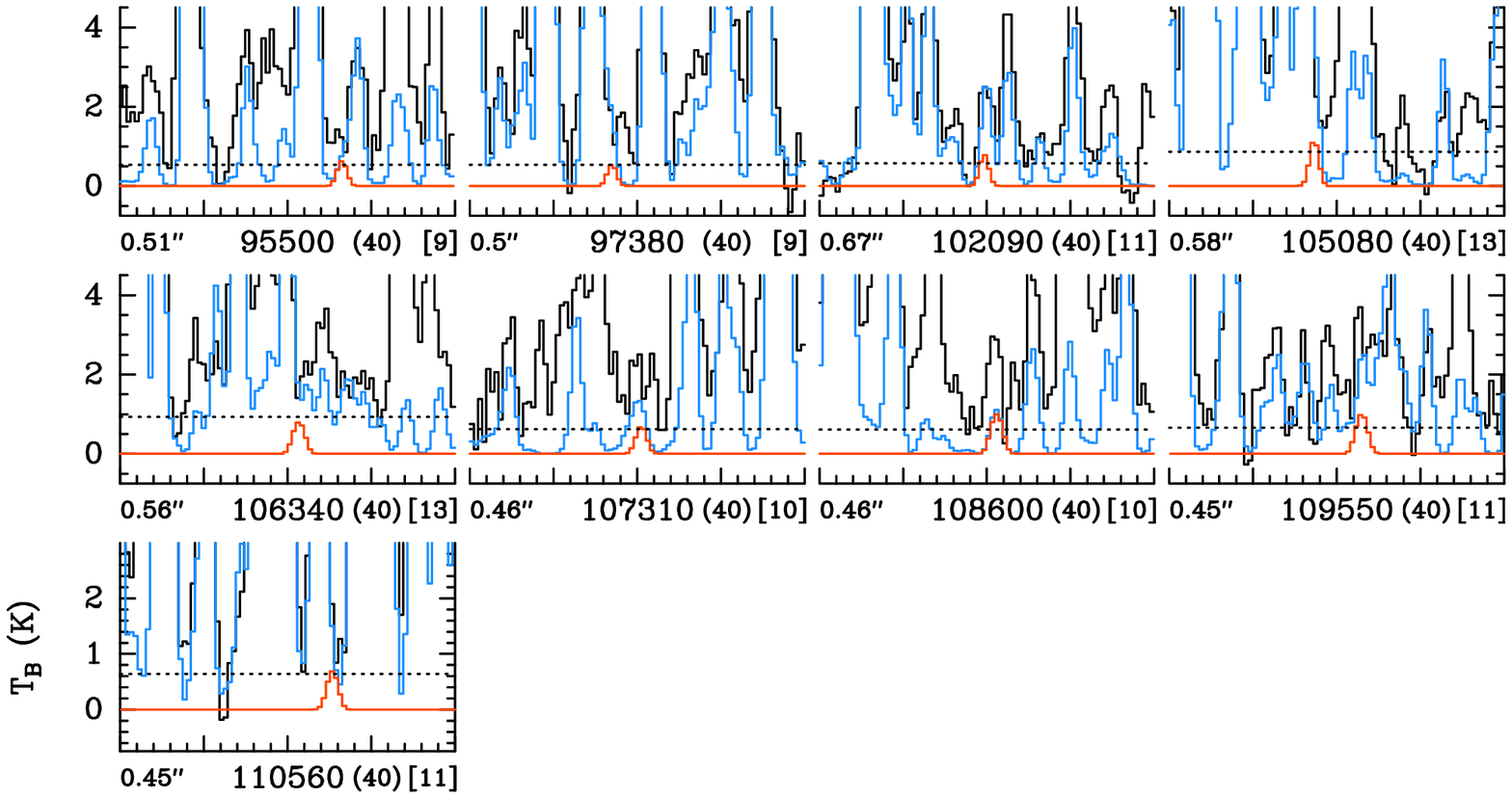}}}
\caption{Same as Fig.~\ref{f:spec_ch3oh_ve0} but for \textit{Aa}-\textit{n}-C$_3$H$_7$OH, $\varv$=0.}
\label{f:spec_c3h7oh-n-Aa_ve0}
\end{figure*}

\begin{figure*}
\centerline{\resizebox{0.82\hsize}{!}{\includegraphics[angle=0]{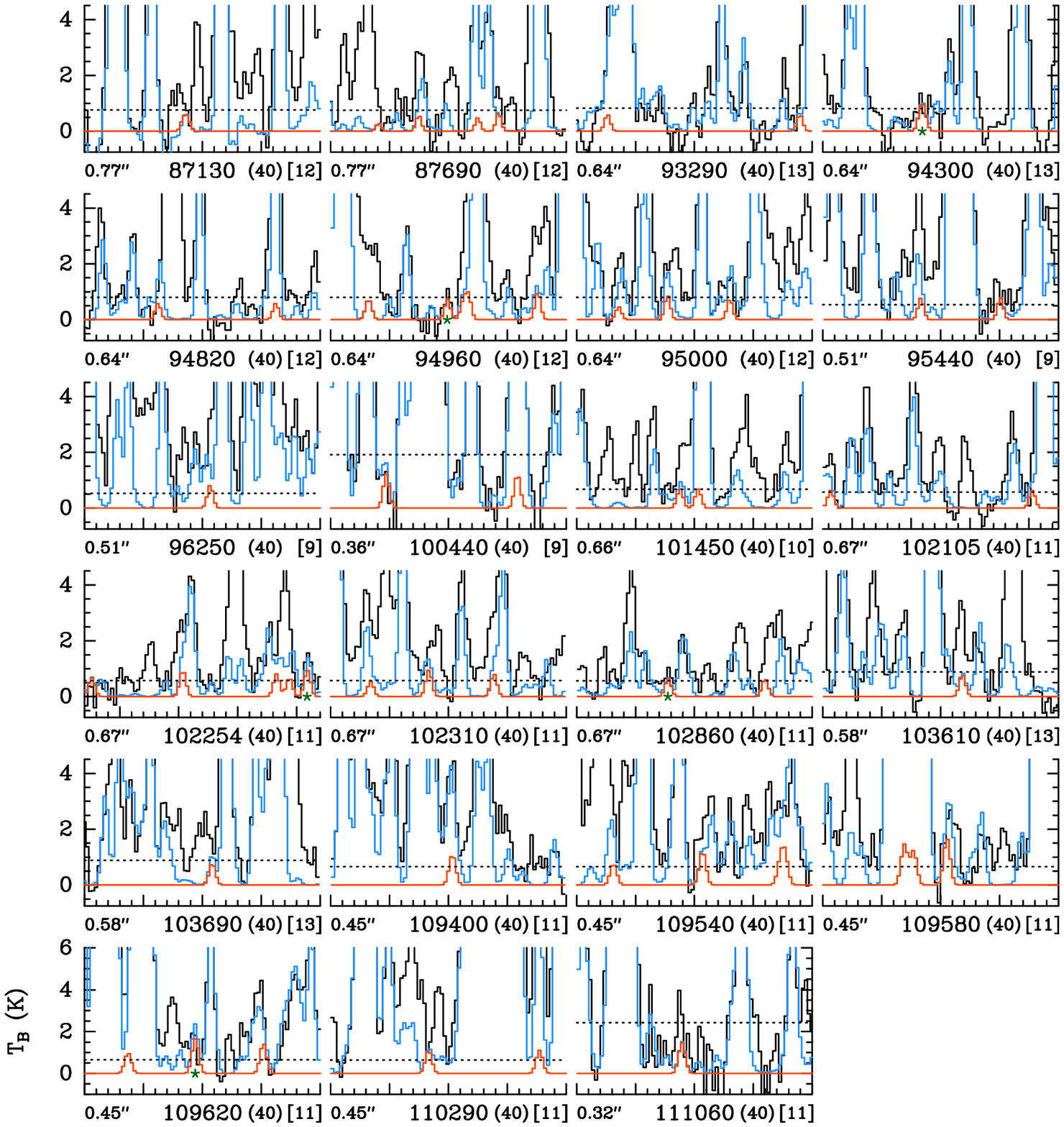}}}
\caption{Same as Fig.~\ref{f:spec_ch3oh_ve0} but for \textit{Ag}-\textit{n}-C$_3$H$_7$OH, $\varv$=0. The green stars mark the transitions that we consider as detected.}
\label{f:spec_c3h7oh-n-Ag_ve0}
\end{figure*}

\clearpage
\section{Complementary figures: Population diagrams}
\label{a:popdiag}

\begin{figure}
\centerline{\resizebox{0.95\hsize}{!}{\includegraphics[angle=0]{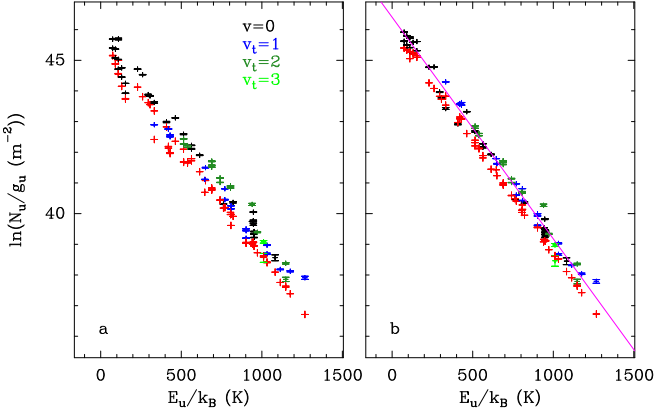}}}
\caption{Population diagram of CH$_3$OH toward Sgr~B2(N2b). The 
observed data points are shown in various colors (but not red) as indicated in 
the upper right corner of panel \textbf{a} while the synthetic populations are 
shown in red. No correction is applied in panel \textbf{a}. 
In panel \textbf{b}, the optical depth correction has been applied to both the 
observed and synthetic populations and the contamination by all other 
species included in the full model has been removed from the observed 
data points. The purple line is a linear fit to the observed populations (in 
linear-logarithmic space).
}
\label{f:popdiag_ch3oh}
\end{figure}

\begin{figure}
\centerline{\resizebox{0.95\hsize}{!}{\includegraphics[angle=0]{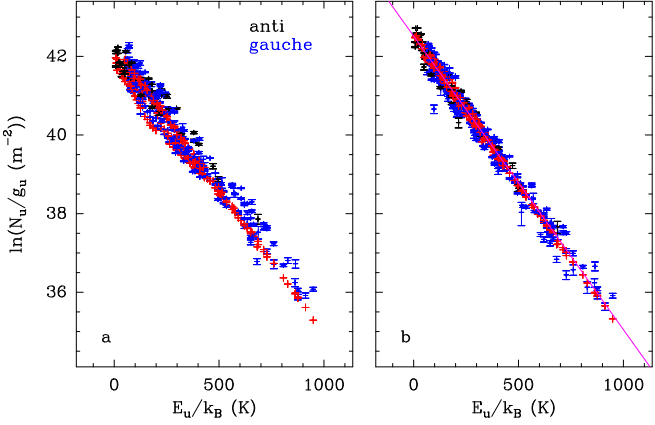}}}
\caption{Same as Fig.~\ref{f:popdiag_ch3oh} for C$_2$H$_5$OH. The 
observed data points of the \textit{anti} and \textit{gauche} conformers of
C$_2$H$_5$OH are shown in black and blue, respectively.
}
\label{f:popdiag_c2h5oh}
\end{figure}

\begin{figure}
\centerline{\resizebox{0.95\hsize}{!}{\includegraphics[angle=0]{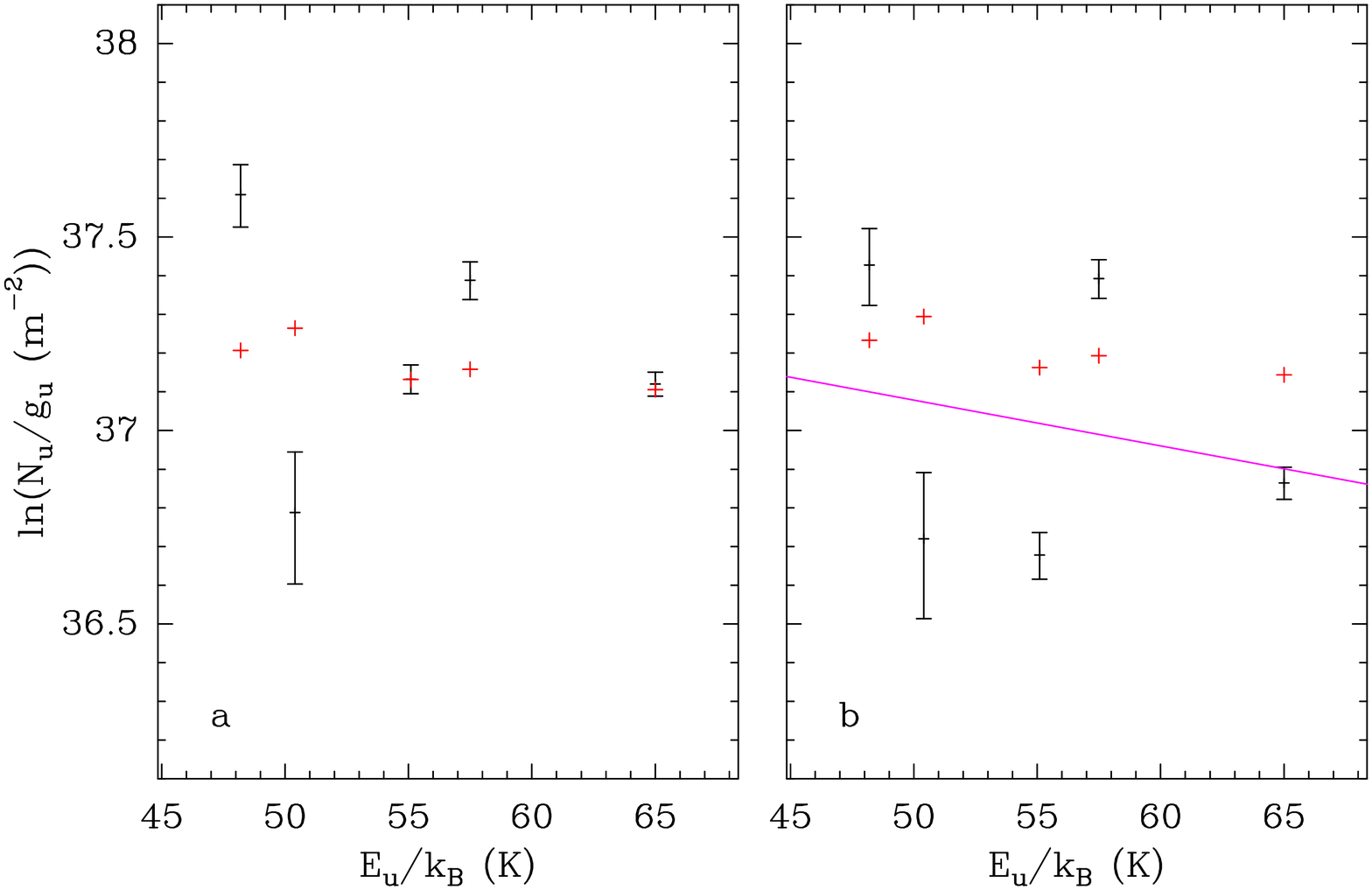}}}
\caption{Same as Fig.~\ref{f:popdiag_ch3oh} for \textit{g}-\textit{i}-C$_3$H$_7$OH.
}
\label{f:popdiag_c3h7oh-i-g}
\end{figure}

\begin{figure}
\centerline{\resizebox{0.95\hsize}{!}{\includegraphics[angle=0]{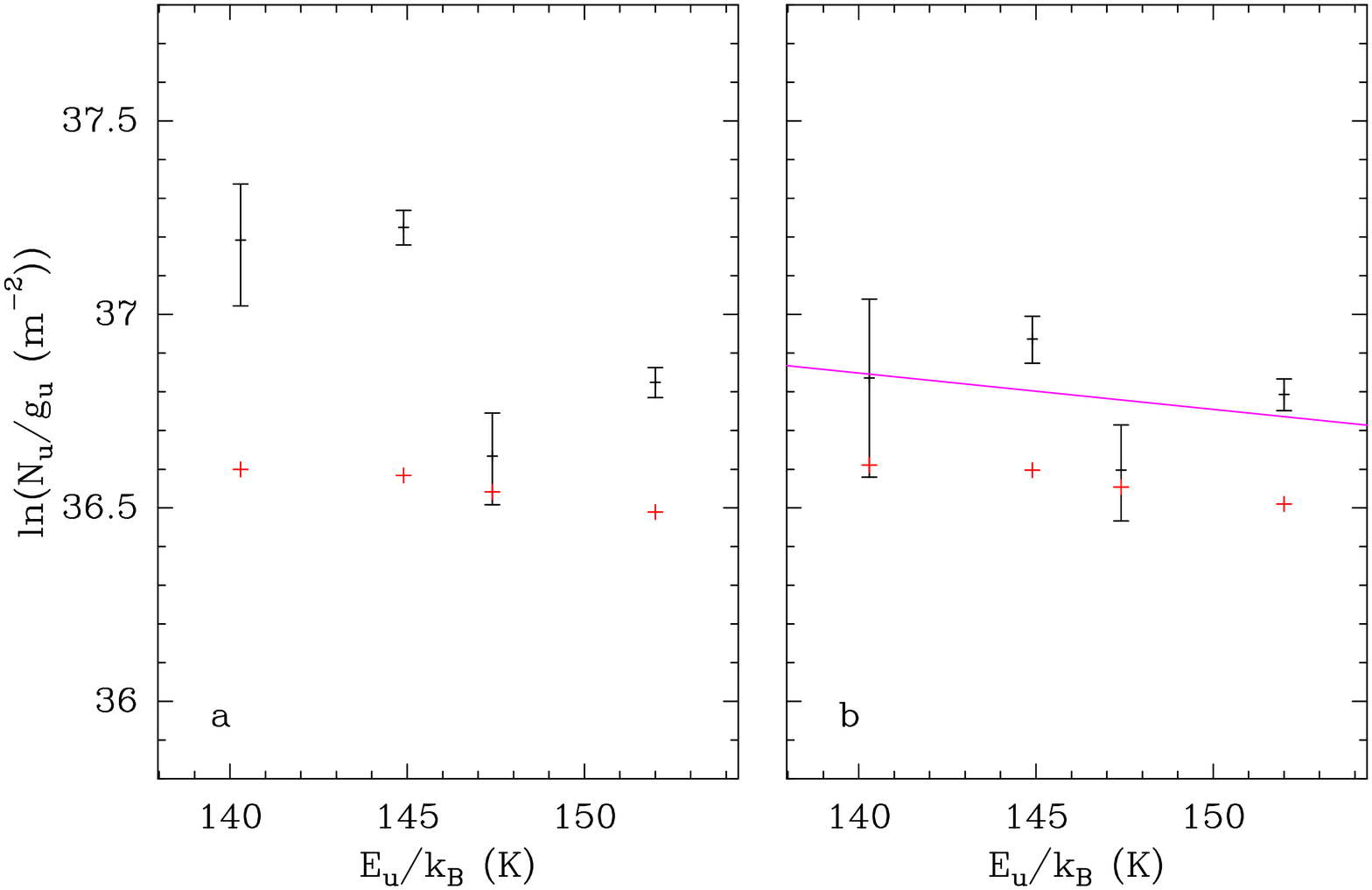}}}
\caption{Same as Fig.~\ref{f:popdiag_ch3oh} for \textit{a}-\textit{i}-C$_3$H$_7$OH.
}
\label{f:popdiag_c3h7oh-i-a}
\end{figure}

\begin{figure}
\centerline{\resizebox{0.95\hsize}{!}{\includegraphics[angle=0]{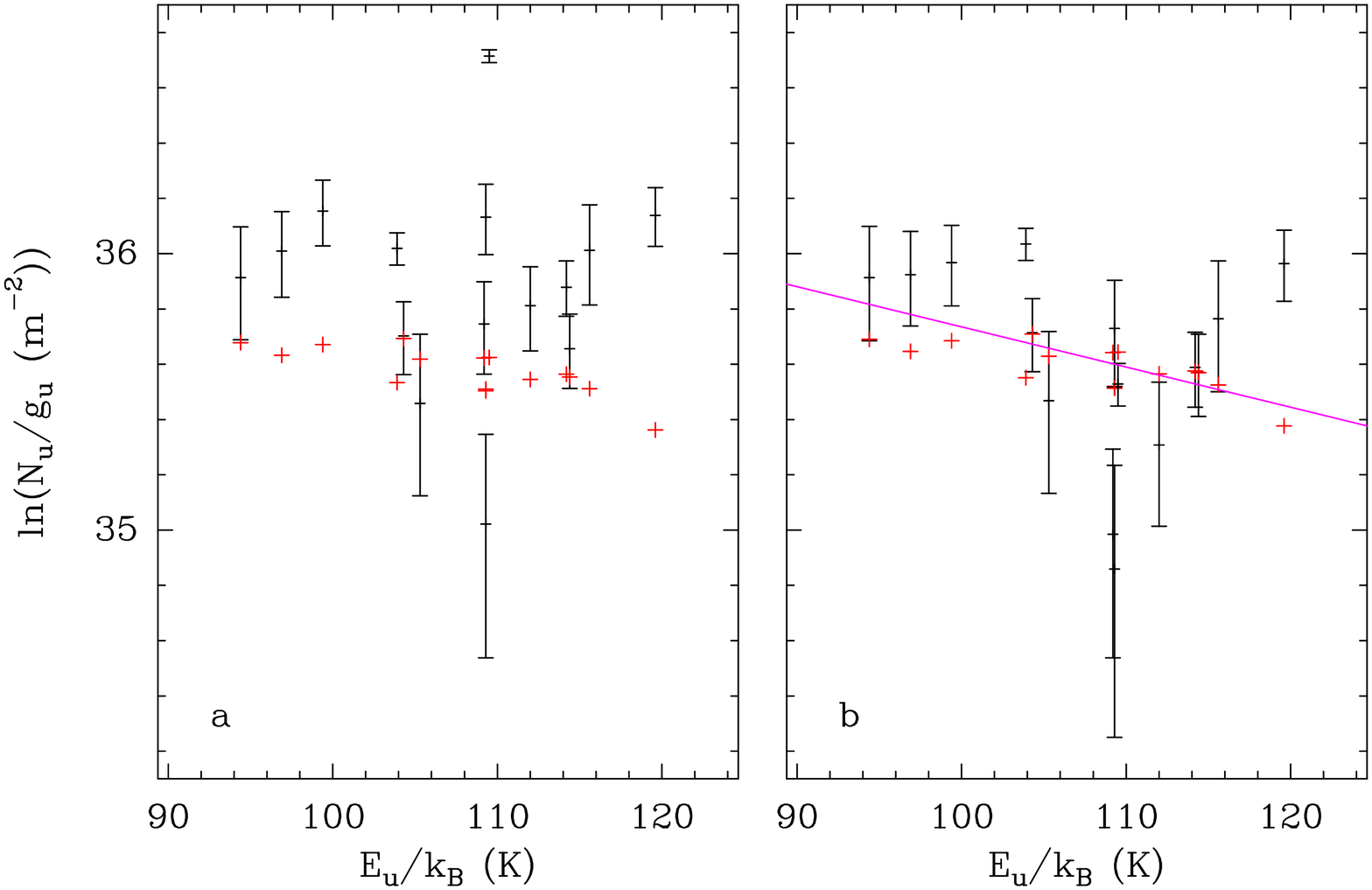}}}
\caption{Same as Fig.~\ref{f:popdiag_ch3oh} for \textit{Gg'}-\textit{n}-C$_3$H$_7$OH.
}
\label{f:popdiag_c3h7oh-n-Ggp}
\end{figure}

\begin{figure}
\centerline{\resizebox{0.95\hsize}{!}{\includegraphics[angle=0]{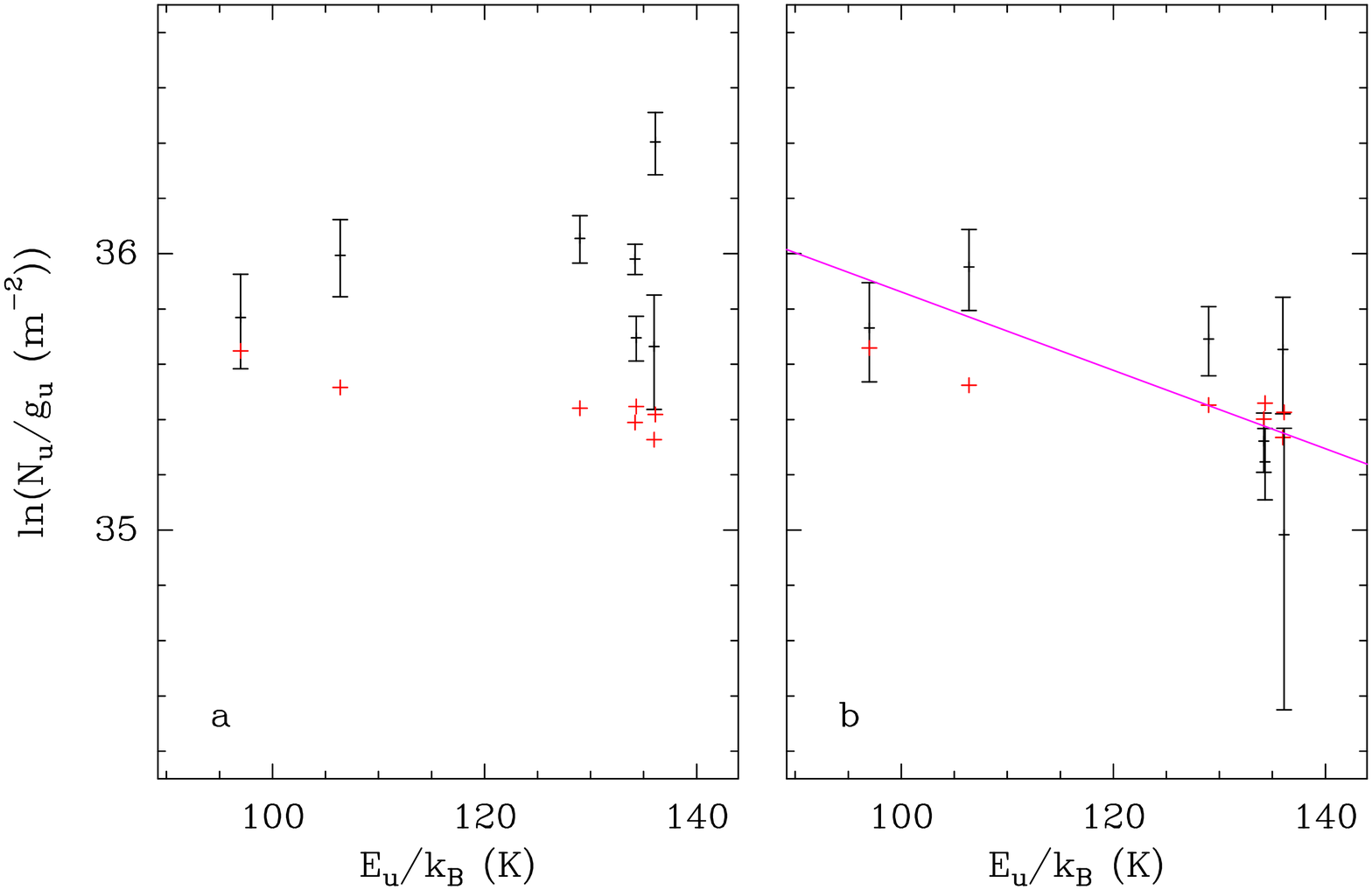}}}
\caption{Same as Fig.~\ref{f:popdiag_ch3oh} for \textit{Ag}-\textit{n}-C$_3$H$_7$OH.
}
\label{f:popdiag_c3h7oh-n-Ag}
\end{figure}

Figures~\ref{f:popdiag_ch3oh}--\ref{f:popdiag_c3h7oh-n-Ag} show the population 
diagrams of CH$_3$OH, C$_2$H$_5$OH, \textit{g}-\textit{i}-C$_3$H$_7$OH, 
\textit{a}-\textit{i}-C$_3$H$_7$OH, \textit{Gg'}-\textit{n}-C$_3$H$_7$OH, and
\textit{Ag}-\textit{n}-C$_3$H$_7$OH, respectively, toward Sgr~B2(N2b).

\clearpage

\clearpage
\section{Tables of chemical model input data}
\label{a:models}

\begin{table*}

\begin{center}
\caption{Grain-surface/ice-mantle reactions involved in the formation of vinyl, ethyl, propyl alcohols, as well as updated reactions related to propyl cyanide production. \citet{Garrod17} provide a more complete list of cyanide-related reactions.}
\label{tab-addition}
\renewcommand{\arraystretch}{1.0}
\small
\begin{tabular}[t]{rrclclcllllrl}
\hline \hline
\# & Reaction &&&&&&& BR & $F_{\mathrm{dir}}$ & $F_{\mathrm{comp}}$ & $E_{A}$ (K) & Ref. \\
\hline

 1&                                                   H  &  +  &                                          C$_2$H$_3$OH  &  $\rightarrow$  &  CH$_3\overset{\centerdot}{\mathrm{C}}$HOH             &   &       &--& 0.375 & 0.25 &  604 & a \\
 2&                                                   H  &  +  &                                          C$_2$H$_3$OH  &  $\rightarrow$  &  $\overset{\centerdot}{\mathrm{C}}$H$_2$CH$_2$OH       &   &       &--& 0.25  & 1.00 & 1850 & a \\
 3&                                          \"{C}H$_2$  &  +  &             $\overset{\centerdot}{\mathrm{C}}$H$_2$OH  &  $\rightarrow$  &  $\overset{\centerdot}{\mathrm{C}}$H$_2$CH$_2$OH       &   &       &--&--&--&--  \\
 4&             $\overset{\centerdot}{\mathrm{C}}$H$_3$  &  +  &             $\overset{\centerdot}{\mathrm{C}}$H$_2$OH  &  $\rightarrow$  &  C$_2$H$_5$OH                                          &   &       &--&--&--&--  \\
 5&       $\overset{\centerdot}{\mathrm{C}}$H$_2$CH$_3$  &  +  &                   $\overset{\centerdot}{\mathrm{O}}$H  &  $\rightarrow$  &  C$_2$H$_5$OH                                          &   &       & 0.5 &--&--&--  \\
 6&       $\overset{\centerdot}{\mathrm{C}}$H$_2$CH$_3$  &  +  &                   $\overset{\centerdot}{\mathrm{O}}$H  &  $\rightarrow$  &  C$_2$H$_4$                                            & + & H$_2$O& 0.5 &--&--&--  \\
 7&                 $\overset{\centerdot}{\mathrm{O}}$H  &  +  &                                           C$_2$H$_4$   &  $\rightarrow$  &  $\overset{\centerdot}{\mathrm{C}}$H$_2$CH$_2$OH       &   &       &--&--&--&    0 & b \\
 8&                 $\overset{\centerdot}{\mathrm{O}}$H  &  +  &                                           C$_2$H$_4$   &  $\rightarrow$  &  $\overset{\centerdot}{\mathrm{C}}$HCH$_2$             & + & H$_2$O&--&--&--& 2990 & c \\
 9&                                                   O  &  +  &         $\overset{\centerdot}{\mathrm{C}}$H$_2$CH$_3$  &  $\rightarrow$  &  C$_2$H$_5\overset{\centerdot}{\mathrm{O}}$            &   &       &--&--&--&--  \\
10&                                                   H  &  +  &             CH$_3\overset{\centerdot}{\mathrm{C}}$HOH  &  $\rightarrow$  &  C$_2$H$_3$OH                                          & + & H$_2$ & 0.5 &--&--&--  \\
11&                                                   H  &  +  &             CH$_3\overset{\centerdot}{\mathrm{C}}$HOH  &  $\rightarrow$  &  C$_2$H$_5$OH                                          &   &       & 0.5 &--&--&--  \\
12&                                                   H  &  +  &       $\overset{\centerdot}{\mathrm{C}}$H$_2$CH$_2$OH  &  $\rightarrow$  &  C$_2$H$_3$OH                                          & + & H$_2$ & 0.5 &--&--&--  \\
13&                                                   H  &  +  &       $\overset{\centerdot}{\mathrm{C}}$H$_2$CH$_2$OH  &  $\rightarrow$  &  C$_2$H$_5$OH                                          &   &       & 0.5 &--&--&--  \\
14&                                                   H  &  +  &            C$_2$H$_5\overset{\centerdot}{\mathrm{O}}$  &  $\rightarrow$  &  C$_2$H$_5$OH                                          &   &       &--&--&--&--  \\
\,\vspace{-2.25mm}\\
15&                                          \"{C}H$_2$  &  +  &       $\overset{\centerdot}{\mathrm{C}}$H$_2$CH$_2$OH  &  $\rightarrow$  &  $\overset{\centerdot}{\mathrm{C}}$H$_2$CH$_2$CH$_2$OH &   &       &--&--&--&--  \\
16&             $\overset{\centerdot}{\mathrm{C}}$H$_3$  &  +  &       $\overset{\centerdot}{\mathrm{C}}$H$_2$CH$_2$OH  &  $\rightarrow$  &  $n$-C$_3$H$_7$OH                                      &   &       &--&--&--&--  \\
17&       $\overset{\centerdot}{\mathrm{C}}$H$_2$CH$_3$  &  +  &             $\overset{\centerdot}{\mathrm{C}}$H$_2$OH  &  $\rightarrow$  &  $n$-C$_3$H$_7$OH                                      &   &       & 0.333 &--&--&--  \\
18&       $\overset{\centerdot}{\mathrm{C}}$H$_2$CH$_3$  &  +  &             $\overset{\centerdot}{\mathrm{C}}$H$_2$OH  &  $\rightarrow$  &  CH$_3$OH                                              & + &C$_2$H$_4$& 0.333 &--&--&--  \\
19&       $\overset{\centerdot}{\mathrm{C}}$H$_2$CH$_3$  &  +  &             $\overset{\centerdot}{\mathrm{C}}$H$_2$OH  &  $\rightarrow$  &  H$_2$CO                                               & + &C$_2$H$_6$& 0.333 &--&--&--  \\
20& $\overset{\centerdot}{\mathrm{C}}$H$_2$CH$_2$CH$_3$  &  +  &                   $\overset{\centerdot}{\mathrm{O}}$H  &  $\rightarrow$  &  $n$-C$_3$H$_7$OH                                      &   &       & 0.5 &--&--&--  \\
21& $\overset{\centerdot}{\mathrm{C}}$H$_2$CH$_2$CH$_3$  &  +  &                   $\overset{\centerdot}{\mathrm{O}}$H  &  $\rightarrow$  &  C$_3$H$_6$                                            & + & H$_2$O& 0.5 &--&--&--  \\
22&                                                   O  &  +  &   $\overset{\centerdot}{\mathrm{C}}$H$_2$CH$_2$CH$_3$  &  $\rightarrow$  &  CH$_3$CH$_2$CH$_2\overset{\centerdot}{\mathrm{O}}$    &   &       &--&--&--&--  \\
23&                                                   H  &  +  & $\overset{\centerdot}{\mathrm{C}}$H$_2$CH$_2$CH$_2$OH  &  $\rightarrow$  &  $n$-C$_3$H$_7$OH                                      &   &       &--&--&--&--  \\
24&                                                   H  &  +  &       CH$_3\overset{\centerdot}{\mathrm{C}}$HCH$_2$OH  &  $\rightarrow$  &  $n$-C$_3$H$_7$OH                                      &   &       &--&--&--&--  \\
25&                                                   H  &  +  &       CH$_3$CH$_2\overset{\centerdot}{\mathrm{C}}$HOH  &  $\rightarrow$  &  $n$-C$_3$H$_7$OH                                      &   &       &--&--&--&--  \\
26&                                                   H  &  +  &    CH$_3$CH$_2$CH$_2\overset{\centerdot}{\mathrm{O}}$  &  $\rightarrow$  &  $n$-C$_3$H$_7$OH                                      &   &       &--&--&--&--  \\
\,\vspace{-2.25mm}\\
27&                                          \"{C}H$_2$  &  +  &             CH$_3\overset{\centerdot}{\mathrm{C}}$HOH  &  $\rightarrow$  &  $\overset{\centerdot}{\mathrm{C}}$H$_2$CH(OH)CH$_3$   &   &       &--&--&--&--  \\
28&             $\overset{\centerdot}{\mathrm{C}}$H$_3$  &  +  &             CH$_3\overset{\centerdot}{\mathrm{C}}$HOH  &  $\rightarrow$  &  $i$-C$_3$H$_7$OH                                      &   &       &--&--&--&--  \\
29&       CH$_3\overset{\centerdot}{\mathrm{C}}$HCH$_3$  &  +  &                   $\overset{\centerdot}{\mathrm{O}}$H  &  $\rightarrow$  &  $i$-C$_3$H$_7$OH                                      &   &       & 0.5 &--&--&--  \\
30&       CH$_3\overset{\centerdot}{\mathrm{C}}$HCH$_3$  &  +  &                   $\overset{\centerdot}{\mathrm{O}}$H  &  $\rightarrow$  &  C$_3$H$_6$                                            & + & H$_2$O& 0.5 &--&--&--  \\
31&                                                   O  &  +  &         CH$_3\overset{\centerdot}{\mathrm{C}}$HCH$_3$  &  $\rightarrow$  &  CH$_3$CH($\overset{\centerdot}{\mathrm{O}}$)CH$_3$    &   &       &--&--&--&--  \\
32&                                                   H  &  +  &   $\overset{\centerdot}{\mathrm{C}}$H$_2$CH(OH)CH$_3$  &  $\rightarrow$  &  $i$-C$_3$H$_7$OH                                      &   &       &--&--&--&--  \\
33&                                                   H  &  +  &      CH$_3\overset{\centerdot}{\mathrm{C}}$(OH)CH$_3$  &  $\rightarrow$  &  $i$-C$_3$H$_7$OH                                      &   &       &--&--&--&--  \\
34&                                                   H  &  +  &    CH$_3$CH($\overset{\centerdot}{\mathrm{O}}$)CH$_3$  &  $\rightarrow$  &  $i$-C$_3$H$_7$OH                                      &   &       &--&--&--&--  \\
\,\vspace{-2.25mm}\\
35&                 $\overset{\centerdot}{\mathrm{C}}$N  &  +  &                                            C$_3$H$_6$  &  $\rightarrow$  &  CH$_3\overset{\centerdot}{\mathrm{C}}$HCH$_2$CN       &   &       &--& 0.375 & -- &    0 & d \\
36&                 $\overset{\centerdot}{\mathrm{C}}$N  &  +  &                                            C$_3$H$_6$  &  $\rightarrow$  &  $\overset{\centerdot}{\mathrm{C}}$H$_2$CH(CN)CH$_3$   &   &       &--& 0.25  & -- &    0 & d \\
37&                 $\overset{\centerdot}{\mathrm{C}}$N  &  +  &                                            C$_3$H$_6$  &  $\rightarrow$  &  C$_3$H$_5$                                            & + &   HCN &--& 0.375 & -- &    0 & d \\
\,\vspace{-2.25mm}\\

38& $\overset{\centerdot}{\mathrm{C}}$H$_2$  &  +  &   C$_2$H$_5$OH  &  $\rightarrow$  &  $\overset{\centerdot}{\mathrm{C}}$H$_2$CH$_2$OH      & + &  CH$_3$  &--& 0.1875 & 0.25 & 3600 & Est. \\
39& $\overset{\centerdot}{\mathrm{C}}$H$_2$  &  +  &   C$_2$H$_5$OH  &  $\rightarrow$  &  $n$-C$_3$H$_7$OH                                     &   &          &--& 0.1875 & 0.25 & 3600 & Est. \\
40& $\overset{\centerdot}{\mathrm{C}}$H$_2$  &  +  &   C$_2$H$_5$OH  &  $\rightarrow$  &  CH$_3\overset{\centerdot}{\mathrm{C}}$HOH            & + &  CH$_3$  &--& 0.125  & 1.00 & 3760 & Est. \\
41& $\overset{\centerdot}{\mathrm{C}}$H$_2$  &  +  &   C$_2$H$_5$OH  &  $\rightarrow$  &  $i$-C$_3$H$_7$OH                                     &   &          &--& 0.125  & 1.00 & 3760 & Est. \\
\,\vspace{-2.25mm}\\

42& $\overset{\centerdot}{\mathrm{C}}$H$_2$  &  +  &   C$_2$H$_5$CN  &  $\rightarrow$  &  $\overset{\centerdot}{\mathrm{C}}$H$_2$CH$_2$CN      & + &  CH$_3$  &--& 0.1875 & 0.25 & 3600 & Est. \\
43& $\overset{\centerdot}{\mathrm{C}}$H$_2$  &  +  &   C$_2$H$_5$CN  &  $\rightarrow$  &  $n$-C$_3$H$_7$CN                                     &   &          &--& 0.1875 & 0.25 & 3600 & Est. \\
44& $\overset{\centerdot}{\mathrm{C}}$H$_2$  &  +  &   C$_2$H$_5$CN  &  $\rightarrow$  &  CH$_3\overset{\centerdot}{\mathrm{C}}$HCN            & + &  CH$_3$  &--& 0.125  & 1.00 & 3760 & Est. \\
45& $\overset{\centerdot}{\mathrm{C}}$H$_2$  &  +  &   C$_2$H$_5$CN  &  $\rightarrow$  &  $i$-C$_3$H$_7$CN                                     &   &          &--& 0.125  & 1.00 & 3760 & Est. \\
\hline
\end{tabular}
\end{center}
\tablefoot{
Dots indicate which atom in a molecule hosts the radical site (i.e. an unpaired electron), where appropriate. Literature references refer to $E_A$ values; $F_{\mathrm{dir}}$ and $F_{\mathrm{comp}}$ values are our own estimates. Branching ratios (BR) are used in case of multiple branches for barrierless, radical-radical reactions. For barrier-mediated reactions, the activation energies and related quantities determine the branching. ``Est.'' indicates a barrier estimate by the authors based on other reactions in the list. 
Dashes indicate an assumed activation energy barrier of zero, or the default values for $F_{\mathrm{dir}}$ and $F_{\mathrm{comp}}$ (i.e. unity). 
\tablefoottext{a}{\citet{Rao11};}
\tablefoottext{b}{\citet{Atkinson97};}
\tablefoottext{c}{\citet{Baulch92};}
\tablefoottext{d}{Based on fit to \citet{Gannon07} gas-phase data, as per \citet{Garrod17}.}
}
\end{table*}

\begin{table*}

\begin{center}
\caption{Selected grain-surface/ice-mantle hydrogen-abstraction and barrier-mediated reactions involved in the formation of vinyl, ethyl, propyl alcohols, as well as updated reactions related to propyl cyanide production. \citet{Garrod17} provide a more complete list of cyanide-related reactions.}
\label{tab-abstraction}
\renewcommand{\arraystretch}{1.0}
\small
\begin{tabular}[t]{rrclclclllrl}
\hline \hline
\# & Reaction &&&&&&& $F_{\mathrm{dir}}$ & $F_{\mathrm{comp}}$ & $E_{A}$ (K) & Ref. \\
\hline
46&                                   H  &  +  &     C$_3$H$_6$  &  $\rightarrow$  &  $\overset{\centerdot}{\mathrm{C}}$H$_2$CH$_2$CH$_3$  &   &          & 0.375 & 0.25 & 1320 & a \\
47&                                   H  &  +  &     C$_3$H$_6$  &  $\rightarrow$  &  CH$_3\overset{\centerdot}{\mathrm{C}}$HCH$_3$        &   &          & 0.25  & 1    &  619 & a \\
48&                                   H  &  +  &     C$_3$H$_6$  &  $\rightarrow$  &  C$_3$H$_5$                                           & + & H$_2$    & 0.375 & 0.25 & 2930 & b \\

49& $\overset{\centerdot}{\mathrm{O}}$H  &  +  &     C$_3$H$_6$  &  $\rightarrow$  &  CH$_3\overset{\centerdot}{\mathrm{C}}$HCH$_2$OH      &   &          & 0.375 & --   &    0 & c \\
50& $\overset{\centerdot}{\mathrm{O}}$H  &  +  &     C$_3$H$_6$  &  $\rightarrow$  &  $\overset{\centerdot}{\mathrm{C}}$H$_2$CH(OH)CH$_3$  &   &          & 0.25  & --   &    0 & d \\
51& $\overset{\centerdot}{\mathrm{O}}$H  &  +  &     C$_3$H$_6$  &  $\rightarrow$  &  C$_3$H$_5$                                           & + & H$_2$O   & 0.375 & 1    &  730 & e \\
\,\vspace{-2.25mm}\\

52&                                   H  &  +  &   C$_3$H$_8$    &  $\rightarrow$  &  $\overset{\centerdot}{\mathrm{C}}$H$_2$CH$_2$CH$_3$  & + &  H$_2$   & 0.75  & 0.25 & 4720 & f  \\
53&                                   H  &  +  &   C$_3$H$_8$    &  $\rightarrow$  &  CH$_3\overset{\centerdot}{\mathrm{C}}$HCH$_3$        & + &  H$_2$   & 0.25  & 1    & 4000 & f  \\

54& $\overset{\centerdot}{\mathrm{O}}$H  &  +  &   C$_3$H$_8$    &  $\rightarrow$  &  $\overset{\centerdot}{\mathrm{C}}$H$_2$CH$_2$CH$_3$  & + & H$_2$O   & 0.75  & 0.25 & 1310 & g \\
55& $\overset{\centerdot}{\mathrm{O}}$H  &  +  &   C$_3$H$_8$    &  $\rightarrow$  &  CH$_3\overset{\centerdot}{\mathrm{C}}$HCH$_3$        & + & H$_2$O   & 0.25  & 1    & 1120 & g \\
\,\vspace{-2.25mm}\\

56&                                   H  &  +  &   C$_2$H$_5$OH  &  $\rightarrow$  &  $\overset{\centerdot}{\mathrm{C}}$H$_2$CH$_2$OH      & + &  H$_2$   & 0.375 & 0.25 & 3770 & h  \\
57&                                   H  &  +  &   C$_2$H$_5$OH  &  $\rightarrow$  &  CH$_3\overset{\centerdot}{\mathrm{C}}$HOH            & + &  H$_2$   & 0.25  & 1    & 2710 & i  \\
58&                                   H  &  +  &   C$_2$H$_5$OH  &  $\rightarrow$  &  C$_2$H$_5\overset{\centerdot}{\mathrm{O}}$           & + &  H$_2$   & 0.375 & 0.25 & 4380 & h  \\

59& $\overset{\centerdot}{\mathrm{O}}$H  &  +  &   C$_2$H$_5$OH  &  $\rightarrow$  &  $\overset{\centerdot}{\mathrm{C}}$H$_2$CH$_2$OH      & + &  H$_2$O  & 0.375 & 0.25 &  889 & E-P  \\
60& $\overset{\centerdot}{\mathrm{O}}$H  &  +  &   C$_2$H$_5$OH  &  $\rightarrow$  &  CH$_3\overset{\centerdot}{\mathrm{C}}$HOH            & + &  H$_2$O  & 0.25  & 1    &   70 & j    \\
61& $\overset{\centerdot}{\mathrm{O}}$H  &  +  &   C$_2$H$_5$OH  &  $\rightarrow$  &  C$_2$H$_5\overset{\centerdot}{\mathrm{O}}$           & + &  H$_2$O  & 0.375 & 0.25 & 1510 & E-P  \\
\,\vspace{-2.25mm}\\

62&                                   H  &  +  &   C$_2$H$_5$CN  &  $\rightarrow$  &  $\overset{\centerdot}{\mathrm{C}}$H$_2$CH$_2$CN      & + &  H$_2$   & 0.375 & 0.25 & 4720 & Est. \\
63&                                   H  &  +  &   C$_2$H$_5$CN  &  $\rightarrow$  &  CH$_3\overset{\centerdot}{\mathrm{C}}$HCN            & + &  H$_2$   & 0.25  & 1    & 4000 & Est. \\

64& $\overset{\centerdot}{\mathrm{O}}$H  &  +  &   C$_2$H$_5$CN  &  $\rightarrow$  &  $\overset{\centerdot}{\mathrm{C}}$H$_2$CH$_2$CN      & + &  H$_2$O  & 0.375 & 0.25 & 1790 & k  \\
65& $\overset{\centerdot}{\mathrm{O}}$H  &  +  &   C$_2$H$_5$CN  &  $\rightarrow$  &  CH$_3\overset{\centerdot}{\mathrm{C}}$HCN            & + &  H$_2$O  & 0.25  & 1    & 1060 & k  \\
\,\vspace{-2.25mm}\\

66&                                   H  &  +  &   C$_3$H$_7$OH  &  $\rightarrow$  & $\overset{\centerdot}{\mathrm{C}}$H$_2$CH$_2$CH$_2$OH & + &  H$_2$   & 0.3   & 0.25 & 3710 & E-P  \\
67&                                   H  &  +  &   C$_3$H$_7$OH  &  $\rightarrow$  &  CH$_3\overset{\centerdot}{\mathrm{C}}$HCH$_2$OH      & + &  H$_2$   & 0.2   & 1    & 3080 & E-P  \\
68&                                   H  &  +  &   C$_3$H$_7$OH  &  $\rightarrow$  &  CH$_3$CH$_2\overset{\centerdot}{\mathrm{C}}$HOH      & + &  H$_2$   & 0.2   & 1    & 3080 & E-P  \\
69&                                   H  &  +  &   C$_3$H$_7$OH  &  $\rightarrow$  &  C$_3$H$_7\overset{\centerdot}{\mathrm{O}}$           & + &  H$_2$   & 0.3   & 0.25 & 4780 & E-P  \\

70& $\overset{\centerdot}{\mathrm{O}}$H  &  +  &   C$_3$H$_7$OH  &  $\rightarrow$  & $\overset{\centerdot}{\mathrm{C}}$H$_2$CH$_2$CH$_2$OH & + &  H$_2$O  & 0.3   & 0.25 &  522 & E-P  \\
71& $\overset{\centerdot}{\mathrm{O}}$H  &  +  &   C$_3$H$_7$OH  &  $\rightarrow$  &  CH$_3\overset{\centerdot}{\mathrm{C}}$HCH$_2$OH      & + &  H$_2$O  & 0.2   & 1    &  125 & E-P  \\
72& $\overset{\centerdot}{\mathrm{O}}$H  &  +  &   C$_3$H$_7$OH  &  $\rightarrow$  &  CH$_3$CH$_2\overset{\centerdot}{\mathrm{C}}$HOH      & + &  H$_2$O  & 0.2   & 1    &  125 & E-P  \\
73& $\overset{\centerdot}{\mathrm{O}}$H  &  +  &   C$_3$H$_7$OH  &  $\rightarrow$  &  C$_3$H$_7\overset{\centerdot}{\mathrm{O}}$           & + &  H$_2$O  & 0.3   & 0.25 & 1210 & E-P  \\
\,\vspace{-2.25mm}\\

74&                                   H  &  +  &$i$-C$_3$H$_7$OH &  $\rightarrow$  &  $\overset{\centerdot}{\mathrm{C}}$H$_2$CH(OH)CH$_3$  & + &  H$_2$   & 0.6   & 0.25 & 3710 & E-P  \\
75&                                   H  &  +  &$i$-C$_3$H$_7$OH &  $\rightarrow$  &  CH$_3\overset{\centerdot}{\mathrm{C}}$(OH)CH$_3$     & + &  H$_2$   & 0.1   & 1    & 3080 & E-P  \\
76&                                   H  &  +  &$i$-C$_3$H$_7$OH &  $\rightarrow$  &  CH$_3$CH($\overset{\centerdot}{\mathrm{O}}$)CH$_3$   & + &  H$_2$   & 0.3   & 0.25 & 6080 & E-P  \\

77& $\overset{\centerdot}{\mathrm{O}}$H  &  +  &$i$-C$_3$H$_7$OH &  $\rightarrow$  &  $\overset{\centerdot}{\mathrm{C}}$H$_2$CH(OH)CH$_3$  & + &  H$_2$O  & 0.6   & 0.25 &  522 & E-P  \\
78& $\overset{\centerdot}{\mathrm{O}}$H  &  +  &$i$-C$_3$H$_7$OH &  $\rightarrow$  &  CH$_3\overset{\centerdot}{\mathrm{C}}$(OH)CH$_3$     & + &  H$_2$O  & 0.1   & 1    &  125 & E-P  \\
79& $\overset{\centerdot}{\mathrm{O}}$H  &  +  &$i$-C$_3$H$_7$OH &  $\rightarrow$  &  CH$_3$CH($\overset{\centerdot}{\mathrm{O}}$)CH$_3$   & + &  H$_2$O  & 0.3   & 0.25 & 2040 & E-P  \\
\,\vspace{-2.25mm}\\

80&                                   H  &  +  &   C$_3$H$_7$CN  &  $\rightarrow$  & $\overset{\centerdot}{\mathrm{C}}$H$_2$CH$_2$CH$_2$CN & + &  H$_2$   & 0.3   & 0.25 & 3710 & E-P  \\
81&                                   H  &  +  &   C$_3$H$_7$CN  &  $\rightarrow$  &  CH$_3\overset{\centerdot}{\mathrm{C}}$HCH$_2$CN      & + &  H$_2$   & 0.2   & 1    & 3080 & E-P  \\
82&                                   H  &  +  &   C$_3$H$_7$CN  &  $\rightarrow$  &  CH$_3$CH$_2\overset{\centerdot}{\mathrm{C}}$HCN      & + &  H$_2$   & 0.2   & 1    & 3080 & E-P  \\

83& $\overset{\centerdot}{\mathrm{O}}$H  &  +  &   C$_3$H$_7$CN  &  $\rightarrow$  & $\overset{\centerdot}{\mathrm{C}}$H$_2$CH$_2$CH$_2$CN & + &  H$_2$O  & 0.3   & 0.25 & 1000 & Est. \\
84& $\overset{\centerdot}{\mathrm{O}}$H  &  +  &   C$_3$H$_7$CN  &  $\rightarrow$  &  CH$_3\overset{\centerdot}{\mathrm{C}}$HCH$_2$CN      & + &  H$_2$O  & 0.2   & 1    &  800 & Est. \\
85& $\overset{\centerdot}{\mathrm{O}}$H  &  +  &   C$_3$H$_7$CN  &  $\rightarrow$  &  CH$_3$CH$_2\overset{\centerdot}{\mathrm{C}}$HCN      & + &  H$_2$O  & 0.2   & 1    &  800 & Est. \\
\,\vspace{-2.25mm}\\

86&                                   H  &  +  &$i$-C$_3$H$_7$CN &  $\rightarrow$  &  $\overset{\centerdot}{\mathrm{C}}$H$_2$CH(CN)CH$_3$  & + &  H$_2$   & 0.6   & 0.25 & 3710 & E-P  \\
87&                                   H  &  +  &$i$-C$_3$H$_7$CN &  $\rightarrow$  &  CH$_3\overset{\centerdot}{\mathrm{C}}$(CN)CH$_3$     & + &  H$_2$   & 0.1   & 1    & 3080 & E-P  \\

88& $\overset{\centerdot}{\mathrm{O}}$H  &  +  &$i$-C$_3$H$_7$CN &  $\rightarrow$  &  $\overset{\centerdot}{\mathrm{C}}$H$_2$CH(CN)CH$_3$  & + &  H$_2$O  & 0.6   & 0.25 & 1000 & Est. \\
89& $\overset{\centerdot}{\mathrm{O}}$H  &  +  &$i$-C$_3$H$_7$CN &  $\rightarrow$  &  CH$_3\overset{\centerdot}{\mathrm{C}}$(CN)CH$_3$     & + &  H$_2$O  & 0.1   & 1    &  800 & Est. \\

\hline
\end{tabular}
\end{center}
\tablefoot{
Dots indicate which atom in a molecule hosts the radical site (i.e. an unpaired electron), where appropriate. Literature references refer to $E_A$ values; $F_{\mathrm{dir}}$ and $F_{\mathrm{comp}}$ values are our own estimates. ``E-P'' indicates an activation energy calculated using the Evans-Polanyi relation; see \citet{Garrod13}. ``Est.'' indicates a barrier estimate by the authors based on other reactions in the list.
\tablefoottext{a}{\citet{Curran06};}
\tablefoottext{b}{\citet{Tsang92};}
\tablefoottext{c}{\citet{Atkinson97};}
\tablefoottext{d}{\citet{Thomsen09};}
\tablefoottext{e}{\citet{Tsang91};}
\tablefoottext{f}{\citet{Baldwin79};}
\tablefoottext{g}{\citet{Hu97};}
\tablefoottext{h}{\citet{Sivaramakrishnan10};}
\tablefoottext{i}{\citet{Olm16};}
\tablefoottext{j}{\citet{Atkinson01};}
\tablefoottext{k}{based on fitting to data from \citet{Sun08}.}
}
\end{table*}

\begin{table*}
\begin{center}
\caption{Physical quantities of new or related chemical species.}
\label{tab-quantities}
\renewcommand{\arraystretch}{1.0}
\small
\begin{tabular}[t]{lrrl}
\hline \hline
Species                                                & $E_{\mathrm{des}}$ & $\Delta H_{f}$(298 K) & Notes \\
                                                       &    (K)                                &  (kcal mol$^{-1}$) \\
\hline
H$_2$O                                                 &     4815      &      --57.80     & \smallskip $E_{\mathrm{des}}$: Jin et al. (in prep.) \\

CH$_3$OH                                               &     5534      &      --48.00     & \smallskip $E_{\mathrm{des}}$: \citet{Garrod08}, \cite{Garrod13} \\

$\overset{\centerdot}{\mathrm{C}}$H$_2$CH$_2$OH        &     4950      &       --5.70     & \\
CH$_3\overset{\centerdot}{\mathrm{C}}$HOH              &     4950      &      --12.91     & \\
C$_2$H$_5$OH                                           &     5400      &      --56.23     & \smallskip $E_{\mathrm{des}}$: Jin et al. (in prep.) \\

$\overset{\centerdot}{\mathrm{C}}$H$_2$CH$_2$CH$_2$OH  &     5675      &      --12.28     & $\Delta H_{f}$ based on C$_3$H$_8$ -- $\overset{\centerdot}{\mathrm{C}}$H$_2$CH$_2$CH$_3$ \\
CH$_3\overset{\centerdot}{\mathrm{C}}$HCH$_2$OH        &     5675      &      --14.18     & $\Delta H_{f}$ based on C$_3$H$_8$ -- CH$_3\overset{\centerdot}{\mathrm{C}}$HCH$_3$  \\
CH$_3$CH$_2\overset{\centerdot}{\mathrm{C}}$HOH        &     5675      &      --14.18     & $\Delta H_{f}$ based on C$_3$H$_8$ -- CH$_3\overset{\centerdot}{\mathrm{C}}$HCH$_3$ \\
CH$_3$CH$_2$CH$_2\overset{\centerdot}{\mathrm{O}}$     &     5462      &       --9.00     & \smallskip \\
$n$-C$_3$H$_7$OH                                       &     6125      &      --61.20     & \smallskip \\

$\overset{\centerdot}{\mathrm{C}}$H$_2$CH(OH)CH$_3$    &     5675      &      --16.27     & $\Delta H_{f}$ based on C$_3$H$_8$ -- $\overset{\centerdot}{\mathrm{C}}$H$_2$CH$_2$CH$_3$  \\
CH$_3\overset{\centerdot}{\mathrm{C}}$(OH)CH$_3$       &     5675      &      --18.17     & $\Delta H_{f}$ based on C$_3$H$_8$ -- CH$_3\overset{\centerdot}{\mathrm{C}}$HCH$_3$  \\
CH$_3$CH($\overset{\centerdot}{\mathrm{O}}$)CH$_3$     &     5462      &       --9.00     & $\Delta H_{f}$ based on $n$-C$_3$H$_7\overset{\centerdot}{\mathrm{O}}$ \smallskip \\
$i$-C$_3$H$_7$OH                                       &     6125      &      --65.19     & \smallskip \\

$\overset{\centerdot}{\mathrm{C}}$H$_2$CH$_2$CH$_3$    &     5637      &       +23.90     & $E_{\mathrm{des}}$: \citet{Garrod17} \\
CH$_3\overset{\centerdot}{\mathrm{C}}$HCH$_3$          &     5637      &       +22.00     & $E_{\mathrm{des}}$: \citet{Garrod17} \\                       
C$_3$H$_8$                                             &     6087      &      --25.02     & \smallskip\smallskip $E_{\mathrm{des}}$: \citet{Garrod17} \\

CH$_3$CN                                               &     6150      &       +17.70     & $E_{\mathrm{des}}$: \citet{Bertin17}; used by \citet{Garrod22} \\
C$_2$H$_5$CN                                           &     6875      &       +12.71     & $E_{\mathrm{des}}$: \citet{Garrod22} \\
$n$-C$_3$H$_7$CN                                       &     7600      &        +7.46     & $E_{\mathrm{des}}$: \citet{Garrod22} \\
$i$-C$_3$H$_7$CN                                       &     7600      &        +5.44     & $E_{\mathrm{des}}$: \citet{Garrod22} \\
\hline
\end{tabular}
\end{center}
\tablefoot{
$E_{\mathrm{des}}$ is the desorption energy and $\Delta H_{f}$ the enthalpy of 
formation.
Dots indicate which atom in a molecule hosts the radical site (i.e. an unpaired electron), where appropriate. As in previous models, binding energies are representative of physisorption on an amorphous water ice surface.  Enthalpies of formation are obtained from the NIST WebBook database; where not available, estimates were adopted as described in the Notes column.}
\end{table*}


\end{appendix}

\end{document}